\pdfoutput=1
\documentclass[9pt,twocolumn,letterpaper]{extarticle}
\usepackage{sweepannarxiv}
\newcommand{\TopK}{\operatorname{TopK}}
\hypersetup{pdftitle={SweepANN: A GPU-Accelerated Large-K Approximate Vector Search for Drug Discovery},pdfauthor={Zhenyu Zhang, Qi Lin, Nan Zhang, Zhichao Cao}}
\begin{document}
\titleblock{SweepANN: A GPU-Accelerated Large-$K$ Approximate Vector Search for Drug Discovery}%
{Zhenyu Zhang$^{1}$ \qquad Qi Lin$^{1}$ \qquad Nan Zhang$^{2}$ \qquad Zhichao Cao$^{1}$\\[3pt]
{\small $^{1}$Arizona State University, Tempe, AZ, USA \qquad $^{2}$Independent Researcher}\\[2pt]
{\small\texttt{zzhan641@asu.edu} \quad \texttt{qlin36@asu.edu} \quad \texttt{nanzhang1213@gmail.com} \quad \texttt{Zhichao.Cao@asu.edu}}}%
{DrugCLIP places small molecules and protein pockets in one embedding space, turning genome-wide drug discovery into large-$K$ retrieval by vector proximity. A campaign fixes over 20K pockets in advance for one library of over 500M molecules. Each pocket keeps $K=500$K candidates, 0.1\% of the collection, versus 100 in ANN benchmarks. Missed hits never get tested, so mean recall must reach 99\% relative to exact search. Large answers make IVF probe every list holding a result and graph search track $K$ or more candidates, so costs grow with $K$. On our CPUs and GPUs, neither index family reaches the target without losing most of its edge over scanning.

SweepANN partitions compressed vectors across GPUs and scans each resident partition. One matrix multiplication scores each block for a whole query batch, so each block is read a single time per batch. Several plans perform the $K$-dependent selection and merging of each query's $K$ answers, each returning an exact top\nobreakdash-$K$ under compressed scoring. SweepANN fixes batch size for block reuse and runs the fastest measured plan whose state fits beside the library for the requested $K$. The window plan retains each query's best $K$ from sampled blocks as its seed. GPUs exchange each seed's minimum score once, and only rows above this cutoff enter fixed-capacity buffers. Overflow triggers exact recomputation, and CUDA Graphs replay the scan on every GPU.

Four NVIDIA RTX PRO 6000 Blackwell GPUs answer all 26,562 pockets with a resident time of about 61 seconds, excluding library loading, and 99.6\% mean recall relative to exact FP32 search. Our library follows DrugCLIP's published construction. DrugCLIP reports roughly 24 hours with eight A100 GPUs in its own campaign over its own library. On 500M rows, the window plan runs \textbf{1.41\texttimes} faster than SweepANN's grouped plan, a single selection within the same memory. Across 25M--500M rows and $K$ of 10K--500K, all GPU IVF configurations meeting the recall target take \textbf{8.2--24\texttimes} SweepANN's resident time. Every tested CAGRA configuration misses the target on 25M rows at $K\ge100$K. The plans serve two public vector datasets and two more embedding models. On DEEP1B's 1B vectors across eight GPUs, the baseline is SweepANN's exact full-score scan of uncompressed FP32 rows. At $K=100$K, the window plan cuts resident time by 5.4\texttimes{} and per-GPU peak memory by 6.0\texttimes{}, losing 1.0 point of mean recall. At $K=500$K, the factors are 3.2\texttimes{} and 5.3\texttimes{} for a 0.7-point loss.}
\section{Introduction}
\begin{figure}[t]
\centering
\includegraphics{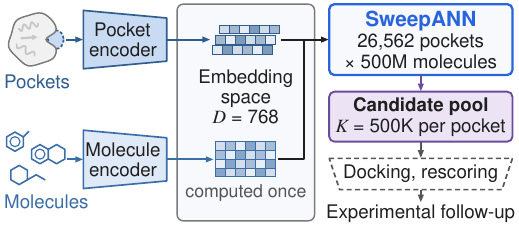}
\caption{Vector retrieval in virtual screening. Learned encoders map protein pockets and small molecules into one 768-dimensional embedding space~\cite{gao2023drugclip,jia2026genome}, and the molecule vectors are computed once and reused. SweepANN keeps the top $K$ molecules of each pocket query as a candidate pool, which later stages such as docking and rescoring narrow for experimental follow-up. Drawings are schematic.}
\Description{Pockets and molecules pass through two encoders into vectors of 768 dimensions. SweepANN searches 500M molecule vectors for each of 26,562 pocket queries and keeps 500K molecules per pocket, and the candidate pool goes on to docking and rescoring.}
\label{fig:screening-pipeline}
\end{figure}

Vector search is becoming a central operation in drug discovery. A key step searches libraries of billions of molecules for candidates likely to bind a disease-relevant protein. Docking predicts how each molecule fits into the protein's binding site, and evaluating every molecule this way is computationally expensive~\cite{lyu2019ultralarge,gorgulla2020virtualflow}. Fast learning-based screening therefore ranks the candidates first and selects a smaller pool for downstream evaluation. For example, DrugCLIP maps protein pockets and candidate molecules into a shared
embedding space and retrieves promising molecules for each pocket
through vector similarity search~\cite{gao2023drugclip}
(Figure~\ref{fig:screening-pipeline}).
Its genome-wide screening campaign searched more than 500M
molecular vectors for each of more than 20K
pockets~\cite{jia2026genome}.
This exhaustive search took about 24 hours on eight A100
GPUs~\cite{jia2026genome,genomescreenCard}. Target pockets and research
priorities change, and every repeated campaign pays this cost again.

Screening also expects more from vector search than approximate nearest-neighbor (ANN) search commonly delivers. Where ANN search usually returns the top $K=10$ or 100 most similar vectors per query, screening keeps a broad candidate pool for later docking, rescoring, and clustering. A molecule left out of the pool is never tested, so our workload keeps $K=500$K molecules per pocket, 0.1\% of the library, at 99\% mean recall or better. The same library serves many campaigns (Section~\ref{sec:problem}), and libraries keep growing. ZINC-22 reported over 37B purchasable compounds, over 4.5B of them in ready-to-dock 3D~\cite{tingle2023zinc22}. At a fixed fraction of the library, the answers grow with it.

ANN indexes are the established answer to fast vector search, and for most workloads they are the right one. A graph index such as HNSW or CAGRA, or an inverted-file (IVF) index, visits a small part of the library per query. It answers in milliseconds and is reused across queries~\cite{malkov2020hnsw,ootomo2024cagra,johnson2019billion}. Our measurements confirm these advantages where answers are small. At $K=100$, an IVF index
reaches 99\% recall $16\times$ faster than probing every list (Section~\ref{sec:ann-motivation}). \textbf{\emph{The question of this
paper is what happens when the answer itself becomes extremely large
(i.e., large\nobreakdash-$K$).}}

Large answers shrink the part of the library that an index can skip. A graph search must hold at least $K$ candidates, and an IVF search must probe enough lists to contain them. At $K=100$K on 10M rows, IVF-Flat at 99\% recall is only $3.7\times$ faster than probing every list. With product quantization, the IVF index misses the target even with every list probed. On 25M rows, no tested CAGRA configuration reaches 99\% recall at $K=100$K. Even worse, construction and storage recur with every re-encoded library. The 50M-row HNSW index alone takes two hours to build and 156~GiB to store, and it covers only 10\% of our library.

These measurements point to a different design, a GPU scan that scores every row at any $K$, as DrugCLIP's own campaign did. A scan has full coverage, needs no index, and reads the library in a regular pattern: a batch of queries shares each read of the library through one matrix multiplication. Compressed to INT8, our 500M rows take 387~GB and stay resident within the 408~GB of four GPUs. Scoring the compressed rows is then the only approximation. What grows with the answer is the cost of maintaining it. Consider a scan that merges each block's top $K$ into a running answer. Its batch time more than doubles from $K=1$K to $K=500$K, although it reads and scores the same rows (Section~\ref{sec:ann-insights}).

We present \emph{SweepANN}, a GPU system for large\nobreakdash-$K$ approximate vector
search with high recall for drug discovery. SweepANN scans the library instead of building an index. It keeps one compressed row partition of the library, a \emph{shard}, resident in each GPU's memory. Each GPU scores the same batch of pocket queries against every block of its shard with one matrix multiplication per block. SweepANN then merges each pocket's candidates across GPUs into its global top\nobreakdash-$K$ answer (Figure~\ref{fig:overview}). At large $K$, the scan stays fast only if those candidates are cheap to keep in the memory left beside the library. SweepANN has four plans for keeping them (Figure~\ref{fig:plan-space}). The index measurements of Sections~\ref{sec:ann-motivation} and~\ref{sec:published-baselines} favor a scan at large $K$. \textbf{All four plans are SweepANN's designs} for that scan, so comparing them isolates the cost of keeping the answers. \textbf{The baselines are ANN indexes}, large\nobreakdash-$K$ result collectors, and DrugCLIP's released script.

The plans differ in how often they select and in how much state they keep beside the library. The \textbf{full-score plan} stores every score and selects once, so it is feasible only when the batch's whole score matrix fits. SweepANN's \textbf{per-tile plan} selects after every block, which is slow at large $K$. Its \textbf{grouped plan} selects once per group of blocks, but its score buffer grows with the group. The \textbf{window plan} instead scores a sample of blocks first and retains each query's top $K$ in that sample as a \emph{seed}. The seed's lowest score then becomes a cutoff that passes only the later rows that can still enter the answer. All plans return the same top\nobreakdash-$K$ of the compressed scores, so they differ only in time and memory. SweepANN first fixes the batch size, which sets how many queries share each block read. It then rules out the plans whose state does not fit beside the library and runs the one measured fastest at the requested $K$.

Recent large\nobreakdash-$K$ systems take other routes. They make an index cheaper at collecting and reranking many candidates~\cite{yin2026bbc} or run large\nobreakdash-$K$ retrieval inside an OLAP engine~\cite{wu2026tengine}. SweepANN instead scores every row and uses the memory left beside the library to select as rarely as that memory allows.

Three ideas make the window plan fit beside the library and run fast. 1) \emph{Fused scoring}, which also serves the full-score plan, reconstructs compressed values inside the scoring kernel and multiplies them with the pocket batch. It stores no expanded copy of the block, so the resident library leaves room for the per-query candidates (Section~\ref{sec:fused-scoring}). 2) \emph{Selection per window} selects once per group of blocks, called a \emph{window}, instead of after every block. Between selections, it keeps only the rows above the cutoff in a fixed-capacity buffer, and the cutoff is exchanged once across GPUs (Sections~\ref{sec:selection}--\ref{sec:seed-capacity}). 3) \emph{Replay} captures each GPU's operations in two CUDA Graph segments, one before and one after the cutoff exchange. It replays them for successive query batches, which reduces the CPU work of submitting the scan (Section~\ref{sec:execution}).

SweepANN meets the screening target at campaign scale. On four RTX PRO 6000 GPUs, it answers every pocket in about one minute of resident time at 99.63\% mean recall against exact FP32 search. Resident time starts from the loaded library and ends with the answers left in GPU memory, global merging included. Four GH200 GPUs complete the same campaign in similar time (Section~\ref{sec:campaign}). From $K=10$K, SweepANN beats the GPU index baselines at the same 99\% target. On 25M to 500M rows, the GPU IVF configurations that reach the target take $8.2$--$24\times$ SweepANN's resident time. CAGRA, the fastest tested method at $K=1$K on 25M rows, is slower than SweepANN at $K=10$K (Section~\ref{sec:published-baselines}). Among SweepANN's own plans at 500M rows, the window plan fits a batch of 64 queries. It delivers $2.51\times$ the throughput of the full-score plan at that plan's best batch size, $B=16$. At equal batch, it is $1.10\times$ faster (Section~\ref{sec:physical-plan-crossover}).

The same plans serve two other embedding models (Section~\ref{sec:model-generalization}) and two public vector-search collections (Section~\ref{sec:general}). On DEEP1B, 1B image descriptors on eight GPUs, the window plan reaches 99.0\% and 99.3\% mean recall at $K=100$K and 500K. It takes 48 and 96~ms of resident time per batch of 64 queries. That is 5.4 and 3.2 times faster than the full-score plan on the uncompressed FP32 rows, an exact scan. The window plan also needs 6.0 and 5.3 times less peak memory per GPU. On 100M Cohere text embeddings of 1,024 dimensions, the window and full-score plans both exceed 99.5\% mean recall at $K=100$K and 500K. With 12.5M rows per GPU, the full-score plan is the faster one, and the plan rule picks it on the calibration queries.

The paper makes three contributions. First, an analysis of large\nobreakdash-$K$ batched scans, in which maintaining the answers is the cost that grows with $K$. It leads to a plan rule that decides by the memory beside the library first and by measured cost second. Second, the window plan. It retains each query's top $K$ rows in a sample as a seed, and the lowest seed score is a cutoff that passes later rows into fixed-capacity buffers. A survivor law sizes these buffers for rows in random order. Our library is shuffled by file, not by row, so it uses an empirical margin instead. A batch that overflows is recomputed exactly, and each GPU replays its scan with fused scoring as CUDA Graphs. At 500M rows on four GPUs, the window plan is $1.41\times$ faster than SweepANN's widest grouped plan in the same memory. Third, an evaluation that shows where indexes miss the recall target or lose their speed advantage at large $K$, and which plan is fastest in each tested setting. It includes complete 500M-row campaigns on two GPU architectures.

\section{Background and Expectations}
\label{sec:background}

\subsection{AI for Drug Discovery}
\label{sec:drug-discovery}

Most small-molecule drugs exert their effects by modulating disease-relevant proteins. In target-based discovery, researchers select and validate a target and then identify initial hits, compounds with measurable binding or functional activity. They optimize promising chemical series into candidates for preclinical development~\cite{paul2010rd}. Machine learning has been applied across this workflow, including target discovery, structure prediction, molecular design, and property prediction~\cite{vamathevan2019ml,jumper2021alphafold}. This paper focuses on learning-based virtual screening for hit identification. Such screening ranks the compounds of large molecular libraries against a given protein binding pocket to prioritize candidates for downstream evaluation.

Large screening libraries now contain billions of molecules~\cite{irwin2020zinc20,gorgulla2020virtualflow}. Structure-based screening evaluates candidates using the three-dimensional structure of a protein binding site~\cite{sadybekov2023computational}. Molecular docking searches for plausible binding poses within the pocket and scores their fit~\cite{trott2010vina}. Docking screens of 170M~\cite{lyu2019ultralarge} and over 1B molecules~\cite{gorgulla2020virtualflow} have identified potent ligands, but their computational cost grows with both library size and the number of targets. This cost motivates a fast first stage of screening that selects the candidate pool. Docking, rescoring, and chemical-diversity selection then narrow this pool to tens or hundreds of compounds for laboratory testing~\cite{lyu2019ultralarge,gorgulla2020virtualflow}.

Contrastive dual-encoder models turn this first stage into vector search
(Figure~\ref{fig:screening-pipeline}). DrugCLIP trains separate encoders
to map protein pockets and matching molecules to similar
vectors~\cite{gao2023drugclip}. Molecular embeddings do not depend on the pocket, so they are precomputed once and reused across pocket queries. Online screening is left with query encoding, similarity scoring, and top\nobreakdash-$K$ selection. DrugCLIP's genome-wide
application screened more than 500M molecules against more than 20K
pockets from approximately 10K human proteins, using
AlphaFold2-predicted structures~\cite{jia2026genome,genomescreenCard}.
ConPLex~\cite{conplex2023} and ConGLUDe~\cite{conglude2026} use different
target representations but follow the same retrieval pattern.

\subsection{What Screening Expects of Large-\texorpdfstring{$K$}{K} Search}
\label{sec:problem}
A top\nobreakdash-$K$ query returns the $K$ library vectors with the highest scores. Vector search systems are built and benchmarked for small answers, the 10 or 100 nearest images or documents~\cite{aumuller2020annbenchmarks,simhadri2022bigann}. We call a query \textbf{\emph{large\nobreakdash-$K$}} when its answer holds 100K or more vectors (K, M, and B denote thousands, millions, and billions). The search keeps and updates this answer as it reads the library, so a large answer is also a large amount of state to maintain.

Virtual screening needs large answers because its first stage only proposes candidates. Its scores are fast but imprecise predictions, so
it keeps a broad pool to retain true hits and to give the later stages diverse candidates~\cite{lyu2019ultralarge,gorgulla2020virtualflow,sadybekov2023computational}. \textbf{The pool is the only way into docking}. An active ranked beyond $K$, or dropped by an approximate search, is lost to every later stage. Screening therefore asks for a large $K$ and a nearly complete answer.
DrugCLIP's released screening code keeps the top 100K molecules of
every pocket~\cite{drugclipCode}. The main workload of this paper keeps $K=500$K, five times that setting, so that the pool stays broad as libraries grow. The plan
comparisons also report $K=100$K, and
Sections~\ref{sec:batch-reuse-controls} and~\ref{sec:plan-map} evaluate answer sizes from 1K to 5M.

Each library molecule is encoded once into a vector of $D$ numbers, stored as a table row whose position is its ID. A pocket becomes a query vector of the same $D$ dimensions. The search returns the IDs and scores of the $K$ best-matching rows, which feed docking and the later stages. The score of a row is its inner product with the query vector, so a screening query is a top\nobreakdash-$K$ maximum inner product search ($D=768$ for DrugCLIP). A genome-wide screen issues one query for every pocket of every protein against the same library. We call the queries of one screen a \emph{campaign} (Figure~\ref{fig:screening-pipeline}).

\paragraph{Problem definition.}
The library $X$ holds $N$ vectors $x_0,\ldots,x_{N-1}$ in
$\mathbb{R}^D$, and $i$ is the row ID of $x_i$. A query is a vector
$q\in\mathbb{R}^D$, and $s(q,x_i)$ denotes the score by which rows are
ranked. For a set of rows $V$, $\TopK(V,q)$ denotes $K$ distinct rows of
$V$ with the highest scores, returned with those scores. The exact answer
is $R^*(q)=\TopK(X,q)$ under $s(q,x_i)=q\cdot x_i$ computed in FP32, and
we call it the \emph{FP32 reference}.

The answer has set semantics: it may be returned in any order, and when
several rows tie with the $K$-th score, any of them may complete it.
Formally, a campaign is a set $Q$ of query vectors over one library, and its result is the answer of every $q\in Q$, exactly $K$ score-ID pairs each. The library changes only when molecules are added or re-encoded. It is therefore prepared once and reused by many campaigns, whose queries are known in advance and processed in batches. In the workload of this paper
(Section~\ref{sec:setup}), $N=500$M, $D=768$, $|Q|=26{,}562$, and
$K=500$K. As FP32 vectors, the library occupies 1.54~TB. An exhaustive campaign scores 13.3 trillion query-molecule pairs and returns 13.3B score-ID pairs, 79.7~GB in total.

\paragraph{Approximate answers.}
Exact answers are not required. Later stages rescore the pool more accurately, so a pool that holds nearly all exact answers serves the same purpose. An approximate search returns $K$ distinct rows $\hat R(q)$ per query. We measure its quality by recall against the FP32 reference, $\operatorname{Recall@}K(q)=|\hat R(q)\cap R^*(q)|/K$, and report the mean and the minimum over queries. Recall ignores which molecules are active, so we also measure the enrichment of known actives near the top (Section~\ref{sec:setup}).

\paragraph{Expectations.}
The workload sets three bars. The first is $K=500$K per query at 99\% mean recall or better. The second is the cost of whole, repeated campaigns of known queries against one library, including preparation that recurs with every re-encoding. The third is delivery of every answer to the host and to durable storage. The next section measures how well ANN indexes, the established tool for fast vector search, meet the first two bars.

\section{Where ANN Indexes Work, and Where Large Answers Stop Them}
\label{sec:ann-motivation}
ANN indexes speed up vector search by scoring only part of the library.
A graph index such as HNSW~\cite{malkov2020hnsw} links each vector to
near neighbors and answers a query by walking the graph from entry
points. An inverted-file (IVF) index, as implemented in Faiss~\cite{johnson2019billion}, clusters
the vectors into lists and scans only the lists whose centroids are
closest to the query. Both can keep the original FP32 vectors (Flat) or store compressed ones to save
memory: SQ8 stores one byte per component, and PQ48 encodes each vector in 48 bytes. This section asks whether an index can still skip work as $K$ grows. It measures CPU indexes on 1M to 50M rows of our library (Section~\ref{sec:setup}), and Section~\ref{sec:published-baselines} adds the GPU indexes on 25M to 500M rows.

\begin{figure*}[t]
\centering
\includegraphics{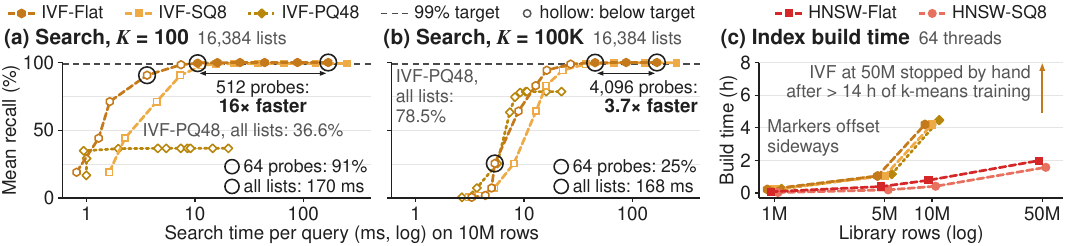}
\caption{ANN indexes on CPUs (Faiss, 64 threads, 1K pocket queries). (a,b) Mean recall against search time per query on 10M rows over all tested probe counts. Hollow markers are below 99\% mean recall. Rings mark IVF-Flat at 64 probes, at the fewest probes reaching 99\% mean recall, and at all lists. Arrows give the time ratio of the last two. IVF-PQ48 stays below the target with every list probed, so its remaining error comes from its codes. (c) Build time of the 17 completed builds, with markers offset sideways so that coinciding builds stay visible. The 50M-row IVF builds were stopped by hand before finishing.}
\Description{Two recall-versus-search-time plots for IVF-Flat, IVF-SQ8 and IVF-PQ48 on 10M rows show that reaching 99 percent mean recall takes 512 probes at K=100 but 4,096 probes at K=100K, so the gain over scanning all lists falls from 16 to 3.7 times, and IVF-PQ48 stays below the target even with every list probed. A third plot shows build hours growing with library rows, with the 50M IVF builds stopped by hand after more than 14 hours.}
\label{fig:index-limits}
\end{figure*}

\subsection{What ANN Indexes Do Well}
\label{sec:ann-credit}
An index pays off when the few rows it visits contain the answer. This is the common case for the small answers that indexes are designed and benchmarked for. On 10M rows, IVF-Flat reaches 99\% mean recall at $K=100$ by probing 512 of its 16,384 lists. It takes 10.59~ms per query, $16\times$ faster than scanning every list (Figure~\ref{fig:index-limits}(a), Table~\ref{tab:ann-quality-budget}). On one GPU, CAGRA answers a batch
of 32 queries at $K=1$K in 10.9~ms on 25M rows at 99\% recall
(Section~\ref{sec:published-baselines}). HNSW builds finish in tens of minutes at 10M rows and then serve every later query. Compressed variants fit libraries that the original vectors would not. These are the regimes an index serves well: answers of tens to a few thousand rows, latency-bound single queries, hosts without accelerators, and libraries that rarely change. There, a scan cannot compete, since its cost per query is the whole library at any $K$. The next subsection measures what remains of this advantage as the answer grows.

\begin{table}[t]
\centering
\caption{Fastest CPU IVF settings (64 threads) reaching mean recall
$\geq99\%$ on the 10M-row library with 1K pocket queries.
Each cell gives probes / milliseconds per query.}
\label{tab:ann-quality-budget}
\begin{tabular}{@{}lrr@{}}
\toprule
Representation & $K=100$ & $K=100$K\\
\midrule
IVF-Flat &512 / 10.59&4,096 / 45.23\\
IVF-SQ8 &1,024 / 22.87&4,096 / 67.73\\
IVF-PQ48 &\multicolumn{2}{c}{No tested setting reaches the recall target}\\
\bottomrule
\end{tabular}
\end{table}

\subsection{Query Performance as Answer Size Grows}
\label{sec:ann-query}
At a fixed recall target, a larger answer forces an IVF index to probe more of its lists. Figure~\ref{fig:index-limits}(a,b) compares two answer sizes on a 10M-row library: $K=100$, typical of ANN benchmarks, and $K=100$K. The queries are the first 1K of the campaign's pockets in stored order. Each curve traces mean recall against CPU time per query as more lists are probed. At the fixed target of 99\% mean recall, Table~\ref{tab:ann-quality-budget} lists the fastest tested setting of each IVF index (Appendix~\ref{app:ann-studies}). Compared with $K=100$, the answer at $K=100$K needs $8\times$ as many probes and about $4\times$ the query time with IVF-Flat. With IVF-SQ8, it needs $4\times$ the probes and about $3\times$ the time, and IVF-PQ48 reaches the target at neither size. Probing all 16,384 lists of IVF-Flat scores every row and takes about 170~ms per query at either answer size. An IVF index saves only the lists it skips. The share it must probe to keep 99\% recall grows with $K$, to a quarter of its lists at $K=100$K in the tested grid.

Graph and GPU indexes follow the same trend. HNSW needs a 100K-candidate search list (\texttt{efSearch}) to reach 99.7\% recall at $K=100$K on 10M rows. It then takes 7.39~s per query on four threads with Faiss's default bounded queue and 0.49~s with a heap queue (Appendix~\ref{app:ann-studies}). Among the GPU indexes on 25M rows, IVF-Flat is slower than brute force from $K=10$K. At $K=500$K, its fastest setting that keeps 99\% recall probes every list (Section~\ref{sec:published-baselines}).

\subsection{Why ANN Indexes Struggle with Large\texorpdfstring{\nobreakdash-$K$}{-K}}
\label{sec:ann-factors}
A large answer works against all three sources of an index's speed: a short candidate list, a small visited region of the library, and, if compressed, cheap approximate scores. More hardware only splits the added work, since each query must still return $K$ entries. High recall adds a second constraint: every result in this paper is held to 99\% mean recall against the FP32 reference (Section~\ref{sec:problem}). An index must reach this target while it skips rows, whereas a scan scores every row and loses recall only to compression (Section~\ref{sec:quality}). The paragraphs below take answer size and search coverage first, then the cost of building and storing an index, and then compression.

\paragraph{Answer size.} A search must collect at least $K$ distinct candidates. A graph search keeps at least $K$ entries in its sorted search list, and an IVF search must probe enough lists to hold most of the answer. The work of collecting candidates therefore grows with $K$ itself, not with the answer's share of the library. On 25M rows, the fastest CAGRA configuration that reaches 99\% recall takes 10.9~ms per batch of 32 queries at $K=1$K and 83.2~ms at $K=10$K. Even the larger answer is only 0.04\% of the library.

\paragraph{Search coverage.} A fixed search budget covers too little of the library for large answers. With 64 probes, IVF-Flat's mean recall drops from about 91\% at $K=100$ to 25\% at $K=100$K. The probed lists return only about 36K valid IDs per query, far short of the $K$ rows the answer needs.

\begin{figure}[t]
\centering
\includegraphics{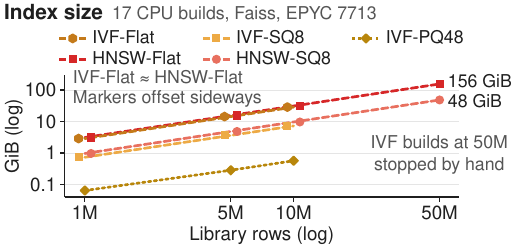}
\caption{Index size of the 17 completed untuned CPU builds (Faiss 1.13.2, EPYC 7713, 64 threads). The three IVF builds at 50M rows were stopped by hand before finishing. Series that coincide are drawn with their markers offset sideways.}
\Description{Index size against library rows on log scales for five CPU indexes, reaching about 156 GiB for HNSW-Flat and 48 GiB for HNSW-SQ8 at 50M rows, with the 50M IVF builds stopped before finishing.}
\label{fig:app-index-size}
\end{figure}

\paragraph{Index construction and storage costs.}
\label{sec:ann-build}
Building and storing an index recur with every re-encoded library, and both costs are large. We build five indexes with Faiss~\cite{johnson2019billion}: IVF-Flat, IVF-SQ8, IVF-PQ48, HNSW-Flat, and HNSW-SQ8 (17 completed untuned builds, Figure~\ref{fig:index-limits}(c)). At 10M rows, each IVF build takes four to four and a half hours, mostly k-means training. At 50M rows, HNSW-Flat takes about two hours to build, and the IVF builds were stopped by hand after more than 14 hours without finishing. The 50M-row HNSW-Flat index occupies about 156~GiB (HNSW-SQ8 about 48~GiB), for 10\% of our library (Figure~\ref{fig:app-index-size}).

\paragraph{CPU index builds.}
The 17 builds of Figures~\ref{fig:index-limits}(c) and \ref{fig:app-index-size} ran with Faiss
1.13.2~\cite{johnson2019billion} on EPYC 7713 processors with 64
threads. Build time covers training and insertion, not loading or
serialization. IVF list counts run from 4,096 at 1M rows to 16,384 at 10M and 32,768 at 50M, and the builds were not tuned. HNSW uses $M=32$ and an
\texttt{efConstruction} of 200. IVF training keeps the Faiss defaults: 10
coarse-clustering iterations with at most 256 training points per list,
and 25 PQ48 codebook iterations. This k-means training of the list centroids dominates an IVF build. At 10M rows it ran at about 87~GFLOP/s, counted by the operations of default k-means. The three IVF variants pay the same clustering cost.

\paragraph{Compression and refinement.} Compressed vectors make an index smaller (Figure~\ref{fig:app-index-size}) but perturb its scores. SQ8 still reaches the 99\% target at $K=100$K (Table~\ref{tab:ann-quality-budget}), whereas PQ48 does not. Even with every list probed, the 10M IVF-PQ48 index reaches only about 78.5\% at $K=100$K, so the remaining error comes from the representation, not from the search. Recovering it needs a finer representation or \emph{refinement}, which rescores more than $K$ candidates per query with the original vectors. The cost of refinement grows with the answer.
CAGRA's vector-PQ mode on 1M rows (1.464~GB of device memory) reaches
97.50\% mean recall at $K=100$K and 60.66\% at 500K. Host refinement with the FP32 vectors, tested up to $K=10$K, meets the target at every tested size. Even FP16 vectors cost recall. An IVF-Flat index of FP16 vectors on the 25M rows of Section~\ref{sec:published-baselines} accumulates in FP16. On the 32 held-out queries, it reaches 96.60\%, 97.70\%, 98.67\%, and 99.22\% mean recall at $K=1$K, 10K, 100K, and 500K. It thus misses the target up to $K=100$K.

\paragraph{One-bit codes with rescoring.}
One-bit codes reach the target only after rescoring many more candidates than the answer holds. The test uses a native one-bit RaBitQ index~\cite{gao2024rabitq} on the same 25M rows (Faiss, inner product, unquantized queries), scored on the CPU. From its codes alone, it returns 75.7\% of the FP32 top-100K and 81.7\% of the top-500K. These are means over the 32 held-out queries of Section~\ref{sec:setup}, with minima of 65.3\% and 74.0\%. Rescoring the top 400K and 2M code candidates with the FP32 vectors gives 99.56\% and 99.96\% mean recall at $K=100$K and 500K (Appendix~\ref{app:ann-studies}).

\subsection{Observations and Insights}
\label{sec:ann-insights}
\begin{figure}[t]
\centering
\includegraphics{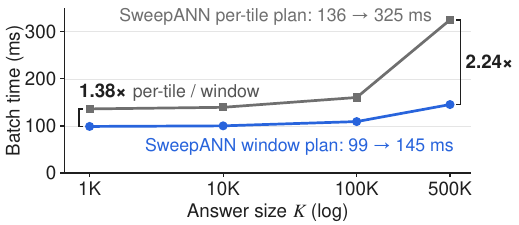}
\caption{Batch time against $K$ for two SweepANN plans, per-tile and window, on 500M rows (four GPUs, 64 held-out queries in batches of $B=64$). Both plans replay as CUDA Graphs with the same fused scorer and global merge. Points show medians of three runs, whose ranges are smaller than the markers, and brackets give per-tile over window time.}
\Description{Batch time against answer size on 500M rows: the per-tile plan grows from 136 to 325 ms and the window plan from 99 to 145 ms, so their ratio grows from 1.38 to 2.24.}
\label{fig:answer-size-cost}
\end{figure}
\begin{figure*}[t]
\centering
\includegraphics{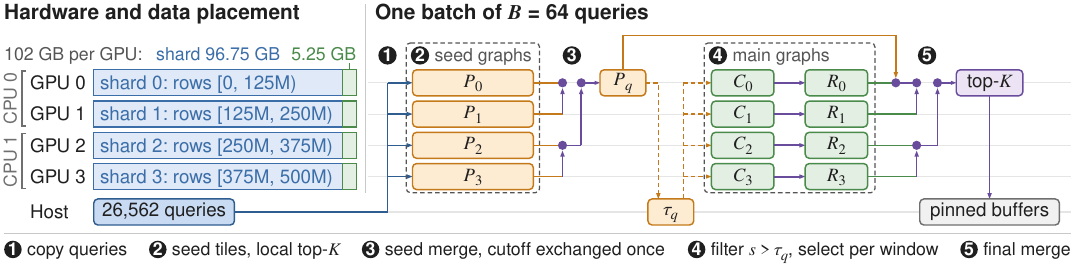}
\caption{SweepANN's architecture. Left: four GPUs with memory drawn to scale, each holding one compressed 125M-row shard and, beside it, the CUDA runtime and all query state. Right: one query batch in time order, with steps numbered as in Section~\ref{sec:architecture}. Labels mark the local seeds $P_j$, the seed $P_q$, the cutoff $\tau_q$, the buffers $C_j$, and the running answers $R_j$. Dashed frames are the two CUDA Graph segments, and dashed lines carry the cutoff exchanged once through the host.}
\Description{Left, four GPU memory bars drawn to scale with a 96.75-GB shard and 5.25 GB beside it, two GPUs per CPU socket and a host row with 26,562 queries. Right, one batch in five numbered steps on the same GPU rows: query copy, seed tiles to local top-K, seed merge and cutoff sent through the host, main scan that filters, buffers and selects per window, and a final merge whose answer goes to pinned host buffers.}
\label{fig:overview}
\end{figure*}
Three observations from these measurements carry into the design. First, an index saves only
the rows it skips, and across the tested indexes its measured advantage
shrinks as the answer grows. On 10M rows, the advantage of IVF-Flat at 99\% recall over
probing every list falls from $16\times$ at $K=100$ to $3.7\times$ at
$K=100$K, an answer of 1\% of the library. On a second CPU server, an exact flat scan already beats IVF-Flat with a quarter of its lists, taking 141.0 against 160.7~ms per query at $K=100$K (Section~\ref{app:setups}). Every IVF, GPU graph, and large\nobreakdash-$K$ collector configuration we measured misses the 99\% target at large $K$ or loses most or all of its advantage over a full scan.

\begin{figure*}[t]
\centering
\includegraphics{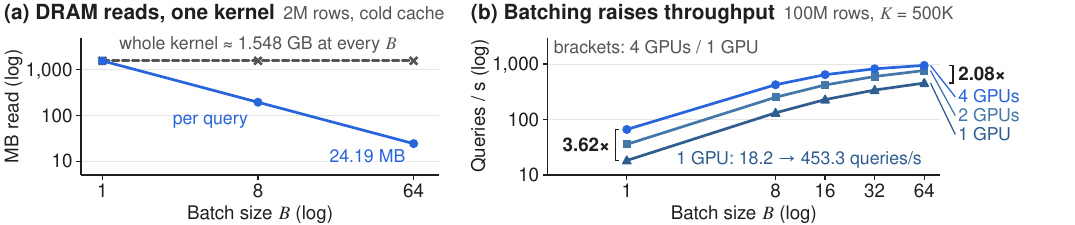}
\caption{Why batching pays. (a) Cold-cache DRAM reads of one production scoring kernel over 2M rows (Nsight replay), for the whole kernel and per query. Points at $B=1$ and 64 average two independent profiles. (b) Throughput of the window plan against batch size on 100M rows at $K=500$K (64 held-out queries) with one, two, or four GPUs and a fixed 16M-row sample. Brackets give four-GPU over one-GPU throughput at $B=1$ and $B=64$.}
\Description{Two panels. (a) DRAM reads of one scoring kernel against batch size on a log scale: the whole kernel reads about 1.548 GB at every batch size, so the reads per query fall to 24.19 MB at B=64. (b) Throughput against batch size on 100M rows for one, two and four GPUs: on one GPU, batching raises throughput from 18.2 to 453.3 queries per second, and four GPUs are 3.62 times faster than one at B=1 and 2.08 times at B=64.}
\label{fig:batch-reuse}
\end{figure*}

Second, a scan does not rely on skipping rows and needs no index. On a GPU, one read of each library block serves the whole query batch (Figure~\ref{fig:batch-reuse}(a)), so the queries of a batch share the cost of reading the library. A scan is fast only if the library stays in device memory. Compression to about one byte per component keeps it there at high recall (Sections~\ref{sec:ann-factors}, \ref{sec:fused-scoring}, and~\ref{sec:quality}).

Third, one cost grows with $K$ for index and scan alike: selecting and
maintaining $K$ entries for every query of a batch in the memory left
beside the library. Consider a compressed scan on four GPUs that merges the top $K$ of each block of rows into a running answer, the per-tile plan of Section~\ref{sec:architecture}. On all 500M rows, it takes 136~ms per batch of 64 queries at $K=1$K and 325~ms at $K=500$K. It reads and scores the same rows at both sizes (Figure~\ref{fig:answer-size-cost} and Section~\ref{sec:batch-reuse-controls}). The simplest plan keeps every score and selects once. Wherever it fits at a setting's batch, it is also the fastest at the largest tested $K$ (Section~\ref{sec:plan-map}). At the campaign's batch it does not fit. For $B=64$ queries, the FP16 scores of a 125M-row shard take 16~GB against the 5.25~GB beside it. The memory beside the library therefore decides which plan can run.

\begin{takeaway}
An index saves only the rows it skips. On 10M rows, IVF-Flat's advantage at 99\% recall over probing every list falls from $16\times$ at $K=100$ to $3.7\times$ at $K=100$K. Maintaining the answers is the cost that grows with $K$ for index and scan alike. On 500M rows, the \textbf{per-tile plan} takes 136~ms per batch at $K=1$K and 325~ms at $K=500$K.
\end{takeaway}

\section{SweepANN Design}
\label{sec:design}

\subsection{Overall Architecture}
\label{sec:architecture}

\begin{figure*}[t]
\centering
\includegraphics{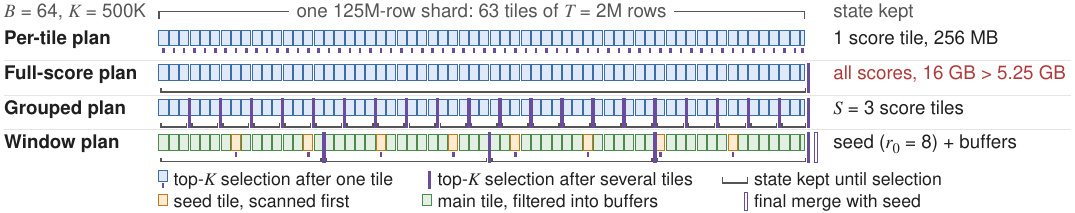}
\caption{One batch ($B=64$, $K=500$K) under SweepANN's four plans on the 63 tiles ($T=2$M rows) of one 125M-row shard. The grouped plan selects once per $S=3$ tiles, and the window plan uses $r_0=8$ seed tiles and $G=16$ tiles per window. The full-score plan's 16~GB of scores exceeds the 5.25~GB beside the shard.}
\Description{Four rows draw the same 63 tiles of one shard: the per-tile plan selects after every tile, the full-score plan keeps all scores (16 GB, more than the 5.25 GB beside the shard) and selects once, the grouped plan selects after every three tiles, and the window plan scans eight seed tiles first and selects at the ends of four windows before a final merge with the seed.}
\label{fig:plan-space}
\end{figure*}
Following the observations of Section~\ref{sec:ann-insights}, SweepANN answers large\nobreakdash-$K$ queries on the GPUs of one server by scanning a compressed library that stays resident in GPU memory (Figure~\ref{fig:overview}). On the 500M-row campaign of Section~\ref{sec:campaign}, this design takes 61.04~s of resident time, about 2.3~ms amortized per query. Three ideas make the scan fast. First, a fused kernel scores each block of rows once for the whole query batch, straight from the INT8 codes. Second, a seed of $K$ actual rows gives every query a cutoff, and only the rows above it reach the query's fixed-capacity buffer. Selection then runs once per window of blocks, and the state beside the shard stays bounded. Third, each GPU's work for a batch replays as two CUDA Graph segments, one before and one after the cutoff exchange, so a batch costs fewer launches.

\paragraph{Data placement.} The library is split by rows into one \emph{shard} per GPU, 125M rows each. Each GPU reads its shard in \emph{tiles}, blocks of $T=2$M consecutive rows. A shard is stored as INT8 components with FP16 scales and takes 96.75~GB of the 102~GB of device memory that the CUDA runtime reports per GPU (Section~\ref{sec:setup}). The remaining 5.25~GB must hold the CUDA runtime and all query state. This remainder limits how a batch's answers can be kept. Host memory holds the queries and receives the answers.

\paragraph{Four plans for the answer state.}
SweepANN has four plans for keeping the $K$-entry answers of a whole batch beside a shard, and they trade selection work against memory (Figure~\ref{fig:plan-space}). A \emph{full-score plan} stores the complete score matrix and selects once at the end. For a batch of $B$ queries, FP16 scores take $2B$ bytes per row. At $B=64$ the 125M rows of a shard thus need 16~GB, three times the memory available, so this plan must use a smaller batch. A \emph{per-tile plan} stores the scores of one tile only, 256~MB per GPU at this batch. It pays in selection work instead: after each of the 63 tiles of a shard, it merges the tile's top $K$ into a running answer. A \emph{grouped plan} lies between the two. It selects once per group of tiles, with a score buffer that grows with the group. SweepANN's \emph{window plan} uses the seed's cutoff to discard most rows before they reach a buffer. Its selection then runs once per group of tiles with bounded state, although its buffers take more memory than the per-tile plan's (Section~\ref{sec:batch-reuse-controls}).

\paragraph{One query batch.} One read of a tile serves the whole batch. Every GPU compares a batch of $B=64$ queries with its own shard and scores each tile for all $B$ queries with one matrix multiplication. In database terms this is a block nested-loop join: the outer block is the query batch, and the inner relation, the shard, is scanned once per batch. 

The window plan processes a batch in five steps, numbered as in Figure~\ref{fig:overview}. (1) The CPU copies the queries to every GPU. (2) Each GPU scores a few \emph{seed tiles} and keeps the top $K$ rows of each query among them. (3) A merge across the GPUs keeps the best $K$ of these rows as the query's \emph{seed}. The lowest seed score becomes the query's \emph{cutoff}, which is sent to every GPU. (4) Each GPU scores its remaining tiles and appends the rows above the cutoff to the query's \emph{candidate buffer}. After each group of tiles, a \emph{window}, it selects the top $K$ from the buffer and the query's \emph{running answer} and then empties the buffer. The running answer holds the best $K$ rows of the earlier windows. (5) A final merge combines the seed with the running answers of all GPUs into the query's top\nobreakdash-$K$ answer, which stays in GPU memory or is copied to the host. The CPU submits steps 2 and 4 on each GPU as two CUDA Graphs. A CUDA Graph is a recorded sequence of GPU operations that one call launches. Sections~\ref{sec:fused-scoring} to \ref{sec:delivery} give the details. The first four cover the fused scorer, the seed and cutoff, the windows and buffers, and correctness. The rest cover buffer sizing and the plan rule, the cost model, graph replay, and result delivery.

\subsection{Compressed Shards and Fused Scoring}
\label{sec:fused-scoring}
\label{sec:compressed-score}
Compression to about one byte per component keeps the library resident in GPU memory. The loader quantizes every vector symmetrically in groups of 256 components. Each group keeps one FP16 scale and stores its components as rounded, clamped INT8 values. A row of $D=768$ components thus holds three groups and occupies $768+3\cdot2=774$ bytes, so the library takes 387~GB and a shard 96.75~GB. Preparing a library involves no training and no graph construction (Section~\ref{sec:campaign} reports the load time). Each GPU holds a contiguous shard, and adding the shard offset to a local position gives the global row ID.

Fused scoring never stores a dequantized copy of a tile. The Triton~\cite{tillet2019triton} scoring kernel loads the INT8 values and scales of a tile and reconstructs FP16 operands inside the kernel. It multiplies them with the FP16 query batch and accumulates the dot products in FP32. Its output is the \emph{score tile}, the $B\times T$ FP16 scores of the tile, which the compaction kernel reads next (Section~\ref{sec:windows}). A dequantized copy would compete with the answer state for memory: one 2M-row tile in FP16 would take another 3.072~GB, over half of the memory beside the shard. On compressed rows all four plans use this scorer, so they differ only in how they keep the answers.

The resulting \emph{compressed score}, hereafter $s$, is SweepANN's only approximation, because every row is still scored. We call the exact top\nobreakdash-$K$ under $s$, obtained by scoring every row and selecting once, the \emph{compressed reference}. A plan \emph{preserves selection} when it returns the same multiset of selected scores as the compressed reference, each paired with a valid, distinct row ID. The definition separates the effect of the plan from the effect of compression. Sections~\ref{sec:selection} to~\ref{app:theory} show that the window plan preserves selection. Recall against the FP32 reference (Section~\ref{sec:problem}) measures what compression costs in quality.

\subsection{Retained Seed and Cutoff}
\label{sec:selection}

A cutoff lets a buffer cover many more rows than a score buffer in the same memory. Without a cutoff, any row may belong to the answer, so a grouped plan must keep every score of its group. SweepANN therefore first gives each query a cutoff below which no row can improve its answer (Figure~\ref{fig:plan-space}). A buffer then keeps only the rows that can still enter the answer. In the same workspace on all 500M rows, a window thus spans up to 30M rows, whereas a grouped plan spans 6M (Section~\ref{sec:batch-reuse-controls}).

Each shard first scans $r_0$ seed tiles spread evenly over its row range, and their rows form the shard's \emph{sample}. For every query, the shard computes a local top\nobreakdash-$K$ over its sample, with global row IDs. A hierarchical merge across GPUs combines these local results into the global seed $P_q$. The seed holds the $K$ sample rows with the highest scores, which are kept with their scores until the final merge. The lowest seed score is the cutoff for filtering later rows:
\[
\tau_q=\min_{i\in P_q}s(q,x_i).
\]
The \emph{main scan} then scores every remaining tile, a \emph{main tile}, and keeps a row only if $s(q,x_i)>\tau_q$. Seed tiles and main tiles partition the library, so each row is scored exactly once per query.

Main-scan rows that pass this filter are the query's \emph{survivors}, $U_q$. The theorem below shows that the seed and the survivors suffice for the answer.
\begin{theorem}[Selection preservation]\label{thm:preserve}
Let $P_q$ be any $K$ rows of $X$, $\tau_q=\min_{i\in P_q}s(q,x_i)$, and
$U_q=\{i\notin P_q: s(q,x_i)>\tau_q\}$. Then
$\TopK(X,q)\equiv_s\TopK(P_q\cup U_q,q)$, where $\equiv_s$ denotes equal
multisets of selected scores, and any choice among rows tied at the
$K$-th score completes a valid answer.
\end{theorem}
The theorem holds for any choice of $P_q$, so the answer never depends on the sample. Every discarded row is matched or outranked by the $K$ rows of $P_q$, which are scored once per batch and kept for the final merge (proof in Section~\ref{app:theory}). A weaker sample only lowers the cutoff and lets more rows pass, so it changes the cost but not the answer.

\paragraph{Global and local cutoffs.}
The global cutoff is at least as selective as any shard-local cutoff. Let $\tau_j$ be the lowest score of shard $j$'s local top\nobreakdash-$K$ over its sample. The global seed is the top\nobreakdash-$K$ of the union of these local answers, so its cutoff is at least every $\tau_j$. A single exchange before the main scan thus gives every shard this tighter cutoff. No shard then keeps more survivors than its own sample's cutoff would keep.

\begin{algorithm*}[t]
\caption{Retained-seed large\nobreakdash-$K$ retrieval for one query $q$. $\mathcal S_j$ is the sample of shard $j$, $C_j$ its candidate buffer, $R_j$ its running answer, and $\mathcal W_{j,\ell}$ the windows of its remaining rows. $P$, $\tau$, $R_j$, and $C_j$ omit the fixed $q$. The survivors $U_q$ of Section~\ref{sec:selection} are the union of the buffer contents over all windows and shards. Scores are compressed scores $s$.}
\label{alg:screenk}
\begin{tabular}{@{}r>{\raggedright\arraybackslash}p{.94\textwidth}@{}}
1 & For each shard $j$ in parallel, compute
$P_j\leftarrow\TopK(\mathcal S_j,q)$ with global row IDs.\\
2 & Compute and keep $P\leftarrow\TopK(\bigcup_j P_j,q)$, and
set $\tau\leftarrow\min_{i\in P}s(q,x_i)$.\\
3 & For each shard $j$ in parallel, initialize the running answer $R_j$
and an overflow flag $o_j$.\\
4 & \quad For each window $\mathcal W_{j,\ell}$, clear the candidate buffer
$C_j$ and its count.\\
5 & \qquad For each main tile in the window, score every row.\\
6 & \qquad Append rows with $s(q,x_i)>\tau$ to $C_j$ up to capacity $A$, and
set $o_j$ on overflow.\\
7 & \quad At the end of the window, set
$R_j\leftarrow\TopK(R_j\cup C_j,q)$, keeping score-ID pairs.\\
8 & If any $o_j$ is set, recompute the batch that contains $q$ with the
per-tile plan.\\
9 & Otherwise, return $\TopK(P\cup\bigcup_jR_j,q)$ through a hierarchical
GPU merge.\\
\end{tabular}
\end{algorithm*}

\subsection{Window-Based Candidate Maintenance}
\label{sec:windows}
SweepANN scores every main tile but selects only at window ends, and it pays for this deferral with storage. On each GPU, every query has a candidate buffer with room for $A$ score-ID pairs. It also has a running answer, which holds the best $K$ main-scan survivors of the windows completed so far. A window spans $G$ consecutive tile positions and skips the seed tiles among them, which the seed step has already scored. For each main tile, the compaction kernel applies the cutoff to the score tile and appends the survivors to the buffer. At the end of a window, SweepANN selects the top $K$ from the running answer and the buffer and then clears the buffer, reusing its storage. Clearing is what bounds the state, since collecting survivors until the scan ends would need room for every survivor of a query. With $r_0=8$ and $G=16$, the 55 main tiles of a shard fall into four windows of 14, 14, 14, and 13 tiles. The large selection thus runs four times during the main scan instead of once per tile.

Deferring selection to window ends leaves the selected scores unchanged. Each window end selects the top $K$ from the running answer and the buffer. This selection keeps the invariant behind Theorem~\ref{thm:preserve}: every discarded row is matched or outranked by $K$ retained rows (Lemma~\ref{lem:window}, Section~\ref{app:theory}). Merging $P_q$ with the running answers thus returns the same selected scores as the compressed reference. Seed and main-scan rows are disjoint, as are the shard ranges, so the final answer also has distinct IDs. Algorithm~\ref{alg:screenk} gives these steps for one query. A batch runs $B$ queries at once, each with its own row of every buffer.

The cutoff stays fixed for the whole batch. Raising the global cutoff during the scan would need another exchange across GPUs and another CUDA Graph segment. Every survivor of a window must therefore fit in the buffer. A GPU can instead raise its own \emph{local cutoff} to the $K$-th score of any $K$ rows it has scored (Lemma~\ref{lem:window}), which needs no exchange (Section~\ref{app:grouped}). The reported campaigns keep the fixed cutoff.

\paragraph{Overflow and recovery.}
An undersized buffer costs time but never changes the answer. A weak cutoff can let more survivors through than a buffer holds. The compaction kernel counts every survivor, drops the writes beyond capacity, and sets an overflow flag that persists across windows. The host then releases the CUDA Graph state and recomputes the affected batches with the per-tile plan, which needs no candidate buffer.

\subsection{Correctness}
\label{app:theory}
\paragraph{Proof of Theorem~\ref{thm:preserve}.}
A sample row that is not in $P_q$ is matched or outranked by all $K$ rows of $P_q$, by the definition of top\nobreakdash-$K$. A main-scan row that fails the filter has a score of at most $\tau_q$, so again the $K$ rows of $P_q$ are at least as good. Any discarded row can therefore be replaced by a kept row. Ties at the cutoff need no special handling. $P_q$ itself supplies $K$ rows at or above $\tau_q$, so the strict comparison suffices. For example, with $K=2$ and seed scores $[9,7]$, a later 7 is rejected, an 8 is kept, and the final answer is $[9,8]$. The argument holds for any sample of at least $K$ rows.

\begin{lemma}[Window selection and local cutoffs]\label{lem:window}
Replacing any set $V$ of scored rows by $\TopK(V,q)$ at any point of the
scan leaves $\TopK$ of the final union unchanged under $\equiv_s$, and
raising a shard's cutoff to the $K$-th score of any $K$ distinct rows it
has scored preserves Theorem~\ref{thm:preserve}.
\end{lemma}
The argument rests on one invariant: every discarded row is matched or outranked by $K$ distinct retained rows, which lie in $P_q$, a running answer, or a buffer. A row dropped from the top\nobreakdash-$K$ of a set $V$ is dominated by the $K$ rows kept from $V$. A row rejected by a raised cutoff is dominated by the $K$ scored rows that define that cutoff. When a selection later discards one of those $K$ rows, the $K$ rows that dominate it dominate the first row too. The invariant therefore holds to the end of the scan, and merging $P_q$ with the four running answers returns the same selected scores as $\TopK(P_q\cup U_q,q)$. A shard with fewer than $K$ survivors pads its running answer with sentinels, and the seed still gives $K$ real rows.

\subsection{Sizing the Buffers and Choosing the Plan}
\label{sec:seed-capacity}
\paragraph{Memory footprint.}
The window plan's query state follows from four buffers and a remainder. Scores take two bytes and row IDs four. Per GPU, the main buffers are the score tile ($2BT$ bytes), the candidate buffers ($6BA$), the running answers ($6BK$), and an FP16 merge-score buffer ($2B(K+A)$). Selection scratch, counters, graph pools, and merge state make up the remainder, $W_{\mathrm{other}}$. The query state per GPU thus occupies
\[
W_{\mathrm{query}}=2BT+8B(A+K)+W_{\mathrm{other}}
\]
bytes, which must fit beside the shard. The \emph{peak workspace} is the measured peak of memory allocated above the library.

Three parameters set this state and its cost: $B$, $r_0$, and $A$. A large $A$ makes every window selection more expensive, because each one processes a fixed-width buffer of $K+A$ scores. The batch size sets the memory left for each query's buffer, and $r_0$ sets how many survivors that buffer must hold. A larger batch reuses each tile across more queries. But every added query needs $8(A+K)+2T$ bytes and part of $W_{\mathrm{other}}$ per GPU, which leaves less room for survivors. SweepANN therefore budgets the three in a fixed order. It first fixes $B$ by resident throughput on calibration batches (at $K=500$K, 64 on four PRO 6000 GPUs and 256 on eight). It then chooses $r_0$ and $A$ in the remaining memory among measured pairs (Section~\ref{sec:capacity}).

\paragraph{Sizing the buffer.}
The sample size trades unfiltered work against buffer capacity. The sample is scanned before any cutoff exists, so every sample row goes through top\nobreakdash-$K$ selection. More seed tiles add this unfiltered work, but they raise the cutoff and leave fewer survivors, so the capacity $A$ depends on $r_0$ and $G$. For rows in random order, the following law bounds the mean number of survivors per window. \begin{proposition}[Random-order survivor law]\label{prop:law}
If the rows are in random order and the scores are fixed, the sample
rows are a uniform subset of any query's ranks, and the number of
survivors in a window of $N_w$ main-scan rows has mean at most
$N_wK/(N_0+1)$, with equality for distinct scores, where $N_0$ counts the
sample rows of all GPUs (Appendix~\ref{app:seed-capacity-robustness} gives its
distribution).
\end{proposition}
With 2M-row tiles on four GPUs, the campaign's widest windows span 28M rows at $r_0=8$ and 30M at $r_0=4$. The law expects 218,750 and 468,750 survivors in them. Capacities 1.8\% and 1.4\% above these means, 222,605 and 475,273, bound the chance that any window of the campaign overflows by $10^{-6}$. The bound is a union bound over every query and window, and Section~\ref{app:replication} tests these capacities on two permutations.

Our library is shuffled by file, not by row, so the law's random-order assumption does not hold. Its fullest windows hold 2.3 and 1.8 times these means. A library permuted once at load could use the law's capacities. For our order we use an empirical margin instead: twice the random-order mean, rounded up to a multiple of 128,000. Two configurations use this margin. \emph{Stable} uses $r_0=8$ and $G=16$ with $A=512{,}000$. \emph{Optimized} uses $r_0=4$ and $G=16$ with $A=1{,}024{,}000$. It spends Stable's free-memory headroom on a larger buffer so that it can scan fewer seed tiles. At $B=64$, doubling $A$ adds $\Delta W=8B\,\Delta A=262.144$~MB of buffer and merge-score storage per GPU, which Section~\ref{sec:capacity} confirms. Section~\ref{sec:cost-model} gives a cost model of each plan's selections and compactions per shard.

\paragraph{Choosing the plan.}
The plans of Figure~\ref{fig:plan-space} differ in the state they keep beside the shard. The full-score plan keeps $2BN_j$ bytes of scores on a shard of $N_j$ rows, and a grouped plan of $S$ tiles keeps $2BST$. The window plan keeps $8B(A+K)$ bytes of buffers and answers beside one score tile. SweepANN chooses among them as a query optimizer does. At the batch size fixed above, a plan whose state does not fit beside the shard is excluded. Among the remaining candidates, the plan fastest on the calibration queries at the requested $K$ wins. Every plan returns the compressed-score top\nobreakdash-$K$, so the choice among the plans that fit can rest on cost alone. Two regularities hold in the tested settings (Section~\ref{sec:plan-map}, Table~\ref{tab:plan-map}). First, wherever the score matrix fits at a setting's batch, keeping all scores is fastest at the largest tested $K$. At $K=100$K, the faster plan depends on the model and the rows per GPU. Second, where the matrix does not fit, as for DrugCLIP at 500M rows on four GPUs, the window plan is the fastest candidate. It runs ahead of the widest grouped span in its workspace (Section~\ref{sec:batch-reuse-controls}).

\begin{takeaway}
Memory decides first and cost second. At the fixed batch, a plan whose state does not fit beside the shard is excluded, and the measured fastest of the rest wins. Under random row order, buffers 1.8\% and 1.4\% above the law's means bound the overflow chance by $10^{-6}$. Our file-shuffled library fills its fullest windows to 2.3 and 1.8 times these means, so SweepANN sizes its buffers by an empirical margin instead.
\end{takeaway}

\subsection{Cost Model}
\label{sec:cost-model}
The cost model counts each plan's selections and compactions per shard, the maintenance work that grows with $K$. Let $m$ be the number of tiles per shard and $L(n,K)$ the cost of selecting $K$ score-ID pairs from $n$ candidates. Scoring and the final merge across shards are common to all full-library plans, so the model omits them. A per-tile plan selects from each tile and merges the result into the running answer, so its maintenance cost per shard is about
\[
\mathcal M_{\mathrm{tile}}
=m\bigl(L(T,K)+L(2K,K)\bigr).
\]
The window plan pays the same selection cost on its $r_0$ seed tiles. On each main tile it pays a filter-and-compact cost $F(T)$ instead, and it runs one wide selection at each of the $w=\lceil m/G\rceil$ window ends:
\[
\begin{split}
\mathcal M_{\mathrm{window}}={}&r_0\bigl(L(T,K)+L(2K,K)\bigr)\\
&+(m-r_0)F(T)+wL(K+A,K),
\end{split}
\]
plus one global seed merge and cutoff exchange per batch. The window plan thus replaces $m-r_0$ large selections with compaction and $w$ wider selections. It pays for this trade with buffer memory in $W_{\mathrm{query}}$. Without a cutoff, the grouped plan selects once per group of $S$ tiles. Its cost is about $(m/S)\bigl(L(ST,K)+L(2K,K)\bigr)$, and it needs a $2BST$-byte score buffer.

The window plan saves time in two ways. It runs fewer large selections, which Section~\ref{sec:batch-reuse-controls} measures with both plans replayed and against a grouped plan. Graph replay also cuts the host's cost of launching the scan's many small kernels (Section~\ref{sec:replay-eval}). The selection term is calibrated from measurements: on one PRO~6000, $L$ interpolated from 24 measured synthetic selections predicts eight held-out cases within 8\% (Appendix~\ref{app:seed-capacity-robustness}).

\paragraph{Worked example.}
The campaign configuration shows the plan rule, the law, and the cost model at work. Each of the four GPUs holds a 125M-row shard of 63 tiles with 5.25~GB beside it, and the batch is fixed at $B=64$ for $K=500$K. The rule first excludes the full-score plan, whose scores would take 16~GB per shard. A grouped plan needs 256~MB of scores for each tile of its group. The matched control in Section~\ref{sec:batch-reuse-controls} caps the grouped plan at the window plan's workspace peaks. The widest span that fits under those caps is three tiles. The window plan keeps one score tile beside $8B(A+K)$ bytes of buffers and answers. Under Optimized, the four named terms of $W_{\mathrm{query}}$ take 1.04~GB of the merging GPU's 2.57-GB peak workspace (Section~\ref{sec:campaign}).

The law then sizes the window plan's buffer. Optimized scans $r_0=4$ seed tiles per shard, so the sample holds $N_0=32$M rows over the four GPUs. Its widest windows span 30M main-scan rows, where Proposition~\ref{prop:law} expects 468,750 survivors under random row order. A capacity of 475,273, 1.4\% above this mean, bounds the chance that any window of the campaign overflows by $10^{-6}$. Our file-shuffled library fills its fullest windows to 1.8 times this mean. The margin rule therefore doubles the mean and rounds up to a multiple of 128,000, which gives $A=1{,}024{,}000$. For Stable, $r_0=8$ gives $N_0=64$M, a mean of 218,750 for its 28M-row windows, and $A=512{,}000$.

The cost model then counts the large selections of the plans that fit. Per shard, the per-tile plan selects after each of its 63 tiles, and the three-tile grouped plan selects $K$ from up to 6M scores 21 times. With $r_0=4$ and $G=16$, the window plan runs only four seed and four window selections, and it filters and compacts its other main tiles. The held-out batch times follow these counts. They fall from 327.6~ms for the per-tile plan and 206.6~ms for the three-tile span to 146.2~ms for the window plan, which therefore wins at this setting. The full-score plan, excluded at $B=64$, runs best at $B=16$, and the window plan at $B=64$ delivers $2.51\times$ its throughput (Section~\ref{sec:physical-plan-crossover}).

\subsection{Two-Segment Graph Replay}
\label{sec:execution}
The window plan makes the work around the scoring kernel cheaper, but each batch still consists of many small operations on four GPUs: scoring, compaction, buffer resets, and window selections. Eager execution launches them one by one from the host, with idle gaps between them (Section~\ref{sec:replay-eval}). CUDA Graphs replay a captured sequence instead. One dependency limits capture: the cutoff of a batch is known only after every GPU has finished its seed tiles and the seed has been merged. SweepANN therefore splits each batch into two segments at the cutoff exchange. It captures each segment as one CUDA Graph per GPU, a seed graph before the exchange and a main graph after it (Figure~\ref{fig:overview}).

The host writes the queries into fixed device buffers and launches the four seed graphs, each computing its shard's local top\nobreakdash-$K$, $P_j$. An eager hierarchical merge on the GPUs forms the global seed $P_q$ and extracts $\tau_q$ for each query. The executor copies the cutoffs through the host into fixed device buffers. It then launches the four main graphs and an eager final merge, so a batch takes eight graph replays. Capture uses one stream per GPU and fixed addresses for the query, cutoff, candidate, and merge buffers. A GPU's seed and main graphs share one memory pool beside the shard. Between batches only the buffer contents change, so the same graphs replay for every batch.

\FloatBarrier

\begin{table*}[t]
\caption{Map of the evaluation. Each row gives the question that a subsection or section answers and its setting (rows, GPUs, $B$, $K$, queries). It also gives the headline finding as printed there and the figures and tables with the evidence. The main server has four PRO 6000 GPUs. GPU batch and campaign times are resident.}
\label{tab:results-map}
\footnotesize
\renewcommand{\arraystretch}{0.92}
\setlength{\tabcolsep}{3pt}
\begin{tabular}{@{}>{\raggedright\arraybackslash}p{.035\textwidth}>{\raggedright\arraybackslash}p{.175\textwidth}>{\raggedright\arraybackslash}p{.215\textwidth}>{\raggedright\arraybackslash}p{.38\textwidth}>{\raggedright\arraybackslash}p{.14\textwidth}@{}}
\toprule
Sec. & Question & Setting & Finding & Evidence\\
\midrule
\ref{sec:campaign} & How fast does SweepANN serve the whole campaign? & 500M rows, main server, $B=64$, $K=500$K, all 26,562 pockets & Optimized takes 61.04~s at 99.63\% mean FP32 recall, and our earlier per-tile implementation takes $5.91\times$ longer. Host delivery adds 0.09~s, and four GH200 GPUs take a median 60.59~s. & Figure~\ref{fig:app-replay}(a), Table~\ref{tab:stages}\\
\addlinespace[2pt]
\ref{sec:batch-reuse-controls} & How do batch size, answer size, and grouped selection change the window plan's gain? & 100M rows on one, two, and four GPUs, $B=1$ to 64. 500M rows on four GPUs, $B=64$, $K=1$K to 500K, 64 held-out queries. Eight-GPU server, full campaign at $B=256$ & Batching raises one-GPU throughput from 18.2 to 453.3 queries/s. The window plan's speedup over the per-tile plan grows from $1.38\times$ at $K=1$K to $2.24\times$ at $K=500$K. At 500M rows the widest grouped plan in the same workspace takes $1.41\times$ longer. On the eight-GPU server the campaign takes 40.3~s. & Figures~\ref{fig:batch-reuse} and~\ref{fig:answer-size-cost}, Tables~\ref{tab:grouped} and~\ref{tab:largek-8gpu}\\
\addlinespace[2pt]
\ref{sec:physical-plan-crossover} & When does keeping every score beat the window plan? & 500M rows, four PRO 6000 or four GH200 GPUs, $K=500$K, 64 held-out queries. 1M to 100M rows on one GPU & At 125M rows per GPU the full-score plan is best at $B=16$. The window plan fits $B=64$ and delivers $2.51\times$ its throughput, $1.10\times$ at equal batch and $1.89\times$ on GH200. On one GPU at 100M rows the full-score plan takes 130.5 against 141.3~ms but needs 13.32 instead of 2.38~GB. & Figures~\ref{fig:memory-boundary} and~\ref{fig:app-plan-crossover}\\
\addlinespace[2pt]
\ref{sec:capacity} & How should the sample and the buffer share the memory beside the shard? & 500M rows, main server, $B=64$, $K=500$K, full campaign & Optimized halves Stable's seed tiles and doubles its buffer. Its peak workspace grows by 262.2~MB, matching the 262.144~MB predicted, for 5.68\% lower resident time. Stable's fullest window holds 505,512 of its 512,000 entries. & Figure~\ref{fig:app-seed-window}, Table~\ref{tab:capacity}\\
\addlinespace[2pt]
\ref{app:replication} & How much do seed placement, query order, and row order matter? & 500M rows, $K=500$K, full campaigns on the main server, and Stable and Optimized on permuted rows on the eight-GPU and fourth servers & All four query orders stay within 0.53\% of the original time. One of four random placements overflows and takes 390.0~s with recovery under the fixed cutoff and 79.6~s under the local cutoff. On permuted rows the law's capacities hold every window. & Tables~\ref{tab:replication}, \ref{tab:res-24}, and~\ref{tab:counts-cells}\\
\addlinespace[2pt]
\ref{sec:replay-eval} & What does graph replay save? & 500M rows, main server, Stable, $B=64$, $K=500$K, five batches & Replay cuts the median batch time from 249.8 to 157.7~ms, a 36.9\% reduction. Kernel launches per batch fall from 1,868 to 121 plus eight graph replays. & Figure~\ref{fig:app-replay}(b,c)\\
\addlinespace[2pt]
\ref{sec:published-baselines} & How do the indexes and the collectors compare at 99\% mean recall? & GPU indexes on 25M rows (one GPU, $B=32$, 32 held-out queries), 100M rows (four GPUs), and 500M rows (four and eight GPUs). CPU collectors on 1M and 10M rows & CAGRA misses the target at $K=100$K and 500K even in a 140-point grid. On 25M rows IVF-Flat takes 19 to 24 times as long as the window plan. On 500M rows and eight GPUs, IVF-SQ8 takes $11.6\times$ SweepANN's resident time at both $K$. At 1M rows and $K=100$K, BBC takes 53.1~ms per query and an exact FlatIP scan 54.5~ms. & Figures~\ref{fig:gpu-index-comparison} and~\ref{fig:app-cpu-collectors}, Table~\ref{tab:res-35}\\
\addlinespace[2pt]
\ref{sec:quality} & Do the answers still serve screening? & Complete Optimized campaign. 117 targets of DUD-E and LIT-PCBA, 231 pocket queries & Every query returns the score multiset of the unpruned top\nobreakdash-$K$, over 13.3B returned pairs. EF1\% is unchanged for 106 of 117 targets, and the top 0.1\% keeps 947 of the 949 reference hits. & Figure~\ref{fig:app-enrichment}\\
\addlinespace[2pt]
\ref{sec:model-generalization} & Do the plans carry over to other embedding models? & ConGLUDe (256-d) and ConPLex (1024-d), 100M to 500M rows, four or eight GPUs, $B=32$ and 64, $K=100$K and 500K, 32 held-out queries & Mean recall exceeds 99.6\% in all four 100M and 250M settings. For ConGLUDe at $K=100$K, the window plan's time relative to the full-score plan falls from 1.42 at 100M rows to 0.98 at 250M and 0.82 at 500M. At $K=500$K the full-score plan is fastest in every setting. & Figure~\ref{fig:app-other-models}, Tables~\ref{tab:plan-map} and~\ref{tab:model-pressure}\\
\addlinespace[2pt]
\ref{sec:plan-map} & Which plan wins where? & All settings of Table~\ref{tab:plan-map}. Eight-GPU server, 500M rows, $B=64$, $K=1$M to 5M & Where the score matrix fits, the full-score plan is fastest at the largest tested $K$. At $K=5$M it takes 258.5~ms per batch against 483.2 for the window plan, which then allocates more beyond the library, 22.8 against 15.0~GB. & Tables~\ref{tab:plan-map} and~\ref{tab:largek-8gpu}\\
\addlinespace[2pt]
\ref{sec:general} & Do the plans serve public vector-search collections? & DEEP1B (1B rows, 96 dimensions) and Cohere (100M rows, 1,024 dimensions), eight PRO 6000 GPUs, $B=32$ and 64, $K=1$K to 500K, 64 held-out queries. DEEP1B also on two and four GPUs and at $B=128$ and 256 (256 held-out queries) & On DEEP1B at $B=64$ the window plan takes 48.1~ms at $K=100$K against 77.0 for the full-score plan. At $K=100$K and 500K it is 5.4 and 3.2 times faster than the full-score plan on the uncompressed FP32 rows, for 1.0 and 0.7 points of mean recall. The full-score plan is faster in all four Cohere cells. & Tables~\ref{tab:deep1b-grid}, \ref{tab:deep1b-batch}, and~\ref{tab:cohere-grid}, Figure~\ref{fig:general-frontier}\\
\addlinespace[2pt]
\ref{sec:extension} & What serves a library beyond aggregate GPU memory? & 1B logical rows on eight GPUs. 100M rows with a cuVS IVF-PQ index on one A40, 32 queries, $K=500$K & Two copies of the DrugCLIP library, 1B logical rows of 774 bytes each, stay resident with 4,408~MiB free per GPU. On the A40, one buffer of the candidate vectors would need 53.67~GB with the index. The bounded refiner completes the request, host delivery included, in 1,083.5~s at a 27.08-GB peak. It reaches 99.93\% mean recall. & Section~\ref{app:replicated1b}, Appendix~\ref{app:ann-studies}\\
\bottomrule
\end{tabular}
\end{table*}

\subsection{Implementation and Result Delivery}
\label{sec:delivery}
SweepANN is written in Python on top of PyTorch~\cite{paszke2019pytorch} tensor and selection operators, Triton kernels, CUDA Graphs, and peer-to-peer transfers. All plans select with PyTorch's GPU top\nobreakdash-$K$ operator without sorting its output. They use the 64-bit positions it returns to gather the 32-bit row IDs. Each batch ends with a global result in GPU memory, which a GPU consumer reads in place. A host consumer receives score-ID batches through two pinned host buffers that it owns. Host memory thus holds only two batches of the 79.7~GB result (Section~\ref{sec:problem}). A writer publishes durable output after flushing its files (Section~\ref{sec:campaign}).

\section{Evaluation}
\label{sec:evaluation}
After the setup (Section~\ref{sec:setup}), Section~\ref{sec:campaign} serves the 500M-row campaign end to end. Sections~\ref{sec:batch-reuse-controls} to~\ref{sec:replay-eval} then take that result apart by comparing SweepANN's plans with one another. Section~\ref{sec:published-baselines} holds the index and collector baselines to the same 99\% target, and Sections~\ref{sec:quality} to~\ref{sec:plan-map} check answer quality, other embedding models, and the plan rule. Table~\ref{tab:results-map} gives the question, setting, headline finding, and evidence of each subsection and of Sections~\ref{sec:general} and~\ref{sec:extension}.

\subsection{Experimental Setup}
\label{sec:setup}
\paragraph{Hardware and software.}
The main server has two 16-core AMD EPYC 9115 CPUs, 755~GiB of host memory, and four RTX PRO 6000 Blackwell Server Edition GPUs on PCIe, two per socket. Each GPU has a nominal 96~GB of memory, 102.0~GB as the CUDA runtime reports. A second deployment on DeltaAI uses four GH200 GPUs with 102.1~GB per device and runs the same library, queries, and Optimized configuration. Most eight-GPU runs use a third server with eight PRO 6000 GPUs, called the eight-GPU server below. A fourth server has eight PRO 6000 GPUs and two AMD EPYC 9B14 CPUs. It runs the 1B logical rows of Section~\ref{app:replicated1b} (two copies of the library), the second row permutation and counts-only campaigns of Appendix~\ref{app:seed-capacity-robustness}, and the eight-GPU runs of Section~\ref{sec:general}. Sizes are decimal GB and MB. Host memory and the Faiss index sizes of Section~\ref{sec:ann-motivation} are in GiB, and free memory is in MiB, as the driver reports it. On the main server, SweepANN runs on Python 3.10, a PyTorch 2.11 nightly for CUDA 12.8, and Triton 3.6.

\paragraph{Library and queries.}
DrugCLIP's genome-wide study reports the sources, property filters, and size of its 500M-molecule library but not the molecules themselves. We therefore rebuilt one following that description. Applied to ZINC20~\cite{irwin2020zinc20} and to the Enamine REAL compounds prepared by VirtualFlow~\cite{gorgulla2020virtualflow}, its filters pass 621M molecules. The released DrugCLIP model encoded them as vectors with $D=768$ components (explained below). To form the deployed library, we shuffled the encoded files of each source in a fixed pseudorandom order, interleaved them to keep each source's share, and took the first 500M rows. Rows keep their order within a file. The 500M prefix spans 3,005 of the 3,701 files. The 1M- to 100M-row libraries of Section~\ref{sec:ann-motivation} and of the controlled experiments are prefixes of the same order. The queries are the 26,562 pockets of DrugCLIP's genome-wide campaign~\cite{jia2026genome}, several per protein, which we encoded with the released DrugCLIP model. Pocket encoding and any pooling of a protein's pockets happen outside the search and its timing.

\paragraph{Campaign measurements.}
A campaign serves all of these pockets at $B=64$, in 415 full batches and a padded last batch of two queries. We run Stable and Optimized over the whole sequence and repeat Optimized after reloading the library. \emph{Resident time}, the main time measure, is the sum of the per-batch wall-clock intervals for query staging, both scan segments, the cutoff exchange, global merging, and overflow checks. It ends with the global result on the GPUs. Library loading and quantization, graph capture, pinned-buffer allocation, and instrumentation between batches lie outside these intervals. Separate runs measure delivery to host memory and durable output to storage.

\paragraph{Baselines and controlled experiments.}
\textbf{The baselines are the indexes and the collectors} of Sections~\ref{sec:ann-motivation} and~\ref{sec:published-baselines} and DrugCLIP's released script (Section~\ref{sec:campaign}). \textbf{The four plans are SweepANN's}: they share the fused INT8 scorer and the global merge and differ in how they keep the answer state. The full-score plan runs at its best measured setting, chosen from two storage variants over a batch sweep (Section~\ref{sec:physical-plan-crossover}). The grouped plan runs at the widest span that fits within the window plan's peak workspace (Section~\ref{sec:batch-reuse-controls}). A gap between two plans therefore comes only from how they keep the answers. In the campaign, the reference point is our earlier implementation of the per-tile plan on the same platform, library, queries, and $K$ (Section~\ref{sec:campaign}). The replay study, the dense FP16 and FP32 variants of SweepANN's full-score plan, and the external cuVS brute-force scan~\cite{cuvsBruteForceDocumentation} are described below.

The controlled experiments use the library prefixes or all 500M rows. Their fixed held-out query sets hold 32 to 256 pockets, and Table~\ref{tab:setups} lists them per experiment (Appendix~\ref{app:experiment-index}). Tunable plans are first frozen on separate calibration queries. The exceptions are Stable and Optimized, which were chosen from campaign measurements (Section~\ref{sec:capacity} and Limitations in Section~\ref{sec:conclusion}), and the $S=7$ control of Section~\ref{app:grouped}. In these controlled experiments, the window plan chooses among three configurations at 1M to 50M rows and tunes only its buffer capacity at 500M rows. CAGRA uses its fastest tested configuration that reaches 99\% mean recall. Dense and full-score plans need no tuning. Table~\ref{tab:setup-summary} summarizes the main settings, and Table~\ref{tab:control-formats} gives each plan's output format and query staging. Latency in these experiments ends with the results in GPU memory.

\paragraph{Metrics.}
We judge the outputs by selection preservation, recall, and screening enrichment. In the controlled 100M and 500M experiments and in Section~\ref{sec:model-generalization}, every result of a compressed plan preserves selection against the compressed reference of Section~\ref{sec:fused-scoring}. Section~\ref{sec:quality} checks the campaign itself. Recall (Section~\ref{sec:problem}) is measured on each plan's actual output against the FP32 reference over the same rows, so it includes the cost of compression. Screening quality is the enrichment factor at the top one percent (EF1\%). It divides the concentration of known active molecules in the top 1\% of the ranking by their concentration in the benchmark pool.

\begin{table}[t]
\caption{Main settings of the evaluation. Table~\ref{tab:setups} in Appendix~\ref{app:experiment-index} lists the setup of each experiment. P, G, H: PRO 6000, GH200, H200.}
\label{tab:setup-summary}
\footnotesize
\renewcommand{\arraystretch}{0.9}
\begin{tabular}{@{}>{\raggedright\arraybackslash}p{.29\columnwidth}@{\hspace{4pt}}>{\raggedright\arraybackslash\hangindent=0.8em\hangafter=1}p{.67\columnwidth}@{}}
\toprule
Setting & Description\\
\midrule
Main server (4P) & Four RTX PRO 6000 Blackwell Server Edition GPUs on PCIe, two per socket, 102.0~GB each as the CUDA runtime reports, two 16-core AMD EPYC 9115 CPUs, 755~GiB of host memory\\
Eight-GPU server (8P) & Eight PRO 6000 GPUs\\
Fourth server (8P) & Eight PRO 6000 GPUs and two AMD EPYC 9B14 CPUs, for the run on 1B logical rows, the second row permutation and counts-only campaigns, and the eight-GPU runs on the public collections\\
DeltaAI (4G) & Four GH200 GPUs with 102.1~GB per device, one node with NVLink between every pair\\
Other GPUs & Four H200 GPUs (4H), one A100, and two A40s\\
CPU servers & Faiss indexes on 1M--50M rows, collectors with one thread, CPU controls with 32 cores and 24 threads\\
DrugCLIP library & 500M rows, $D=768$, 26,562 pocket queries, 1M- to 100M-row prefixes\\
Other models & ConGLUDe and ConPLex on 100M, 250M, and 500M rows\\
Public collections & DEEP1B with 1B rows (8P, also 2P and 4P) and Cohere with 100M rows (8P)\\
Sizes & Decimal GB and MB, host memory and Faiss index sizes in GiB, free memory in MiB\\
Resident time & Ends with the global result on the GPUs\\
Host and storage & Separate runs measure host delivery and durable output\\
\bottomrule
\end{tabular}
\end{table}

\begin{table}[t]
\caption{Output formats and query staging in the controlled GPU experiments. cuVS brute force and CAGRA return sorted answers, whereas IVF-Flat and the compressed plans do not.}
\label{tab:control-formats}
\footnotesize
\renewcommand{\arraystretch}{0.9}
\begin{tabular}{@{}>{\raggedright\arraybackslash}p{.39\columnwidth}>{\raggedright\arraybackslash}p{.26\columnwidth}>{\raggedright\arraybackslash}p{.25\columnwidth}@{}}
\toprule
Plan & Scores / IDs & Query staging\\
\midrule
Dense FP16 &FP16 / int64&Before timing\\
Dense FP32 and cuVS brute force &FP32 / int64&Before timing\\
cuVS CAGRA &FP32 / uint32&Before timing\\
cuVS IVF-Flat &FP32 / int64&Before timing\\
Compressed plans &FP16 / int32&Included in call\\
\bottomrule
\end{tabular}
\end{table}

\begin{figure*}[t]
\centering
\includegraphics{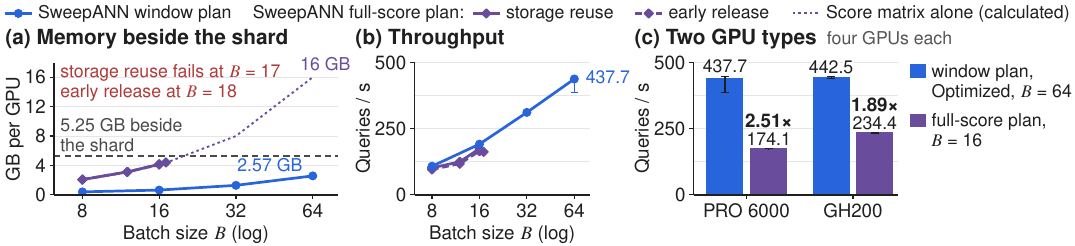}
\caption{The score matrix limits SweepANN's full-score batch size (500M rows, four GPUs, $K=500$K, 64 held-out queries, global merge included). The full-score plan runs in two versions, storage reuse (solid) and early release (dashed). (a) Per-GPU memory beside the shard: the window plan's peak workspace (Optimized), the peak allocation of both full-score versions, and the score matrix by formula. The two versions have the same peak up to $B=16$, so their lines coincide there. (b) Throughput on PRO 6000 GPUs against batch size, with whiskers for three runs of the window plan. (c) Window plan at $B=64$ (left bar of each pair) against full-score at $B=16$ (right bar) on PRO 6000 and GH200 GPUs, medians and ranges of three runs.}
\Description{Left, per-GPU memory beside the shard against batch size: the window plan stays below 2.6 GB, and the calculated full-score matrix reaches 16 GB. The dashed line marks 5.25 GB beside the shard. Storage reuse fails at B=17 and early release at B=18. Failed runs have no measured peak. Middle, the window plan's throughput rises to 437.7 queries per second at B=64 while the full-score curves end at B=16 and 17. Right, bars on two GPU types with the window plan 2.51 and 1.89 times faster.}
\label{fig:memory-boundary}
\end{figure*}

\begin{figure*}[t]
\centering
\includegraphics{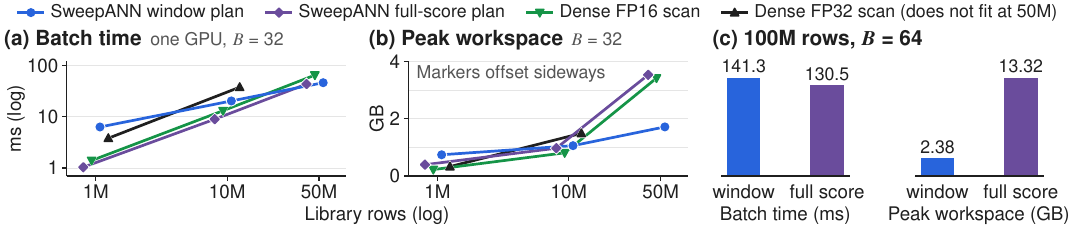}
\caption{SweepANN's full-score, window, and dense plans on smaller libraries (one GPU, $K=500$K). (a) Batch time and (b) peak workspace beside the library at $B=32$ on 1M, 10M, and 50M rows for 32 held-out queries. Dense FP32 vectors do not fit at 50M. (c) Batch time and peak workspace on 100M rows at $B=64$. Dense plans stage queries before timing, and compressed plans include the query copy. In (a,b), markers at each library size are offset sideways so that coinciding points stay visible.}
\Description{Batch time and peak workspace against library rows (1M, 10M, 50M) for the window, full-score, dense FP16 and dense FP32 plans on one GPU, and bars at 100M rows where the full-score plan takes 130.5 ms with 13.32 GB and the window plan 141.3 ms with 2.38 GB.}
\label{fig:app-plan-crossover}
\end{figure*}

\paragraph{Scores, plans, and query sets.} DrugCLIP scores a pocket against a molecule by averaging the cosine similarities of six independently trained models. Concatenating their normalized 128-dimensional outputs gives one vector with $D=768$ components. The inner product of two such vectors, divided by six, equals that average. The replay study of Section~\ref{sec:replay-eval} runs the window plan both eagerly and with graphs, timing five batches after warmup on one library load. The dense FP16 and dense FP32 plans scan the uncompressed vectors with matrix multiplication and select the top $K$ once. The external cuVS brute-force scan~\cite{cuvsBruteForceDocumentation} does the same in FP32. The held-out query sets of the controlled experiments are 32 pockets for Figures~\ref{fig:app-plan-crossover}(a,b) in Section~\ref{sec:batch-reuse-controls} and 64 pockets for Figures~\ref{fig:answer-size-cost}, \ref{fig:memory-boundary}, and \ref{fig:app-plan-crossover}(c). 

\paragraph{Tiles and platforms.}
Tile size varies with the experiment. The DrugCLIP runs score 2M-row tiles on 500M rows (Section~\ref{sec:architecture}) and 1M-row tiles on 100M rows (Figures~\ref{fig:batch-reuse}(b) and \ref{fig:app-plan-crossover}(c), Section~\ref{app:grouped}). The four-GPU 100M runs of Section~\ref{app:gpu-variants} are the exception and use 2M-row tiles, as do the other models' runs in Figure~\ref{fig:app-other-models} (Section~\ref{sec:model-generalization}). Section~\ref{app:replication} uses the campaign setting, and the row-order control of Appendix~\ref{app:seed-capacity-robustness} uses 25M rows and 1M-row tiles on one GPU. The four GH200 GPUs share one node, with NVLink between every pair.

\paragraph{CPU controls on other servers.}\phantomsection\label{app:setups}
Two further controls on the 10M rows ran on other CPU servers. The first, with 32 cores, retrains the coarse quantizer with spherical $k$-means (20 iterations, at most 64 points per list, 16.7~min). On 200 of the 1K queries, the retrained quantizer selects the same probe counts as the default one (Table~\ref{tab:ann-quality-budget}) for IVF-Flat and IVF-SQ8 at both $K$. On the other 800, both quantizers meet the 99\% mean target and differ in search time by at most 1.2\%. The second control, with 24 threads, compares an exact flat scan with IVF-Flat at $K=100$K. The flat scan of the 10M rows takes 141.0~ms per query. IVF-Flat takes 160.7~ms with a quarter of its lists (99.87\% mean recall, 17 of 1K queries below 99\%) and 562.8~ms with all of them (Appendix~\ref{app:experiment-index}). CPU time per query is batch time divided by batch size.

\begin{figure*}[t]
\centering
\includegraphics{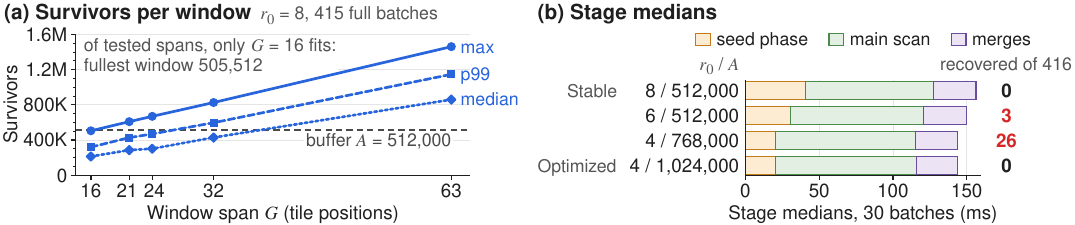}
\caption{Sample size and window span of SweepANN's window plan (500M rows, four GPUs, $B=64$, $K=500$K). (a) Survivors per window against the span $G$ at $r_0=8$: median, p99, and maximum over every query, shard, and window of the 415 full batches. The dashed line is Stable's buffer. (b) Stacked stage medians (seed phase, main scan, merges) over the first 30 batches for four of the six compared $(r_0,A)$ settings. The right column counts the batches of a full 416-batch pass that need recovery.}
\Description{Left, survivors per window against window span: among the tested spans, only the span of 16 tiles keeps the fullest window (505,512) under the buffer of 512,000. Right, stacked stage medians for four seed and buffer settings with 0, 3, 26 and 0 of 416 batches needing recovery.}
\label{fig:app-seed-window}
\end{figure*}
\subsection{End-to-End Campaign}
\label{sec:campaign}
Optimized completes the 500M-row campaign in 61.04~s of resident time, and after a fresh library reload it reproduces this time within 0.2\%. Our earlier implementation, the campaign's reference point, runs the per-tile plan eagerly from Python and takes 360.56~s on the same hardware, $5.91\times$ longer. Both keep the library compressed and fuse dequantization into scoring. Two controlled experiments isolate what selecting once per window and graph replay each contribute to this gap. With both plans replayed, the window plan is $2.24\times$ faster than the per-tile plan at $K=500$K (Section~\ref{sec:batch-reuse-controls}). Replay alone shortens the window plan's batch time by 36.9\% (Section~\ref{sec:replay-eval}). The two factors do not multiply to the whole gap, because replay was measured on the window plan, which issues fewer launches than the per-tile plan.

DrugCLIP's released screening script~\cite{drugclipCode}, one of the baselines, follows the per-tile plan. We ran it unchanged on the four PRO 6000 GPUs of the main server, with our 768-dimensional embeddings as input. It searched the first 25M rows for the first 1K pockets at $K=100$K. One process of the script takes 187.1~s from start to exit at its default batch of 4 queries, 172.4~s at a batch of 32, and 199.2~s at 64. Its FP32 scores return at least 99.999\% of the FP32 top-100K on eight checked queries. SweepANN on the same rows and queries ($B=4$, $r_0=1$, $G=2$, $A=1{,}024{,}000$) takes 7.4~s from main entry to its first delivered pass. This interval includes loading the INT8 rows, quantized beforehand, as well as initialization, graph capture, and warm-up. Each pass then takes 1.6~s with delivery (1.2~s resident) at 99.67\% mean recall on the eight checked queries (minimum 99.52\%). Appendix~\ref{app:campaign-delivery} gives the batch sweep of the script and the matched SweepANN runs.

\begin{figure*}[t]
\centering
\includegraphics{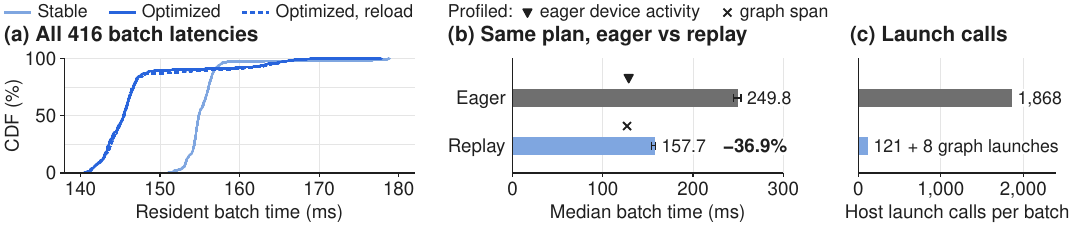}
\caption{Campaign batches and graph replay (four PRO 6000 GPUs, 500M rows, $B=64$, $K=500$K). (a) Distribution of resident time over all 416 batches. (b) Stable's window plan run eagerly or as two replayed graph segments: median and range of five batches on one library load. A triangle marks the profiled eager device activity, and a cross marks the graph span. (c) Host launch calls per batch, where each graph launch executes its captured kernels.}
\Description{Left, the cumulative distribution of all 416 batch times of Stable, Optimized and its reload. Middle, the same plan eager at 249.8 ms against replayed graphs at 157.7 ms, 36.9 percent less. Right, 1,868 kernel-launch calls per batch eager against 121 plus eight graph launches.}
\label{fig:app-replay}
\end{figure*}

Within the campaign, batch latency varies little. The latency distributions of Optimized and its reload nearly coincide, and Optimized's 99th-percentile batch latency is 167.02~ms. Stable's batches are slower because its seed phase scans twice as many tiles (Figures~\ref{fig:app-seed-window}(b) and~\ref{fig:app-replay}(a)). Loading and graph capture lie outside resident time and are paid once per library load. For Optimized, loading and quantizing the library takes 502.2~s, and graph capture takes 1.4~s.

\paragraph{Delivery to the host and to storage.}
Delivering the answers to host memory adds little to resident time. In this test, a host consumer receives every batch, checks that the scores are finite and the IDs valid, and computes a SHA-256 hash. Times are medians of three campaigns and exclude the initial buffer allocation. With the pinned double buffers of Section~\ref{sec:delivery}, the CPU processes one result while the GPUs compute the next. The complete run then takes 60.72~s, 0.09~s more than the 60.63~s of resident time of the same three campaigns. A serial consumer takes 118.08~s with pinned and 136.61~s with pageable buffers. The two pinned host buffers take 384~MB. Durable output costs more. In one run, writing the complete 79.686-GB result durably takes 156.30~s, including transfer, validation, hashing, and file publication. The durable writer stores the scores and row IDs with the query identities.

\paragraph{Second platform.}
The campaign runs as fast on the second platform. On four GH200 GPUs, the median resident time over three runs is 60.59~s, similar to that of the main server. One durable output takes 137.05~s instead of 156.30~s.

\paragraph{Output quality.}
The campaign returns complete answers that meet the recall target. Every query of the complete Optimized output has exactly $K$ distinct valid row IDs with finite scores. Against the FP32 reference, the output reaches 99.63\% mean and 98.52\% minimum recall, and four of the 26,562 queries fall below 99\%. Every saved answer contains the FP32 top-100K, so every reference row that an answer misses ranks below that prefix. The reference lists the exact FP32 top 1M rows of each query over the full 621M-row library. At least $K$ of them lie among the 500M deployed rows. Recomputing the reference over the deployed rows with TF32 disabled, for 64 held-out queries, changes each query's recall by at most $4\times10^{-6}$ (Appendix~\ref{app:campaign-delivery}). Section~\ref{app:grouped} reports the recall on pockets of 256 further proteins, chosen at random.

\paragraph{Stages of a 500M-row batch.}\phantomsection\label{app:stages}
The two platforms reach similar batch times by different routes: the GH200 GPUs spend longer in the main graph and less in the merges. Table~\ref{tab:stages} splits one batch of the window plan into four synchronized segments, with Optimized, $B=64$, $K=500$K, and the queries of Figure~\ref{fig:memory-boundary}(c). Each column of the table is one instrumented pass. On GH200, the main segment is 17.3~ms longer, and the two merge segments are together 17.0~ms shorter. Three uninstrumented runs on the same queries have medians of 146.2~ms on PRO 6000 and 144.6~ms on GH200 GPUs (Figure~\ref{fig:memory-boundary}(c)). The seed merge and the final merge follow the PCIe topology of the main server. They first combine the two GPUs of each socket and then the two results on GPU~0. In the final merge, GPU~0 first combines its running answer with the seed. Scores and row IDs move together, and selection runs on the GPUs. Measured per GPU as CUDA-event spans, the main graph takes 95.2--95.9~ms on each PRO 6000 GPU and 112.0--113.2~ms on each GH200. Table~\ref{tab:stages} instead reports the synchronized host interval, which includes graph submission and synchronization. The platforms run PyTorch 2.11 with CUDA 12.8 and PyTorch 2.13 with CUDA 13.0, respectively. The measurement does not split the main graph into scoring, compaction, and window selection, nor the merges into transfers and computation.

\begin{table}[t]
\caption{Stages of one instrumented 500M-row batch of Optimized ($B=64$, $K=500$K), as synchronized host intervals in ms.}
\label{tab:stages}
\footnotesize
\renewcommand{\arraystretch}{0.9}
\begin{tabular}{@{}lrr@{}}
\toprule
Segment & 4 PRO 6000 & 4 GH200\\
\midrule
Query copy and seed graph &20.5&18.2\\
Seed merge and cutoff exchange &12.6&5.7\\
Main graph &96.0&113.3\\
Final merge and overflow check &16.7&6.6\\
\midrule
Batch &145.9&143.9\\
\bottomrule
\end{tabular}
\end{table}

\paragraph{Per-GPU memory.}
The merging GPU has little memory to spare in the campaign. On GPU~0 under Optimized at $B=64$, the four named terms of $W_{\mathrm{query}}$ take 1.04~GB of its 2.57-GB peak workspace. They are the score tile, the candidate buffers, the running answers, and the merge-score buffer. Selection scratch, counters, graph pools, and merge state ($W_{\mathrm{other}}$) take the rest, and 158~MiB is free in the timed PRO 6000 runs (Appendix~\ref{app:campaign-delivery}).

\begin{takeaway}
Optimized, a \textbf{window plan} configuration, answers all 26,562 pockets over 500M rows in 61.04~s of resident time at 99.63\% mean FP32 recall. That is $5.91\times$ faster than our earlier eager implementation of the \textbf{per-tile plan}. Double-buffered delivery to the host adds only 0.09~s, and four GH200 GPUs take a median 60.59~s.
\end{takeaway}

\subsection{Batch Size, Answer Size, and Grouped Selection}
\label{sec:batch-reuse-controls}
The campaign's speed starts with batching, because one read of a tile serves every query of the batch. Figure~\ref{fig:batch-reuse}(b) fixes 100M rows and $K=500$K and runs the 64 held-out queries at $B=1,8,16,32,64$, with the sample fixed at 16M rows in total. On one GPU, throughput rises from 18.2 to 453.3 queries/s. Four GPUs reach 943.0 queries/s at $B=64$, which is $2.08\times$ the one-GPU rate, against $3.62\times$ at $B=1$. Scaling across GPUs is sublinear at large batches. Each GPU still maintains a $K$-entry answer per query for its shard, and the global merge combines one such answer per GPU.

The shared read is visible in the profile. Under Nsight Compute profiling with cold caches, one scoring kernel over 2M rows reads about 1.548~GB of DRAM at $B=1$, 8, and 64 alike (Figure~\ref{fig:batch-reuse}(a)). Reads per query thus fall to 24.19~MB at $B=64$. At that batch, profiled outside the throughput runs, the kernel moves 1,400.5~GB/s. That is 87.7\% of the DRAM peak against 41.5\% of peak compute, so memory traffic, not arithmetic, limits it.

The window plan's advantage over the per-tile plan grows with the answer size, from $1.38\times$ at $K=1$K to $2.24\times$ at $K=500$K (Figure~\ref{fig:answer-size-cost}). This control varies $K$ on all 500M rows with four GPUs and $B=64$. Both plans replay as CUDA Graphs with the same fused scorer and the full global merge, and the window plan also builds its seed and exchanges the cutoff. 

At twice the campaign's answer size, the window plan still fits beside the shard. It serves $K=1$M on the 500M rows ($B=32$, the 64 held-out queries, four PRO 6000 GPUs), where the full-score plan does not fit at $B=32$ (Section~\ref{sec:physical-plan-crossover}). With $A=1{,}920{,}000$, it takes 125~ms per batch resident and 183~ms to the host, without overflow. Every answer contains all of the FP32 top-500K. With $A$ set to the law's mean of 937,500, both batches overflow, and each completes in 2.4~s including recovery (Appendix~\ref{app:campaign-delivery}). Each window run is one pass, and recall at $K=1$M itself is not measured.

\paragraph{Larger answers on eight GPUs.}
Where the score matrix fits, keeping every score stays fastest even for the largest tested answers. On the eight-GPU server (62.5M-row shards), the full-score, window, and per-tile plans ran all 500M rows at $B=64$ and $K=1$M, 2M, and 5M. Each ran five timed repetitions on the 64 held-out queries (Table~\ref{tab:largek-8gpu}). No run needed a smaller fall-back batch, and no buffer overflowed. The full-score plan is fastest at every $K$, and the per-tile plan allocates least and is slowest. Every plan's output has the score multiset of a reference whose $K$-th score was computed over every tile. Every returned pair was also recomputed from the compressed rows and read back by an independent CPU check. Recall counts matching row IDs against the first $K$ rows of an FP32 top-5M reference computed on the CPU. The reference divides the inner products by six, does not normalize, and breaks ties by row ID. Memory is the largest per-GPU PyTorch allocator peak over setup, capture, warm-up, and serving, library included. The library alone takes 48.38~GB. The plans are fixed, not selected on calibration queries (Appendix~\ref{app:campaign-delivery}).

\begin{table}[t]
\caption{Answer sizes of $K=1$M to 5M on 500M rows (eight PRO 6000 GPUs, $B=64$, 64 held-out queries). Medians of five repetitions in ms, resident and to host arrays. Largest per-GPU allocated (Alloc.) and reserved (Resv.) peaks in GB, library included. Mean and minimum recall in \%.}
\label{tab:largek-8gpu}
\footnotesize
\renewcommand{\arraystretch}{0.86}
\setlength{\tabcolsep}{3pt}
\begin{tabular}{@{}rlrrrrrr@{}}
\toprule
$K$ & Plan & Resident & Host & Alloc. & Resv. & Mean & Min.\\
\midrule
1M & Full score & 116.7 & 467.4 & 57.78 & 60.36 & 99.67 & 99.35\\
1M & Window & 151.9 & 495.1 & 52.93 & 56.17 & 99.67 & 99.35\\
1M & Per-tile & 288.1 & 628.2 & 51.58 & 54.53 & 99.67 & 99.35\\
2M & Full score & 152.5 & 832.9 & 59.19 & 64.32 & 99.71 & 99.42\\
2M & Window & 228.1 & 922.3 & 57.48 & 62.90 & 99.71 & 99.42\\
2M & Per-tile & 502.5 & 1,187.0 & 54.52 & 60.42 & 99.71 & 99.42\\
5M & Full score & 258.5 & 1,976.9 & 63.42 & 76.26 & 99.76 & 99.51\\
5M & Window & 483.2 & 2,035.8 & 71.15 & 80.53 & 99.76 & 99.51\\
5M & Per-tile & 850.2 & 2,257.5 & 62.46 & 74.77 & 99.76 & 99.51\\
\bottomrule
\end{tabular}
\end{table}

\paragraph{Grouped selection in the same workspace.}
Without a cutoff, a scan can still select less often by scoring several tiles into one buffer, as the grouped plan does (Section~\ref{sec:architecture}). At 100M rows, grouping alone obtains nearly all of the window plan's gain, but at 500M rows it does not. On one GPU with 100M rows, $B=64$, $K=500$K, and 1M-row tiles, the full-score plan fits and is the measured choice (Section~\ref{sec:physical-plan-crossover}). There we capped the grouped plans at the window plan's peak allocated workspace, 2.38~GB (Table~\ref{tab:grouped}). The largest tested span within this cap, 7M rows, takes 145.1~ms per batch on the 64 held-out queries. The window plan, timed in the same session, takes 142.1~ms (141.3~ms in the runs of Section~\ref{sec:physical-plan-crossover}). The per-tile plan, which selects after every tile, takes 389.2~ms. To host memory, the same three plans take 199.0, 199.2, and 443.3~ms.

On all 500M rows and four GPUs, with the window plan's memory peaks as caps, the largest fitting span is three 2M-row tiles. On the held-out queries it takes 206.6~ms per batch against 146.2~ms for the window plan, $1.41\times$ longer, and 261.9 against 200.2~ms to the host. The gap persists in two further controls, described below. One uses cuVS selection in place of PyTorch's operator. The other runs on four GPUs of the eight-GPU server, where the gap is $1.31$--$1.49\times$ across the tested workspace budgets. The gap is consistent with the number of large selections each plan runs (Section~\ref{sec:cost-model}). At 500M rows, the grouped plan selects $K$ from up to 6M scores 21 times per shard, against four seed and four window selections for the window plan. At 100M rows, the window plan's 16M-row sample in 1M-row tiles adds seed selections that offset those it saves, so grouping comes close there.

\phantomsection\label{app:grouped}
\paragraph{Grouped selection on 100M rows.}
The 100M-row control behind these numbers uses one PRO 6000 GPU, the first 100M rows of the compressed library, $B=64$, and $K=500$K. Every plan scores 1M-row tiles with the same fused scorer, runs as CUDA Graphs, and returns FP16 scores with 32-bit row IDs. A grouped plan with span $S$ scores $S$ consecutive tiles into one score buffer, selects their top $K$, and merges them into the running answer. The per-tile plan is the special case $S=1$. The window plan keeps the setting of Figure~\ref{fig:app-plan-crossover}(c): a 16M-row sample, $G=16$, and $A=1{,}024{,}000$.

We fixed spans of 1, 2, 4, 8, and 16 tiles in advance. The budget is the window plan's peak allocated workspace on 64 calibration queries, 2.383~GB, and it counts allocated memory. On these calibration queries, the spans of 8 and 16 tiles exceed the budget. After seeing the held-out results of this grid, we added $S=7$. It is the largest span that a memory formula, fitted to the calibration runs, predicts to fit. We measured it on the calibration queries before its held-out run. Both query sets are those of the earlier 100M experiments, so the control does not test fresh queries. Each plan runs in a fresh process with three timed repetitions after warmup, and Table~\ref{tab:grouped} gives the medians. Every held-out repetition of every plan returns the per-tile plan's score multiset with distinct valid IDs, and all returned scores were recomputed from the compressed rows.

\begin{table}[t]
\caption{Grouped selection on 100M rows (one GPU, $B=64$, $K=500$K). Columns give peak allocated (Alloc.) and reserved (Resv.) workspace and the median batch time until the answers are in GPU memory (resident) or in host memory (host).}
\label{tab:grouped}
\footnotesize
\renewcommand{\arraystretch}{0.9}
\begin{tabular}{@{}lrrrr@{}}
\toprule
Plan & Alloc.\ (GB) & Resv.\ (GB) & Resident (ms) & Host (ms)\\
\midrule
Per-tile ($S=1$) &1.67&2.00&389.2&443.3\\
Grouped, $S=2$ &1.73&1.94&231.5&285.9\\
Grouped, $S=4$ &1.98&2.19&168.9&217.5\\
Grouped, $S=7$ &2.37&2.59&145.1&199.0\\
Window &2.38&3.43&142.1&199.2\\
\midrule
\multicolumn{5}{@{}l}{Over the budget, calibration queries only}\\
Grouped, $S=8$ &2.50&2.72&139.7&196.7\\
Grouped, $S=16$ &3.52&4.00&123.8&178.5\\
\bottomrule
\end{tabular}
\end{table}

\paragraph{Grouped selection on 500M rows.}
On all 500M rows and four GPUs, the caps are the window plan's allocated and reserved peaks on each GPU. Spans of one to three tiles fit within them, and a span of four exceeds the reserved cap on three GPUs. The three-tile span of the comparison above was chosen on the calibration queries. In the same runs, the per-tile plan takes 327.6~ms (325~ms in the separate runs of Figure~\ref{fig:answer-size-cost}). All returned scores were recomputed from the compressed rows, and Table~\ref{tab:res-07} in Appendix~\ref{app:campaign-delivery} lists every run with its range.

On four GPUs of the eight-GPU server, the window plan is faster than every grouped span that ran, including spans beyond its caps. It takes 140.5~ms per batch on the held-out queries, against 208.4~ms for the three-tile span under that server's window caps. Four tiles exceed the caps but fit the physical memory and take 192.3~ms. In a separate set of runs, five tiles fit the physical memory and take 184.9~ms against 140.7~ms for the window plan, $1.31\times$ longer. Six tiles failed allocation in the original grid.

Over whole campaigns on those four GPUs ($B=64$, one pass each), the three-tile span and the per-tile plan run within the window plan's caps and without overflow. The window plan takes 57.3~s resident and 112.0~s to the host. The three-tile span takes 85.2 and 137.7~s, $1.49\times$ and $1.23\times$ the window plan's times. The per-tile plan takes 135.4 and 218.8~s, $2.36\times$ and $1.95\times$. All three reach 99.64\% mean and 99.29\% minimum held-out recall. A second window-plan pass takes 57.4 and 128.0~s at the same displayed recall (Appendix~\ref{app:campaign-delivery}).

\paragraph{Local cutoff.}
A local cutoff lets fewer candidates into the buffers on 100M rows but barely changes the batch time. This control keeps the rows, queries, and window plan of the 100M grouped-selection control. At each window end, it raises a query's cutoff to the $K$-th score of the GPU's running answer once that answer holds $K$ real rows. These are distinct main-scan rows, so Lemma~\ref{lem:window} applies to them. The update runs on the GPU in the main graph. On the held-out queries, the raised cutoff lets 97.3M main-scan candidates into the buffers instead of 171.7M. But the largest window count of one query falls only from 621,034 to 613,067, and the median batch takes 142.9 instead of 143.5~ms (196.5 instead of 195.5~ms to the host). The ranges of the three repetitions overlap. A separate set of runs raises the cutoff after each of the 84 main-scan tiles. This lets in 86.7M candidates but takes 274.0~ms per batch against 142.7~ms with the fixed cutoff (329.0 against 197.8~ms to the host). All variants peak at 2.38~GB allocated and 3.43~GB reserved and return the per-tile plan's score multiset.

\paragraph{Selection primitive.}
Replacing PyTorch's top\nobreakdash-$K$ operator with cuVS radix selection speeds the grouped plans and slows the per-tile and window plans. On the 100M held-out queries, in a separate run that times both operators, the 7M-row grouped plan falls from 145.9 to 134.3~ms per batch. The window plan rises from 143.4 to 147.3~ms and the per-tile plan from 391.3 to 410.3~ms. A second control, on all 500M rows of the main server, uses a scorer configuration of its own. There cuVS speeds the three-tile grouped plan by $1.06\times$, from 210.7 to 198.0~ms resident, and slows the window plan from 144.8 to 150.4~ms. The grouped plan then takes $1.32\times$ as long as the window plan, against $1.45\times$ with PyTorch's operator. With cuVS, the window plan exceeds its memory caps at both sizes, as does the 7M-row grouped plan at 100M rows. Every query keeps at least 99.28\% recall against the FP32 reference (Table~\ref{tab:res-34} in Appendix~\ref{app:campaign-delivery}).

\paragraph{Scaling to eight GPUs.}
On all eight GPUs of the eight-GPU server, the window plan serves the whole campaign in 40.3~s of resident time. Each GPU holds a 62.5M-row shard, and the plan uses $r_0=4$, $G=16$, and $A=1{,}024{,}000$, with a 64M-row sample. It runs at $B=256$, the batch size chosen on calibration batches by resident throughput, and serves the 26,562 queries in one pass with no overflow. With synchronous copies after each batch, the answers reach host arrays after 132.3~s. For this run, the 92.0~s between the two endpoints is the copy of each batch's scores and identifiers into independently retained host arrays. These copies move 79.7~GB in total and run after the resident timer stops. The disk writes, hashing, and verification follow and lie outside both endpoints (Appendix~\ref{app:campaign-delivery}). On the 64 held-out queries, the served answers reach 99.64\% mean and 99.28\% minimum recall against the TF32-off FP32 reference.

Two fresh repeats of this campaign on the same server, each verified pair by pair, confirm the resident time. They take 40.28 and 40.53~s resident (40.29~s in the original run) and 134.0 and 152.3~s to host arrays. Both reach 99.64\% mean and 99.28\% minimum recall on the same 64 queries. The resident times of the three complete campaigns thus spread over about 0.25~s, while the host time varies with the synchronous copies. We also drew a further set of 64 pockets at random before computing their recall. On them, both repeats reach 99.63\% mean recall at $K=500$K against the FP32 reference, with a minimum of 99.29\%.

The four-GPU window-plan campaigns on this server took 57.3 and 57.4~s resident (above) at $B=64$. They therefore differ from the eight-GPU campaign in batch size as well as in GPU count. At the same $B=64$ on the held-out queries, eight GPUs take 103.6~ms per batch resident and 270.3~ms to the host, against 140.5 and 245.9~ms on four ($1.36\times$, $0.91\times$).

Doubling the GPUs shortens the per-tile plan's batch more than the window plan's. A paired measurement on the same server runs the window and per-tile plans on four GPUs (125M-row shards) and then on all eight (62.5M-row shards). Each run uses a fresh process with three timed repetitions on the 64 held-out queries. The window plan's settings stay fixed ($r_0=4$, $G=16$, $A=1{,}024{,}000$, four 2M-row seed tiles per shard), so its sample grows from 32M to 64M rows. From four to eight GPUs, the window plan goes from 140.8 to 103.6~ms resident ($1.36\times$) and from 313.5 to 279.4~ms to the host ($1.12\times$). The per-tile plan goes from 331.0 to 181.2~ms resident ($1.83\times$) and from 507.0 to 355.4~ms to the host ($1.43\times$). Resident times agree with the separate runs above within 0.3~ms. Host times differ between the two sets, 313.5 against 245.9~ms on four GPUs, so host ratios are compared within one set only. Every repetition reaches 99.64\% mean and at least 99.28\% minimum recall against the FP32 reference without overflow. The per-tile plan selects once per tile, so its selections per shard halve, from 63 to 32. The window plan keeps four seed tiles and four seed selections per shard, and its windows drop only from four to two. Its seed merge also combines eight lists instead of four, so doubling the GPUs gains it less.

Within the window plan's caps, the fastest grouped span is slower than the window plan at $B=64$, 128, and 256 (held-out sets of 64, 128, and 256 queries). Outside the caps, a 15-tile span ($B=64$) and the full-score plan ($B=128$, 256) are faster. Over the whole campaign at $B=256$, the full-score plan takes 39.5~s resident and 121.6~s to the host at 99.64\% mean and 99.29\% minimum recall. The window plan takes 40.3 and 132.3~s but allocates 58.7~GB per GPU against the full-score plan's 86.3~GB.

\paragraph{Pockets of further proteins.}
Recall also holds on pockets of further proteins. Before any result on these pockets was read, a fixed rule excluded 1,436 proteins by UniProt accession. They include every protein with a pocket among the calibration and held-out queries of the eight-GPU runs. The rule then drew 256 further proteins at random, one pocket each. Against the exact FP32 top-500K, the four-GPU re-run of Section~\ref{sec:quality} at $B=64$ reaches 99.63\% mean and 99.32\% minimum recall. The eight-GPU window campaign above, at $B=256$, reaches 99.63\% and 99.33\%. No query falls below 99\%, and the two recall distributions nearly coincide although the runs differ in GPU count, batch size, and sample (Table~\ref{tab:res-28} in Appendix~\ref{app:campaign-delivery}).

\phantomsection\label{app:replicated1b}
\paragraph{One billion logical rows on eight GPUs.}
Two copies of the library, 1B logical rows, fit on eight GPUs at the main server's load of 125M rows per GPU. The second copy carries row IDs offset by 500M. We loaded them onto the eight GPUs of the fourth server (Section~\ref{sec:setup}), 125M rows per GPU. Copy one is on GPUs 0 to 3 and copy two on GPUs 4 to 7. The 774~GB compressed payload loads in 171.0~s and occupies 96.75~GB per GPU of the 102.0~GB the runtime reports, leaving 4,408~MiB free. Setting up the window plan ($r_0=4$, $G=16$, $A=1{,}024{,}000$) raises the allocation to 97.72~GB per GPU, with a peak of 99.26~GB. After capture and warm-up, the peak is 99.32~GB on GPU 0, which merges. The reserved pool is then 99.41 to 101.07~GB per GPU, with 1,724 to 144~MiB free, the least on GPU 0.

The window plan serves this library without overflow. For the 64 held-out queries at $B=64$ and $K=500$K, three repetitions take 155.0, 155.1, and 167.1~ms resident and 252.4, 250.9, and 265.9~ms to the host. At the same 125M rows per GPU, four GPUs of the eight-GPU server take 140.8~ms resident (above), so twice the rows on twice the GPUs take 10\% longer. The host endpoint is not compared across the two servers, because their CPUs differ. Every returned pair was recomputed from the compressed rows and read back by an independent CPU check. Because every vector appears twice, the exact FP32 top-500K of the doubled library is the original top-250K with each row twice. Against this reference, the served answers reach 99.59\% mean recall, with a minimum of 99.15\%, in all three repetitions. The run thus measures capacity and serving time at 1B logical rows, not a search over 1B distinct compounds.

\begin{takeaway}
On 500M rows, the \textbf{window plan}'s speedup over the \textbf{per-tile plan} grows from $1.38\times$ at $K=1$K to $2.24\times$ at $K=500$K. Grouping alone nearly matches the window plan at 100M rows. At 500M rows, the widest \textbf{grouped plan} in the same workspace takes $1.41\times$ longer, with 21 large selections per shard against eight.
\end{takeaway}

\subsection{Full-Score versus Window Plans: the Memory Boundary}
\label{sec:physical-plan-crossover}
At 125M compressed rows per GPU, the score matrix limits the full-score plan to small batches. This plan uses the same fused scorer as the others, keeps every score, and selects once per shard. We test two versions of it: one reuses the score-matrix storage for the global merge (storage reuse), and the other frees that storage first (early release). Its best setting is $B=16$ with storage reuse: scoring and selection replay as a graph, followed by an eager merge. The plan runs out of memory at $B=17$ with storage reuse and at $B=18$ with early release (Figure~\ref{fig:memory-boundary}(a,b)). The window plan fits at $B=64$.

On the 64 held-out queries and all 500M rows, the full-score plan at $B=16$ serves 174.1~queries/s. The window plan at $B=64$ serves 437.7~queries/s, a $2.51\times$ gain. At equal batch size, $B=16$, which is below the batch SweepANN fixes for tile reuse, the window plan is only $1.10\times$ faster. Most of the gain therefore comes from the larger batch that only the window plan fits. On GH200 the gain is $1.89\times$. There the window plan runs at about the same rate (442.5~queries/s), while the full-score plan is faster than on the PRO 6000 (234.4~queries/s).

Where the score matrix fits beside the library, the full-score plan is fastest at the largest tested answer, at the price of a much larger workspace. On one GPU at 100M rows and $B=64$, it takes 130.5~ms per batch against the window plan's 141.3~ms, but it needs 13.32~GB of peak workspace instead of 2.38~GB (Figure~\ref{fig:app-plan-crossover}(c)). At $K=500$K with 1M to 50M rows (Figure~\ref{fig:app-plan-crossover}(a)), it is fastest at every size. On 25M rows, where the score matrix fits at every $K$, the window plan is faster up to $K=100$K and the full-score plan at $K=500$K (Table~\ref{tab:plan-map}).
\begin{table}[b]
\caption{Measured choice among SweepANN's plans, faster first: window (W) or full score (F), with resident ms per batch (s per campaign in the DrugCLIP 8P, $B=256$ row). No index appears in the table. On DrugCLIP vectors from $K=10$K to 500K, the GPU IVF indexes that reach 99\% take 8.2 to 24$\times$ SweepANN's resident time, and CAGRA misses the target at $K=100$K and 500K (Section~\ref{sec:published-baselines}). Where a cell gives one time, it belongs to the plan selected on the calibration queries. $^*$: F does not fit at the row's $B$, and the ratio is W's throughput at $B=64$ over F's at its fastest setting, $B=16$. $^\dagger$: the second executor of Appendix~\ref{app:matched-models}. P, G, H: PRO 6000, GH200, H200. Both: ConGLUDe and ConPLex. DEEP1B and Cohere: the public image descriptors and text embeddings of Section~\ref{sec:general}.}
\label{tab:plan-map}
\footnotesize
\renewcommand{\arraystretch}{0.92}
\setlength{\tabcolsep}{2.5pt}
\begin{tabular}{@{}lllll@{}}
\toprule
Vectors & Rows (/GPU) & GPUs, $B$ & $K=100$K & $K=500$K\\
\midrule
DrugCLIP & 25M (25M) & 1P, 32 & W & F\\
DrugCLIP & 100M (100M) & 1P, 64 & & F 130.5 / W 141.3\\
DrugCLIP & 100M (25M) & 4P, 32 & W 21.2 & F 27.6 / W 37.1\\
DrugCLIP & 500M (125M) & 4P, 64$^*$ & & W, $2.51\times$ ($1.10\times$ at $B=16$)\\
DrugCLIP & 500M (125M) & 4G, 64$^*$ & & W, $1.89\times$\\
DrugCLIP & 500M (62.5M) & 8P, 256 & & F 39.5 / W 40.3~s\\
DrugCLIP & 500M (62.5M) & 8P, 32 & W 45.9 / F 51.0 & F 59.4 / W 66.3\\
DrugCLIP & 500M (62.5M) & 8P, 64 & \multicolumn{2}{@{}l@{}}{F at $K=1$M--5M (at 5M, F 258.5 / W 483.2)}\\
Both\smash{$^\dagger$} & 100M (25M) & 4P, 32 & F & F\\
Both\smash{$^\dagger$} & 250M (62.5M) & 4P, 32 & W by $<$0.6~ms & F\\
ConPLex\smash{$^\dagger$} & 250M (62.5M) & 4H, 32 & W 74.3 / F 78.7 & F 86.6 / W 92.4\\
ConGLUDe\smash{$^\dagger$} & 500M (125M) & 4P, 32 & W 46.7 / F 56.6 & F 78.2 / W 103.2\\
ConGLUDe\smash{$^\dagger$} & 500M (125M) & 4G, 32 & W 51.7 / F 62.5 & F 71.9 / W 78.3\\
ConPLex\smash{$^\dagger$} & 500M (62.5M) & 8P, 32 & W 68.2 / F 68.3 & F 109.2 / W 160.9\\
ConPLex\smash{$^\dagger$} & 500M (62.5M) & 8P, 64 & & F 199.8 / W 310.2\\
DEEP1B & 1B (125M) & 8P, 32 & W 27.3 / F 42.9 & F 51.1 / W 51.3\\
DEEP1B & 1B (125M) & 8P, 64 & W 48.1 / F 77.0 & F 93.7 / W 95.7\\
Cohere & 100M (12.5M) & 8P, 32 & F 14.8 / W 17.3 & F 23.2 / W 37.6\\
Cohere & 100M (12.5M) & 8P, 64 & F 25.0 / W 29.9 & F 41.1 / W 70.1\\
\bottomrule
\end{tabular}
\end{table}
Dense FP16 scoring competes on the smallest libraries but needs about twice the memory of the compressed rows. On these libraries, the window, full-score, and dense plans all exceed 99.7\% mean recall, so time and memory decide among them.

\begin{takeaway}
At 125M rows per GPU, the \textbf{window plan} fits $B=64$ beside the shard. It delivers $2.51\times$ the throughput of the \textbf{full-score plan} at that plan's best batch, $B=16$, and $1.10\times$ at equal batch. Where the score matrix fits, as on one GPU at 100M rows, the full-score plan is faster but needs 13.32~GB of workspace instead of 2.38~GB.
\end{takeaway}

\subsection{Sample and Buffer Configuration}
\label{sec:capacity}
Trading seed tiles for buffer capacity shortens the campaign. Optimized scans half as many seed tiles as Stable and has twice the buffer capacity (Figure~\ref{fig:app-seed-window}(b)): $r_0=4$ and $A=1{,}024{,}000$, against 8 and 512,000 for Stable. Its peak workspace is 262.2~MB larger, matching the 262.144~MB predicted in Section~\ref{sec:seed-capacity}, and in return its resident time is 5.68\% lower. With the default seed-tile placement, Stable, Optimized, and the reload all finish without overflow or growth in reserved memory. About 398~MiB stays free on GPU~0 under Stable and 158~MiB under Optimized (Table~\ref{tab:capacity}). Stable's buffer comes close to full: its fullest window holds 505,512 of its 512,000 entries (Figure~\ref{fig:app-seed-window}(a)). Section~\ref{app:replication} tests the buffer capacities further, on permuted rows, with capacities fitted on a calibration split, and under random seed-tile placements.

\paragraph{Choosing Stable and Optimized.}
Table~\ref{tab:capacity} lists both configurations. Stable was chosen from the campaign's survivor counts and window-merge costs. Optimized was chosen from six $(r_0,A)$ pairs compared on the first 30 batches (Figure~\ref{fig:app-seed-window}). The figure shows what constrains these choices. Survivors per window grow with the span, so with Stable's buffer only $G=16$ among the tested spans fits every window. A smaller sample with too small a buffer causes recovery: at $K=500$K with $A=768{,}000$, 26 of the 416 batches in Figure~\ref{fig:app-seed-window}(b) need it. A separate recovery exercise with $A=128{,}000$ at $K=100$K also overflows 26 batches, which an untimed pass recomputes with the per-tile plan, exact under $s$. Figure~\ref{fig:app-replay}(a) shows the latency distributions of all 416 batches of Stable, Optimized, and the reload.

\begin{table}[t]
\caption{The Stable and Optimized configurations over the full campaign (500M rows, four PRO 6000 GPUs, $B=64$, $K=500$K).}
\label{tab:capacity}
\small
\begin{tabular}{@{}lrr@{}}
\toprule
Property & Stable & Optimized\\
\midrule
Seed tiles per shard ($r_0$) &8&4\\
Buffer capacity per query ($A$) &512,000&1,024,000\\
Peak workspace (MB) &2,310.2&2,572.4\\
Free memory (MiB, approx.) &398&158\\
Overflow batches / 416 &0&0\\
Growth in reserved memory (MB) &0&0\\
Resident time (s) &64.716&61.042\\
\bottomrule
\end{tabular}
\end{table}

\begin{takeaway}
A smaller sample needs a larger buffer. Optimized halves Stable's seed tiles and doubles its buffer, which costs 262.2~MB more peak workspace, as predicted (262.144~MB), and saves 5.68\% of resident time. Neither configuration overflows in the campaign, although Stable's fullest window holds 505,512 of its 512,000 entries.
\end{takeaway}

\subsection{Robustness to Seed Placement and Query Order}
\label{app:replication}
Query order has little effect on the campaign, whereas seed-tile placement can change buffer pressure enough to cause overflow. The placement of the seed tiles determines the cutoff and, through it, how many survivors the buffers must hold. By default, the seed tiles are evenly spaced over each shard. We fixed four random placements in advance. Each chooses one full tile from each of four consecutive regions of every shard and keeps the sample size, query order, and buffer capacity unchanged. Three placements complete the campaign without overflow, and the fourth overflows in its first pass. Table~\ref{tab:replication} gives the time range of the three completed placements.

\begin{table}[t]
\caption{Seed-tile placement and query order in the 500M-row campaign, all with Optimized (Table~\ref{tab:capacity}) at $K=500$K. Each completed setting has three timed passes over all 26,562 queries, separate from those of Section~\ref{sec:campaign}. Its separately written answer is checked against the FP32 reference for every query. Times are ranges of per-setting medians.}
\label{tab:replication}
\footnotesize
\renewcommand{\arraystretch}{0.9}
\begin{tabular}{@{}lrr@{}}
\toprule
Changed factor & Completed & Resident time (s)\\
\midrule
Seed-tile placement &3/4&62.02--62.30\\
Query order &4/4&61.66--61.98\\
\bottomrule
\end{tabular}
\end{table}

For query order, we keep the default seed tiles and run four orders: the original, its reverse, and two random permutations fixed in advance. All four complete without overflow. Their campaign medians lie within 0.53\% of the original order's time, and the p99 latency of full batches stays near 169~ms. The written answers are mapped back to the original query IDs and checked against the FP32 reference. The rest of this section examines the overflowing placement and the survivor law behind the buffer sizes.

\paragraph{Live recovery with an unfavorable placement.}
Recovery keeps the answers intact but multiplies the resident time. We rerun the overflowing placement as one campaign with live recovery and the same capacity parameters. Each overflowing batch is recomputed with the per-tile plan, and the CUDA Graphs are then recaptured. Of the 416 batches, 106 overflow and are recomputed, and mean recall stays at 99.63\%. Resident time, including recomputation and recapture, is 348.2~s, about 5.7 times the 61.04~s of the default placement in Section~\ref{sec:campaign} (Appendix~\ref{app:seed-capacity-robustness}). Durable output of the complete result takes 451.8~s.

\paragraph{Local cutoff under the unfavorable placement.}
Under this placement, the local cutoff removes the overflow and most of the recovery time. We ran nine instrumented campaigns: three settings, three repetitions each, with Optimized's parameters and the campaign's query order. The overflowing placement with the fixed cutoff takes 390.0~s resident and 415.0~s to the host, recovery included. It overflows in 106 batches per repetition, and its fullest window holds 1,219,067 survivors. With the local cutoff of Section~\ref{app:grouped}, the same placement takes 79.6~s resident and 105.3~s to the host with no overflow, and its fullest window holds 982,679. The default placement with the local cutoff takes 78.6 and 104.0~s. The instrumentation counters add cost to all of these times. All retained pairs match the compressed reference (Table~\ref{tab:res-24} in Appendix~\ref{app:seed-capacity-robustness}).

\paragraph{The random-order law against the data.}
The random-order survivor law (Proposition~\ref{prop:law}) sizes the buffers for permuted rows, but the campaign's file order needs the larger, empirical margin of Section~\ref{sec:seed-capacity}. In file order, Stable's fullest window, 505,512 survivors, is 2.31 times the law's mean of 218,750 for its 28M-row windows (Appendix~\ref{app:seed-capacity-robustness}). On two permutations of all 500M rows, whole campaigns at the law's capacities, 222,605 for Stable and 475,273 for Optimized, complete without overflow. Their fullest windows fill 99.5\% and 99.4\% of these capacities. Fourteen further counts-only campaigns at these capacities overflow no window (Appendix~\ref{app:seed-capacity-robustness}, Table~\ref{tab:counts-cells}).

Capacities fitted to a calibration split of the campaign's queries do not replace that margin either. On a protein-disjoint calibration split, the fitted capacities are 452,221 for Stable and 815,081 for Optimized. Among the 21,240 evaluation queries, 7 windows of 6 queries exceed Stable's fitted capacity, and 5 windows of 5 queries exceed Optimized's. With these capacities, one campaign pass overflows in 6 of its 416 batches for Stable and in 4 for Optimized. The two passes take 102.7 and 91.9~s resident, recovery included (Table~\ref{tab:fitted-capacity} in Appendix~\ref{app:seed-capacity-robustness}).

\begin{takeaway}
Query order barely matters, as all four orders stay within 0.53\% of the original time. But one of four random seed-tile placements overflows and takes 390.0~s resident with recovery under the fixed cutoff. The local cutoff brings that placement to 79.6~s without overflow. In the timed campaigns on permuted rows, the law's capacities hold every window, and the fullest fill 99.5\% and 99.4\% of them.
\end{takeaway}

\subsection{Effect of Graph Replay}
\label{sec:replay-eval}
Graph replay shortens the window plan's batches without changing its device work. Run eagerly with one Python thread per GPU, Stable's window plan has a median batch time of 249.8~ms (245.2--253.0) over five batches. Replay cuts it to 157.7~ms (153.6--158.2), a 36.9\% reduction. The kernel launches per batch fall from 1,868 to 121 plus eight graph replays (Figure~\ref{fig:app-replay}). With less selection work, scoring and compaction dominate the window plan's GPU time, and launch overhead becomes visible. In an eager profile of Stable's main segment, they take 68.5 and 14.5~ms of active device time per GPU, about 94\% of its active time. The device work itself barely changes under replay: the graphs span about 127~ms per batch against 129~ms of device activity in the eager profile. Replay therefore saves host launch overhead, the idle gaps between the kernels that eager execution launches one by one.

\begin{figure*}[t]
\centering
\includegraphics{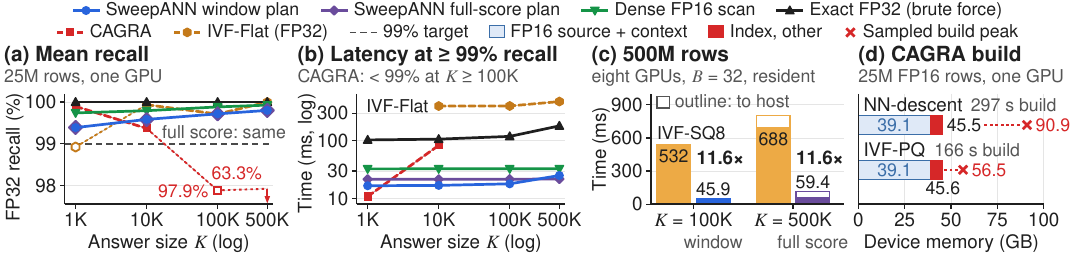}
\caption{GPU indexes against SweepANN's plans as $K$ grows. (a) Mean FP32 recall and (b) median device latency on 25M rows and one GPU ($B=32$, 32 held-out queries), for the configurations selected on calibration queries. Hollow markers are below the 99\% target, and an arrow marks the off-scale CAGRA point. In (a), the full-score markers lie under the window markers. Exact FP32 is cuVS brute force, and Dense FP16 is SweepANN's full-score plan on uncompressed rows. (c) All 500M rows on eight GPUs ($B=32$): resident time per batch of IVF-SQ8 and of the SweepANN plan selected by calibration at each $K$. IVF-SQ8 uses one index per 62.5M-row shard and 1,024 probes. Outlines extend to the host time. (d) Device memory of the two CAGRA graph builders (25M FP16 rows, one GPU). Each bar is the memory in use after construction, CUDA total minus free before query upload. It is split into the FP16 source with the CUDA context (39.1~GB) and the rest. Crosses mark the sampled whole-device build peak (NVML samples, not an allocator peak), 90.9~GB for NN-descent and 56.5~GB for IVF-PQ. The build times, 297 and 166~s, are printed beside the bars.}
\Description{(a) Recall against answer size on 25M rows: all methods stay above the 99 percent target except IVF-Flat at 1K and CAGRA at 100K and 500K (97.9 and 63.3 percent). (b) Latency of the qualifying configurations. (c) On 500M rows and eight GPUs, IVF-SQ8 with one index per shard takes 532 and 688 ms per batch against 45.9 and 59.4 ms for SweepANN, 11.6 times at both answer sizes. (d) Two bars of device memory after building a CAGRA graph on 25M FP16 rows, both about 45.5 GB of which 39.1 GB is the source with the CUDA context, with build peaks of 90.9 GB for NN-descent and 56.5 GB for IVF-PQ.}
\label{fig:gpu-index-comparison}
\end{figure*}
\subsection{ANN Indexes and Large-\texorpdfstring{$K$}{K} Collectors}
\label{sec:published-baselines}
\label{app:gpu-variants}

The preceding subsections compared SweepANN's plans with one another. This subsection compares SweepANN with two families of baselines, GPU indexes and CPU large\nobreakdash-$K$ collectors, all held to the 99\% mean recall target. From $K=10$K on, SweepANN beats the GPU index baselines at this target.

\paragraph{GPU graph search.}
We evaluate the GPU graph index CAGRA~\cite{ootomo2024cagra} through the batched cuVS 26.04 interface on 25M rows and one GPU. It uses the queries and calibration of Section~\ref{sec:setup}, and an exact FP32 scan of these rows gives the reference. Both graph builders, NN-descent~\cite{dong2011nndescent} and IVF-PQ, use FP16 vectors, graph degree 64, and intermediate degree 128. CAGRA is the fastest tested method at $K=1$K (Figure~\ref{fig:gpu-index-comparison}), but it loses that lead as the answer grows. At $K=10$K, its qualifying configuration is already slower than the compressed scans. At $K=100$K and 500K, no tested configuration reaches the recall target, even in a 140-point grid over batches of 32 to 256 queries (Appendix~\ref{app:ann-studies}). Figure~\ref{fig:gpu-index-comparison}(a) shows the best recall these configurations reach.

The graph also costs time and memory to build. At 25M rows, the IVF-PQ builder takes 166~s and NN-descent 297~s, and both leave about 45.5~GB of device memory in use afterwards (Figure~\ref{fig:gpu-index-comparison}(d)).

\begin{figure*}[t]
\centering
\includegraphics{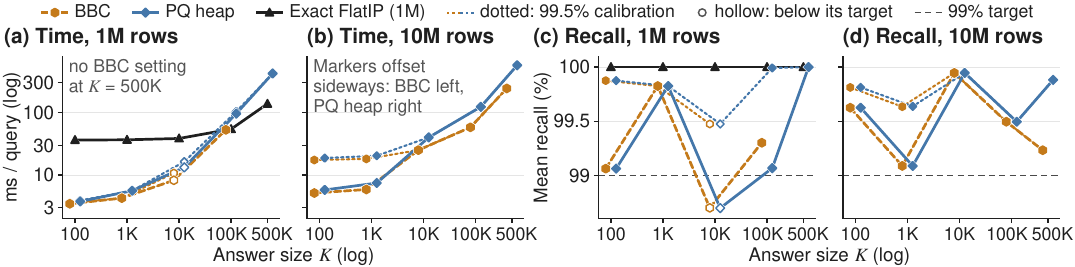}
\caption{Large\nobreakdash-$K$ collectors on one pinned CPU worker (32 held-out queries). BBC and the PQ heap use the same quantized IVF-PQ index. (a,b) Time per query and (c,d) mean recall at 1M and 10M rows. Lines in each method's own style are calibrated to 99\% mean recall, and dotted lines to 99.5\%. Hollow markers miss their calibration target on the held-out queries. BBC points are drawn slightly left and PQ-heap points slightly right of their $K$, so coinciding points stay visible. Where both calibrations select the same setting, one marker shows the 99\% outcome. BBC uses its original sample cap of 5 at 1M rows and an adapted cap of 10 at 10M rows. At 1M rows no BBC configuration qualifies at $K=500$K, nor at 100K for 99.5\%. Rescoring is a separate step.}
\Description{Time per query and held-out recall against answer size at 1M and 10M rows for BBC, the PQ-heap collector and exact FlatIP, with calibration targets of 99 and 99.5 percent and hollow markers where the held-out queries miss them.}
\label{fig:app-cpu-collectors}
\end{figure*}

\paragraph{CPU large\nobreakdash-$K$ collectors.}\phantomsection\label{app:cpu-collectors}
BBC, a bucket-based result collector~\cite{yin2026bbc}, and the plain PQ-heap collector in BBC's repository confirm Section~\ref{sec:ann-motivation} on the CPU. At large $K$, collecting inside a quantized index loses most or all of its advantage over an exact scan. BBC targets the collection and reranking cost of large answers, and its own evaluation uses a 95\% recall target~\cite{yin2026bbc}. Figure~\ref{fig:app-cpu-collectors} tests BBC's released implementation and the PQ heap on one pinned CPU worker. Both collectors share a quantized IVF-PQ index with an inner-product-to-L2 transformation, and both return unordered IDs with approximate scores to host memory. At 1M rows and $K=100$K, BBC calibrated to the 99\% target takes 53.1~ms per query, against 96.6~ms for the PQ heap. Both exceed 99\% mean recall on the held-out queries. BBC thus nearly halves the cost of collecting a large answer inside a quantized index. Yet an exact FlatIP scan on the same worker takes 54.5~ms at this point. At $K=500$K, BBC has no qualifying configuration with its original sample cap, and the exact FlatIP scan completes in 139~ms.

\paragraph{GPU inverted-file search.}
GPU IVF-Flat reaches the recall target from $K=10$K, but it takes far longer than SweepANN's plans. The index (cuVS, 4,096 lists, original FP32 vectors) uses probes calibrated for each $K$. At $K=1$K it misses the target on the held-out queries (98.92\%). From $K=10$K it meets the target (Figure~\ref{fig:gpu-index-comparison}), but a batch takes 398 to 476~ms. That is 2.6 to 3.7 times as long as brute force and 19 to 24 times as long as the window plan. The index holds 82.2~GB on the GPU. Its build takes 13.2~s, GPU copy included, after 159~s to read and hash the FP32 vectors into host memory. The main workload's answer fraction, 0.1\%, corresponds to $K=25$K on these 25M rows. At that size, a batch reaches the host in 400.4~ms with IVF-Flat (99.89\% mean recall). The window and compressed full-score plans take 18.3 and 21.9~ms (both 99.64\%). Section~\ref{sec:ann-factors} covers FP16 IVF-Flat and CAGRA's PQ mode.

The probe counts follow a fixed calibration protocol. The FP32 IVF-Flat runs use the same 32 calibration and 32 held-out queries as the CAGRA runs. Their probe grid is 16, 64, 256, 1,024, 2,048, and 4,096. On the calibration queries alone, they select the fastest setting that meets 99\% mean recall in all three calibration repetitions, and they freeze it before three held-out repetitions. The selected probe counts are 256, 1,024, 1,024, 1,024, and 4,096 at $K=1$K, 10K, 25K, 100K, and 500K, and the last of them probes every list. The grid has no probe counts between 256 and 1,024. The 1K setting passes calibration at 99.14\% but misses the held-out target at 98.92\%. We retain that miss without retuning.

The gap persists when the index is split across GPUs. On 100M rows over four GPUs, we build one index per 25M-row shard (1,024 probes, the fastest calibration setting meeting 99\%). With the answers merged on GPU~0, a 32-query batch with FP32 IVF-Flat takes 426 and 483~ms at $K=100$K and 500K on the held-out queries (99.89\% and 99.66\% mean recall). SweepANN's frozen plans, the window plan at $K=100$K and the full-score plan at 500K, take 21.2 and 27.6~ms (99.64\% and 99.73\%). The index thus takes 20.1 and 17.5 times as long. IVF-SQ8 on the same shards stores one byte per component like SweepANN's codes, but with a different quantizer. It takes 173 and 241~ms resident at 99.59\% and 99.49\% mean recall. These are 8.2 and 8.7 times SweepANN's resident times and 6.7 and 5.6 times its host times (Appendix~\ref{app:ann-studies}).

Letting each method choose its own batch leaves the gap in place. On the same 100M rows and four GPUs, each method took its fastest configuration over $B\in\{32,64,128,256\}$ whose mean recall on 256 calibration queries stayed at or above 99\%. The choice was frozen before 256 held-out queries ran. IVF-SQ8 then runs at $B=32$ for $K=100$K and at $B=256$ for 500K, and SweepANN runs the window plan at $B=64$ and the full-score plan at $B=256$. SweepANN's throughput is $9.90\times$ the index's to the host and $10.38\times$ resident at $K=100$K, and $4.85\times$ and $9.52\times$ at 500K. Both meet the mean target, but the index's minimum recall falls to 96.22\% at $K=500$K, against 99.36\% for SweepANN (Appendix~\ref{app:ann-studies}).

\paragraph{Eight-GPU IVF-SQ8 on 500M rows.}
On all 500M rows of the eight-GPU server, IVF-SQ8 takes $11.6\times$ SweepANN's resident time at both $K$ (Figure~\ref{fig:gpu-index-comparison}(c)). We build one IVF-SQ8 index per 62.5M-row shard (4,096 lists, 1,024 probes at both $K$, selected on the calibration queries). SweepANN runs its calibrated plans, the window plan at $K=100$K and the full-score plan at 500K. A held-out 32-query batch takes 532 and 688~ms resident with the index and 45.9 and 59.4~ms with SweepANN. To the host, it takes 536 and 804~ms against 47.5 and 114.5~ms, $11.3\times$ and $7.0\times$. Both methods reach the target, the index at 99.53\% and 99.58\% mean recall and SweepANN at 99.53\% and 99.64\%, and no held-out query falls below 99.19\%. The index also costs more before and after the search. Preparing the index set took 7,667~s on the main server, against 502.2~s to load and quantize SweepANN's library (Section~\ref{sec:campaign}). The index returns 12 bytes per pair, where SweepANN returns 6 (Appendix~\ref{app:ann-studies}).

\paragraph{Four-GPU IVF-SQ8 on 500M rows.}
In a matched run on the four PRO 6000 GPUs of the main server, the same index is again slower than SweepANN's window plan. It takes $9.8\times$ and $8.0\times$ as long to the host at $K=100$K and 500K, and $9.9\times$ and $8.9\times$ as long resident (Appendix~\ref{app:ann-studies}).

\begin{takeaway}
On 25M rows, CAGRA is fastest at $K=1$K, already slower than the compressed scans at $K=10$K, and misses the 99\% target at $K=100$K and 500K even in a 140-point grid. IVF-Flat meets the target from $K=10$K at 19 to 24 times the \textbf{window plan}'s time. On 500M rows and eight GPUs, IVF-SQ8 takes $11.6\times$ SweepANN's resident time at both $K$.
\end{takeaway}

\subsection{Result Quality}
\label{sec:quality}
Speed matters only if the answers still serve screening. This subsection checks two properties of the answers: that pruning changes no selected score, and that the INT8 codes keep the known actives near the top of the ranking.

\begin{figure}[t]
\centering
\includegraphics{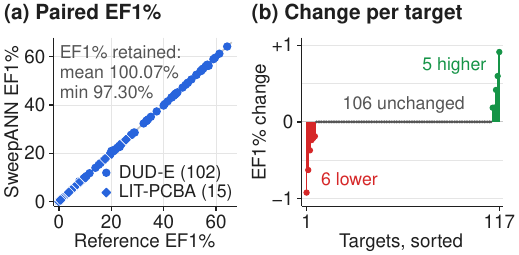}
\caption{Screening enrichment of fused scoring against the FP32 reference on all 117 benchmark targets. (a) Paired EF1\%, with retention over the 111 targets whose reference EF1\% is positive. (b) Signed EF1\% change per target, sorted, all within one EF1\% unit. The 106 unchanged targets include the six with zero enrichment. EF1\% is computed from complete score matrices, and the window plan's selection is checked separately per query.}
\Description{Left, paired EF1\% of SweepANN against the FP32 reference on the diagonal for 117 targets. Right, the sorted per-target change: 6 lower, 106 unchanged and 5 higher, all within one EF1\% unit.}
\label{fig:app-enrichment}
\end{figure}
\begin{figure*}[t]
\centering
\includegraphics{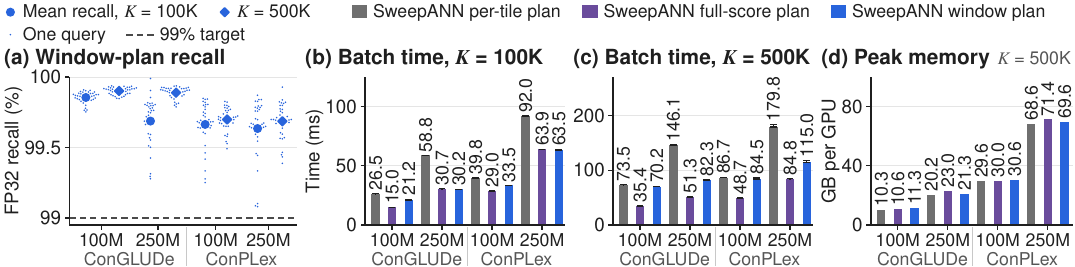}
\caption{Two other embedding models, ConGLUDe (256-d) and ConPLex (1024-d), on the nested 100M and 250M populations (four PRO 6000 GPUs, $B=32$, 32 held-out queries). (a) Strict-ID FP32 recall of the window plan, per query and mean. (b,c) Resident batch time of the per-tile, full-score, and window plans (left to right in each group), median and range of five runs. (d) Peak per-GPU Torch allocation at $K=500$K, library included.}
\Description{Per-query recall of the window plan above 99 percent for ConGLUDe and ConPLex at 100M and 250M rows, and grouped bars of batch time at K=100K and 500K and of peak memory for the per-tile, full-score and window plans.}
\label{fig:app-other-models}
\end{figure*}

\paragraph{Selection preservation.}
The window plan's pruning changes no selected score. A re-run of the complete Optimized campaign on the PRO 6000 returns, for every query, the same score multiset as the top\nobreakdash-$K$ of the unpruned score matrices. The 13.3B returned pairs were recomputed from the tiles with the campaign's kernel configuration. The check found no duplicate IDs, the same $K$-th score, and no buffer overflow. For Stable and the reload, nine batches, the padded last one included, pass the same check. Returned IDs differ only among rows tied at the $K$-th score, as the set semantics of the answer allow. A median of 5,215 rows share a query's $K$-th score (Appendix~\ref{app:campaign-delivery}).

\paragraph{Screening enrichment.}
The INT8 codes leave screening enrichment nearly unchanged. We measured it on all 117 targets of two benchmarks, 102 from DUD-E~\cite{mysinger2012dude} and 15 from LIT-PCBA~\cite{trannguyen2020litpcba} (231 pocket queries). Against the FP32 reference, the fused scoring kernel leaves EF1\% unchanged for 106 targets, lower for six, and higher for five (Figure~\ref{fig:app-enrichment}). Over the 111 targets with a positive reference EF1\%, mean retention is 100.07\% and the minimum is 97.30\%. The other six targets have zero enrichment under both scores. In the top 0.1\%, 947 of the 949 reference hits are kept. A tie-break and operand-precision study on 25M rows (below) attributes the loss against the FP32 reference to the INT8 codes, not to the plan.

\paragraph{Enrichment protocol.}\phantomsection\label{app:quality-details}
The FP32 reference (TF32 disabled) first applies the benchmark's normalization, median and median absolute deviation for DUD-E. It then ranks each molecule by its maximum score over the target's pockets and breaks ties by row ID. EF1\% is computed from the complete score matrices. The window plan also returns the compressed reference for every benchmark query at its top-1\% answer size, with valid, distinct IDs and no overflow. The pools hold 1,433,536 entries with 22,800 actives for the 102 DUD-E targets and 2,807,612 entries with 10,029 actives for the 15 LIT-PCBA targets, counted per target.

\paragraph{Tie breaks and operand precision.}
The recall lost against the FP32 reference comes mostly from the INT8 operands. The study uses the first 25M rows and 32 held-out queries at $K=500$K, with row-ID tie breaks. Mean recall is 99.99\% with uncompressed FP16 operands, 99.82\% with INT8 operands, and 99.80\% after rounding scores to FP16. Ties add a small spread around these values. On average 1,958 rows tie with the $K$-th score, and valid tie choices give 99.75--99.86\%.

\begin{takeaway}
Pruning changes no selected score: for every query of the campaign, the \textbf{window plan} returns the same score multiset as the top\nobreakdash-$K$ of the unpruned score matrices. The INT8 codes, not the plan, cause the loss against FP32, and enrichment survives them. EF1\% is unchanged for 106 of 117 targets, and the top 0.1\% keeps 947 of the 949 reference hits.
\end{takeaway}

\subsection{Other Embedding Models}
\label{sec:model-generalization}
The plans carry over to other embedding models, whose vectors change what the plans cost but not what they return. The seed argument holds for any score function, while the vector width and the score distribution change its cost. We test two more models, each with its released encoder and scoring rule. ConGLUDe uses structure queries and 256-dimensional vectors~\cite{conglude2026}, and ConPLex uses sequence queries and 1024-dimensional vectors~\cite{conplex2023}. Both run on nested populations of 100M and 250M molecule records outside the DrugCLIP library on four PRO 6000 GPUs (Figure~\ref{fig:app-other-models}). Both populations extend to 500M rows, which ConGLUDe serves on four PRO 6000 and four GH200 GPUs and ConPLex on eight PRO 6000 GPUs. ConPLex also runs on 250M rows on four H200 GPUs of NCSA Delta. The two models share the same protein targets, split by gene into 32 calibration and 32 held-out queries. Each has exact FP32 references over the full population, and both run on a second executor that handles any vector dimension. Appendix~\ref{app:other-models} describes the second executor, the model inputs and populations, the measurement, and the survivor load.

\paragraph{Plans and recall.}
SweepANN keeps high recall for both models, and the faster plan shifts with the rows per GPU. On four PRO 6000 GPUs at $B=32$, it keeps more than 99.6\% mean recall on held-out queries in all four 100M and 250M settings. At 100M rows, the score matrix takes 1.6~GB per shard, and the full-score plan is fastest for both models and both answer sizes. At 250M rows, where the matrix takes 4.0~GB per shard, the window plan is faster by less than 0.6~ms at $K=100$K. The full-score plan stays faster at $K=500$K, where the window plan uses about 1.7~GB less memory per GPU.

At 500M rows (Table~\ref{tab:plan-map}), the window plan is fastest at $K=100$K for ConGLUDe on PRO 6000 and GH200 GPUs. It is also fastest for ConPLex on eight PRO 6000 GPUs, by 0.1~ms. At $K=500$K, the full-score plan is fastest in every setting and the per-tile plan slowest (179.8 and 264.9~ms for ConGLUDe on GH200 and PRO 6000). ConGLUDe's narrow vectors leave room for the score matrix even at 500M rows. A 500M-row ConGLUDe shard takes 32.25~GB, a third of a DrugCLIP shard, and its $B=32$ score matrix takes 8~GB. Unlike in the DrugCLIP campaign, this matrix fits beside the shard. ConGLUDe reaches 99.22\% and 99.86\% mean recall at $K=100$K and 500K, with minima of 94.74\% and 99.51\%. The per-tile and full-score plans share this tail, so it comes from compression, not from the window plan.

\paragraph{Fixed costs and larger populations.}
The window plan's fixed costs explain why it loses on the smallest population. At 100M and 250M rows, both models use a 32M-row sample and a buffer of $A=1{,}000{,}000$ pairs per query on four GPUs, at $B=32$ and $K=100$K or 500K. In all four of these settings, the score matrix fits beside the library. At 100M rows, the full-score plan is fastest because it skips seed selection and compaction. The window plan, by contrast, scans a 32M-row sample that is a third of this population. Its fixed candidate and recovery buffers can exceed the 1.6-GB score matrix. As the population grows, the window plan catches up. On PRO 6000 GPUs, ConGLUDe's window-to-full-score time ratio at $K=100$K falls from 1.42 at 100M rows to 0.98 at 250M and 0.82 at 500M. At 500M rows, ConGLUDe has 125M rows per GPU and keeps the smaller populations' sample, buffer, and protocol.

\paragraph{ConPLex on 500M rows.}
For ConPLex on 500M rows and eight GPUs, the window and full-score plans are within 0.1~ms of each other at $K=100$K, and the full-score plan is faster at 500K. Each GPU of the eight-GPU server (Section~\ref{sec:setup}) holds 62.5M rows, as at 250M on four GPUs. The window plan uses $r_0=4$, $G=16$, and $A=1{,}000{,}000$ (a 64M-row sample), at $B=32$ with the same query sets. On the held-out queries, the median 32-query batch over five repetitions takes 68.2~ms resident with the window plan and 68.3~ms with the full-score plan at $K=100$K. At 500K it takes 160.9 against 109.2~ms. The per-tile plan is the slowest at both sizes, at 97.2 and 203.9~ms. Host times vary more across repetitions than resident times. Peak Torch allocation per GPU lies between 66.5 and 72.0~GB for every plan (Table~\ref{tab:res-29} in Appendix~\ref{app:other-models}). The reference is the exact FP32 top\nobreakdash-$K$ over the 500M rows with TF32 disabled.

The three plans return nearly the same recall, 99.61\% and 99.68\% mean held-out recall at $K=100$K and 500K. The minima are 98.91\% and 99.29\% for the window plan and 98.90\% and 99.29\% for the per-tile and full-score plans. Two of the window plan's queries fall below 99\% at $K=100$K. On the 31 nonzero calibration queries, every plan exceeds 99.6\% mean recall at both sizes. All 115.2M returned pairs pass the selection check against the compressed scores (Appendix~\ref{app:other-models}).

A larger batch widens the full-score plan's lead at $K=500$K. A follow-up on the same server and rows runs the full-score and window plans at $B=64$ with unchanged settings, five repetitions each. Each batch holds all 64 calibration and held-out queries. The full-score plan takes 199.8~ms resident and 768.6~ms to the host. The window plan takes 310.2 and 1,280.2~ms, $1.55\times$ and $1.67\times$ as long. Host times again vary across repetitions, from 766.5 to 1,048.7~ms for the full-score plan and from 878.0 to 1,342.9~ms for the window plan. Both plans reach 99.68\% mean held-out recall with a minimum of 99.29\%, and 99.69\% on the 31 nonzero calibration queries. All 320M returned pairs pass the selection check against the compressed scores.

\begin{takeaway}
Other models change what the plans cost, not what they return. For ConGLUDe at $K=100$K, the \textbf{window plan}'s time relative to the \textbf{full-score plan} falls from 1.42 at 100M rows to 0.82 at 500M. At $K=500$K, the full-score plan is fastest in every setting.
\end{takeaway}

\subsection{Plan Choice Across Settings}
\label{sec:plan-map}

Table~\ref{tab:plan-map} collects the measured plan comparisons, and the two regularities of Section~\ref{sec:seed-capacity} were read from them. Its DEEP1B and Cohere rows come from Section~\ref{sec:general} and test the regularities on public collections. The largest answers extend the first regularity. On the eight-GPU server, with 62.5M rows per GPU (Section~\ref{app:grouped}, Table~\ref{tab:largek-8gpu}), the score matrix fits at $B=64$, and the full-score plan is fastest at $K=1$M, 2M, and 5M. At $K=5$M, 1\% of the library, it takes 258.5~ms per batch against 483.2 for the window plan. Both plans return the compressed-score top\nobreakdash-$K$ at 99.76\% mean and 99.51\% minimum FP32 recall. Memory, in contrast, turns against the window plan as $K$ grows. It allocates less beyond the library than the full-score plan at $K=1$M and 2M, but more at 5M (22.8 against 15.0~GB). Its buffers grow with $K$, whereas the score matrix grows with $B$ and not with $K$.

\begin{takeaway}
Where the score matrix fits, as on eight GPUs, the \textbf{full-score plan} is fastest at $K=1$M, 2M, and 5M. At 5M it takes 258.5~ms per batch against 483.2 for the \textbf{window plan}. The window plan's buffers grow with $K$ and the score matrix only with $B$, so at 5M it allocates more beyond the library, 22.8 against 15.0~GB.
\end{takeaway}

\section{General Vector-Search Workloads}
\label{sec:general}
SweepANN's plans are not tied to molecules. The retained seed, the windows, and the merge need only one score per row. They therefore serve any inner-product or Euclidean workload once a fused kernel scores the quantized rows. We ran them on two public collections: DEEP1B, a billion-scale set used by ANN benchmarks~\cite{aumuller2020annbenchmarks,simhadri2022bigann}, and a 1,024-dimensional text-embedding release. Only the scoring kernel changed. A general scorer replaced the DrugCLIP kernel, with one variant per collection. For DEEP1B, the scorer treats the 96 components as one partial quantization group and ends with an epilogue that ranks rows by squared Euclidean distance. The text-embedding variant reads four groups of 256 components. The selection, compaction, window, merge, graph-capture, overflow, and recovery modules are unchanged from Section~\ref{sec:design}. This section first describes the two collections and the protocol. It then reports DEEP1B on eight GPUs, on two and four GPUs, against dense scans, and at larger batches, and it ends with the text embeddings.

\paragraph{Collections.}
DEEP1B~\cite{babenko2016deep1b} holds 1B 96-dimensional image descriptors and 10,000 public queries, ranked by Euclidean distance. The loader quantizes each row into one INT8 group with one FP16 scale and keeps the row's FP32 squared norm. A row takes 102 bytes, and the collection 102,000,000,000 bytes. One PRO 6000 reports 101,974,081,536 bytes of device memory, so the compressed library alone exceeds a single GPU by 25.9~MB before any working memory. A GPU-resident scan needs at least two GPUs. The kernel accumulates the FP16 query against the reconstructed row in FP32, subtracts half the stored squared norm, and rounds to FP16. For a query $q$ and a row $x$, $\lVert q-x\rVert^2=\lVert q\rVert^2-2\,(q\cdot x-\tfrac12\lVert x\rVert^2)$, and $\lVert q\rVert^2$ is the same for every row. The score therefore ranks rows by squared Euclidean distance, up to three roundings: the INT8 rows, the FP16 query and reconstructed rows, and the FP16 score.

The second collection, the Cohere Wikipedia 2023-11 multilingual embedding release~\cite{cohere2023wikipedia}, holds 1,024-dimensional vectors ranked by cosine similarity. The library is the first 1,055 files of the release in file order, 100,000,416 rows. These are all files of the 75 languages from \texttt{ab} through \texttt{eo}, 415 of them English, and the first 17 Spanish files. The queries come from the next file, \texttt{es/0017}, so no query is a library row. Library and query vectors are normalized in FP64 and stored in FP32, and the score is their inner product. Each row is quantized in four groups of 256 components with four FP16 scales, so a row takes 1,032 bytes.

\paragraph{Protocol.}
The two collections share one protocol. For each collection, 128 queries are drawn once from its query pool (PCG64, seed 2026101801), 64 for calibration and 64 held out, before any result is read. The pools are DEEP1B's 10,000 public queries and the 100,000 rows of \texttt{es/0017}. On DEEP1B, the per-tile, grouped (16 tiles), full-score, and window plans run on 2M-row tiles at $B=32$ and 64 and $K=1$K, 10K, 100K, and 500K. The window plan uses $r_0=4$ seed tiles per shard and $G=16$. Its capacity $A$ follows the margin rule of Section~\ref{sec:seed-capacity}, which takes twice the law's random-order mean and rounds it up to a multiple of 128,000. We computed $A$ once for the smallest configuration, two GPUs with a 16M-row sample and windows of at most 32M rows, and kept it on eight GPUs. The resulting capacities are 128,000, 128,000, 512,000, and 2,048,000 for the four answer sizes. With the 64M-row sample of eight GPUs, the rule gives a quarter of these capacities at $K=100$K and 500K. On the text embeddings, the window and full-score plans run at $B=32$ and 64 and $K=100$K and 500K with the same tiles, seed tiles, and capacities. The 103.2~GB of compressed text-embedding rows also need two GPUs, and there the rule gives the same capacities.

The measurement and its checks are the same for both collections. The calibration queries are measured once, the held-out queries three times, and only held-out times are reported. The resident endpoint runs from the query copy to the GPU through local selection, the pairwise merges, and the complete answer on GPU 0. The host endpoint adds the copies of scores (FP16) and 32-bit row IDs into independent host arrays. Where the text gives the host time beyond the resident time, it is the difference of the two medians, computed before rounding. Every returned pair is recomputed from the compressed rows. The reference for recall is the exact FP32 top-$K$ over all rows, computed on the GPUs and read back by an independent CPU check.

\begin{figure*}[t]
\centering
\includegraphics{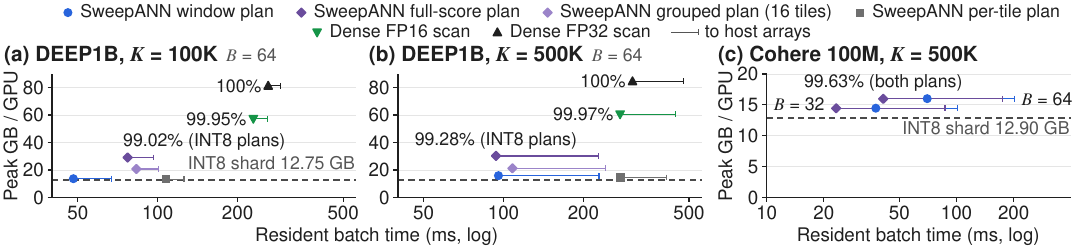}
\caption{Resident time per batch and peak allocated memory per GPU of SweepANN's plans on two public collections, eight GPUs, the same 64 held-out queries, medians of three repetitions. (a,b) DEEP1B (1B rows of 96 dimensions, $B=64$) at $K=100$K and $K=500$K. All plans score every row. The \textbf{window}, \textbf{full-score}, grouped (16 tiles), and per-tile plans scan the INT8 compressed rows. The dense FP32 and FP16 scans are the full-score plan on FP32 and FP16 rows. (c) Cohere 100M (1,024 dimensions) at $K=500$K, the window and full-score plans at $B=32$ and $B=64$, each batch size one row of markers. Labels give the mean FP32 recall, written once where several plans share it. The thin line from each marker ends at the time until the answers are in host arrays. In (c), the two lines of a row overlap, and each ends at its own cap. Dashed lines mark the INT8 shard per GPU, 12.75~GB on DEEP1B and 12.90~GB on Cohere.}
\Description{Three panels of peak allocated memory per GPU against resident time per batch on a log axis. (a,b) DEEP1B at K=100K and 500K: the four INT8 plans share one recall label, 99.02 and 99.28 percent, near the shard line, and the dense FP16 and FP32 scans sit far above it at 99.95, 99.97 and 100 percent with longer resident times. (c) Cohere 100M at K=500K: one row of markers per batch size, the full-score plan left of the window plan in both rows, both at 99.63 percent, above the 12.90 GB shard line.}
\label{fig:general-frontier}
\end{figure*}

\paragraph{DEEP1B on eight GPUs.}
On eight GPUs, the window plan is the fastest plan on DEEP1B up to $K=100$K, and the full-score plan overtakes it at 500K. The compressed collection occupies 12.75~GB per GPU. At $B=64$ (Table~\ref{tab:deep1b-grid}, Figure~\ref{fig:general-frontier}(a,b)), the window plan takes 48.1~ms per batch at $K=100$K, against 77.0 for the full-score plan. At $K=500$K it is 2.0~ms slower, 95.7 against 93.7, but it needs 3.10~GB of peak workspace beyond the shard against 17.47~GB, 5.6 times less. On the calibration queries the same plans are fastest, the window plan up to $K=100$K and the full-score plan at 500K. This pattern matches the first regularity of Section~\ref{sec:seed-capacity}. Of the other two plans, the per-tile plan uses the least workspace at both $K$ of the table. At $K=500$K, it is 2.9 times slower than the window plan. The grouped plan selects once per 16 tiles without the workspace cap of Section~\ref{sec:setup}, so it uses 9.2 and 2.7 times the window plan's workspace.

The same ordering holds at the smaller batch and at smaller answers. At $B=32$, the window plan takes 27.3~ms at $K=100$K (full score 42.9) and 51.3~ms at $K=500$K (full score 51.1). For $K=1$K and 10K at $B=32$, the window plan needs 21.6 and 22.7~ms against 40.8 and 41.4~ms for the full-score plan. At $B=64$ the times are 37.6 and 38.7 against 73.0 and 73.7~ms. From $K=1$K to 100K, the full-score plan's time rises by 5 to 6\%. The window plan's time grows with $K$, as its buffers and window selections do, until the two plans meet at 500K. Copying the answers to the host matters only at large $K$. The host endpoint adds at most 0.14~ms at $K=1$K. At $K=500$K and $B=64$ it adds 133 to 136~ms, the copy of 64 answers of 500K pairs into host arrays.

Recall on DEEP1B rises with $K$ and meets the 99\% mean target from $K=100$K on. For the full-score, grouped, and per-tile plans, mean recall against the exact FP32 top-$K$ is 97.13\%, 98.29\%, 99.02\%, and 99.28\% at $K=1$K, 10K, 100K, and 500K. The minima are 38.7\%, 78.4\%, 97.85\%, and 98.59\%. The window plan differs by less than 0.01 points of mean and at most 0.09 points of minimum recall. INT8 rows of 96-dimensional descriptors cost more recall at small $K$ than the DrugCLIP vectors, which stay above 99\% mean recall from $K=1$K on 25M rows (Figure~\ref{fig:gpu-index-comparison}). No candidate buffer overflowed in any calibration or held-out batch of this grid, so overflow recovery never ran.

\begin{table}[t]
\caption{DEEP1B, 1B rows on eight PRO 6000 GPUs, $B=64$, 64 held-out queries, medians of three repetitions. Resident and host ms per batch, peak workspace per GPU in GB beyond the 12.75~GB shard, and FP32 recall in \% (mean, minimum over queries and repetitions). Every plan returns the same compressed scores. The window plan differs from the others only in which rows it keeps among ties at the $K$-th score. Over the 64 held-out queries, its mean recall agrees with theirs to within 0.01 points.}
\label{tab:deep1b-grid}
\footnotesize
\renewcommand{\arraystretch}{0.9}
\setlength{\tabcolsep}{3.5pt}
\begin{tabular}{@{}rlrrrrr@{}}
\toprule
$K$ & Plan & Resident & Host & Workspace & Mean & Min.\\
\midrule
100K & Window & 48.1 & 67.0 & 0.86 & 99.02 & 97.85\\
100K & Full score & 77.0 & 96.5 & 16.36 & 99.02 & 97.85\\
100K & Grouped (16 tiles) & 83.1 & 100.9 & 7.98 & 99.02 & 97.85\\
100K & Per-tile & 107.5 & 125.6 & 0.55 & 99.02 & 97.85\\
500K & Window & 95.7 & 229.0 & 3.10 & 99.28 & 98.59\\
500K & Full score & 93.7 & 227.8 & 17.47 & 99.28 & 98.59\\
500K & Grouped (16 tiles) & 108.1 & 242.4 & 8.44 & 99.28 & 98.59\\
500K & Per-tile & 275.9 & 411.6 & 1.73 & 99.28 & 98.59\\
\bottomrule
\end{tabular}
\end{table}

\paragraph{Two and four GPUs.}
With fewer GPUs, the window plan keeps its lead over the per-tile plan. On the main server of Section~\ref{sec:setup}, the window and per-tile plans ran on two and four PRO 6000 GPUs for the same 64 held-out queries at $B=64$ and $K=500$K. The compressed library takes 51.0 and 25.5~GB per GPU. The window plan takes 226.9~ms per batch resident on two GPUs and 140.7~ms on four, 1.61 times the throughput. Its peaks are 54.0 and 28.6~GB per GPU, with workspaces of 3.03 and 3.10~GB. The per-tile plan takes 1,012.2 and 524.4~ms, 4.5 and 3.7 times slower, with 1.73~GB of workspace. The window plan's host endpoint gains less from the second pair of GPUs, 683.2 against 611.4~ms, because its copy into host arrays takes 456 and 471~ms. Recall is 99.28\% mean (minimum 98.59\%) in every configuration, and no buffer overflowed. These runs use the same GPU model as the eight-GPU runs on the fourth server (Section~\ref{sec:setup}). Their CPUs and their PyTorch, CUDA, and Triton stack differ, so the two sets are not compared.

\paragraph{Precision frontier.}
Running the full-score plan on the uncompressed rows shows what the INT8 codes buy (Figure~\ref{fig:general-frontier}(a,b)). The dense scans use FP32 rows, 48~GB per GPU, and FP16 rows with FP32 accumulation, 24~GB per GPU. Both serve the same 64 held-out queries at $B=64$. Unlike the dense plans of Table~\ref{tab:control-formats}, they keep and return FP32 scores with 32-bit IDs. They thus copy 8 bytes per returned pair to the host, against 6 for the compressed plans. The FP32 scan, the full-score plan on the uncompressed FP32 rows, returns the exact answer, 100\% recall for every query and repetition. It takes 260.8~ms at $K=100$K and 306.2~ms at 500K resident (290.4 and 475.9~ms to the host), with peaks of 81.4 and 84.6~GB per GPU. The FP16 scan reaches 99.95\% mean recall at $K=100$K (minimum 99.88\%) and 99.97\% at 500K (99.92\%). It takes 229.2 and 274.9~ms (259.9 and 443.3~ms to the host) at peaks of 57.4 and 60.6~GB. The INT8 window plan gives 99.02\% and 99.28\% mean recall in 48.1 and 95.7~ms (67.0 and 229.0~ms to the host) at peaks of 13.6 and 15.8~GB.

In resident time, the window plan is 5.4 and 3.2 times faster than the FP32 scan, with 6.0 and 5.3 times less peak memory. Against the FP16 scan it is 4.8 and 2.9 times faster, with 4.2 and 3.8 times less peak memory. With the plan held fixed, INT8 rows alone make the full-score plan 2.9 to 3.4 times faster than on FP16 or FP32 rows. On these descriptors, the compressed scan thus buys 2.9 to 5.4 times the resident throughput of a dense scan and 1.9 to 4.3 times the host throughput. It costs 0.7 to 1.0 points of mean recall.

\paragraph{Larger batches.}
Larger batches raise resident throughput by only 5 to 7\% for both plans (Table~\ref{tab:deep1b-batch}). We ran the window and full-score plans at $B=64$, 128, and 256 with the capacities above unchanged. Each setting times 256 distinct held-out queries in three complete passes. They are the 64 queries of Table~\ref{tab:deep1b-grid} and 192 more drawn from the remaining public queries (PCG64, seed 2026101802). The $B=64$ rows of Table~\ref{tab:deep1b-batch} are therefore separate measurements from Table~\ref{tab:deep1b-grid}. From $B=64$ to 256, the window plan's resident throughput rises by 7\% at $K=100$K and 5\% at 500K, and the full-score plan's by 6\% and 5\%. The window plan's workspace quadruples, to 3.46 and 12.39~GB, and its peak allocation reaches 16.2 and 25.1~GB per GPU. The full-score plan's workspace, mostly its score matrix, grows with $B$ to 65.4 and 69.9~GB at $B=256$, with peaks of 78.2 and 82.6~GB. At $K=100$K, the window plan is 1.6 times faster than the full-score plan at every $B$, with 18.9 times less workspace. At 500K, the two plans are within 4\% of each other in resident time at every $B$, and the window plan uses 5.6 times less workspace at every $B$.

At the host, larger batches bring the window plan little or no gain, because its copy into host arrays grows at least in proportion to $B$. The window plan's host throughput falls from 1,127 to 977 queries/s at $K=100$K and moves only from 291 to 298 queries/s at 500K. Beyond the resident time, its copy grows in proportion to $B$ at $K=500$K, from 124 to 494~ms, and faster than $B$ at $K=100$K, from 8.8 to 82.3~ms. The full-score plan's copy at $B=256$ and $K=500$K, 1,459~ms, is three times the window plan's. That cell ran in a separate process from the other five. A window run in a separate process gives 365.8~ms resident and 870.5~ms to the host, a copy of 505~ms. The gap thus persists with each plan in its own process. We did not measure the transfer, allocation, and CPU-copy parts of this host time separately.

Recall on the 64 queries of Table~\ref{tab:deep1b-grid}, the only ones with an FP32 reference, is the same at every $B$, 99.02\% and 99.28\% mean. No buffer overflowed. Quadrupling the batch from 64 thus buys 5 to 7\% of resident throughput for four times the window plan's workspace. At the host, the window plan gains 2.4\% at 500K and loses 13\% at 100K.

\begin{table}[t]
\caption{SweepANN's window and full-score plans on DEEP1B (1B rows, eight PRO 6000 GPUs) at $B=64$ to 256: 256 distinct held-out queries timed in three complete passes, medians. Resident and host ms per batch and queries/s, peak allocated memory and workspace per GPU in GB. Recall is measured on the 64 held-out queries of Table~\ref{tab:deep1b-grid} and is the same at every $B$ for both plans, 99.02\% mean at $K=100$K and 99.28\% at 500K. The full-score cell at $B=256$ and $K=500$K ran in a separate process.}
\label{tab:deep1b-batch}
\footnotesize
\renewcommand{\arraystretch}{0.9}
\setlength{\tabcolsep}{2.6pt}
\begin{tabular}{@{}rrlrrrrrr@{}}
\toprule
& & & \multicolumn{2}{c}{Resident} & \multicolumn{2}{c}{Host} & & \\
\cmidrule(lr){4-5}\cmidrule(lr){6-7}
$K$ & $B$ & Plan & ms & queries/s & ms & queries/s & Peak & Workspace\\
\midrule
100K & 64 & Window & 47.9 & 1,335 & 56.8 & 1,127 & 13.6 & 0.86\\
100K & 64 & Full score & 76.7 & 834 & 87.2 & 734 & 29.1 & 16.36\\
100K & 128 & Window & 92.5 & 1,384 & 118.3 & 1,082 & 14.5 & 1.73\\
100K & 128 & Full score & 147.9 & 866 & 188.1 & 681 & 45.5 & 32.71\\
100K & 256 & Window & 179.8 & 1,424 & 262.1 & 977 & 16.2 & 3.46\\
100K & 256 & Full score & 290.1 & 882 & 391.5 & 654 & 78.2 & 65.43\\
500K & 64 & Window & 95.9 & 667 & 219.9 & 291 & 15.8 & 3.10\\
500K & 64 & Full score & 93.4 & 685 & 224.1 & 286 & 30.2 & 17.47\\
500K & 128 & Window & 186.5 & 686 & 439.8 & 291 & 18.9 & 6.19\\
500K & 128 & Full score & 180.1 & 711 & 442.2 & 289 & 47.7 & 34.94\\
500K & 256 & Window & 365.6 & 700 & 859.2 & 298 & 25.1 & 12.39\\
500K & 256 & Full score & 355.9 & 719 & 1,814.5 & 141 & 82.6 & 69.89\\
\bottomrule
\end{tabular}
\end{table}

\paragraph{Cohere 100M\@.}
On the Cohere text embeddings, the full-score plan is the faster plan in all four cells, and both plans clear the recall target. Table~\ref{tab:cohere-grid} and Figure~\ref{fig:general-frontier}(c) compare the window and full-score plans on the 100M text embeddings at $B=32$ and 64, with the FP32 reference for all 64 held-out queries. Each GPU holds 12.5M compressed rows, 12.9~GB. Both plans reach 99.57\% mean recall at $K=100$K and 99.63\% at 500K, with per-query minima of at least 99.23\% and 99.50\%. These recalls exceed those on DEEP1B's 96-dimensional descriptors. The full-score plan is 1.2 times faster at $K=100$K (14.8 against 17.3~ms at $B=32$, 25.0 against 29.9 at $B=64$). At 500K it is 1.6 and 1.7 times faster (23.2 against 37.6 and 41.1 against 70.1~ms). The calibration queries rank the two plans the same way in all four cells.

We attribute the gap to the execution layout, which leaves few rows on each GPU. The full-score plan selects from a small score matrix, 1.6~GB at $B=64$ on 12.5M rows per GPU, within a measured workspace of 1.89~GB at $K=100$K. The window plan's four 2M-row seed tiles cover 64\% of each shard, so its unfiltered seed selection already spans most rows. The cutoff exchange and the window phase then save work on only the remaining 4.5M rows. The window plan still uses less workspace at $K=100$K, 0.43 against 0.95~GB at $B=32$ and 0.86 against 1.89~GB at $B=64$. At 500K it uses slightly more, 3.10 against 3.07~GB at $B=64$. Its buffers keep the capacities set for the 32M-row windows of two GPUs. These capacities are four and 16 times what the margin rule gives at $K=100$K and 500K for windows of 4.5M rows.

Both collections thus follow the first regularity of Section~\ref{sec:seed-capacity}. On both, the score matrix fits beside the shard, and the full-score plan is the fastest at $K=500$K. 

\begin{table}[t]
\caption{SweepANN's window and full-score plans on the Cohere embeddings (100,000,416 rows of 1,024 dimensions, eight PRO 6000 GPUs, 64 held-out queries, medians of three repetitions). Resident and host ms per batch, peak workspace per GPU in GB beyond the 12.9~GB shard, and FP32 recall in \% (mean, minimum over queries and repetitions).}
\label{tab:cohere-grid}
\footnotesize
\renewcommand{\arraystretch}{0.9}
\setlength{\tabcolsep}{3.5pt}
\begin{tabular}{@{}rrlrrrrr@{}}
\toprule
$K$ & $B$ & Plan & Resident & Host & Workspace & Mean & Min.\\
\midrule
100K & 32 & Full score & 14.8 & 16.6 & 0.95 & 99.57 & 99.23\\
100K & 32 & Window & 17.3 & 23.7 & 0.43 & 99.57 & 99.24\\
100K & 64 & Full score & 25.0 & 44.0 & 1.89 & 99.57 & 99.23\\
100K & 64 & Window & 29.9 & 49.3 & 0.86 & 99.57 & 99.24\\
500K & 32 & Full score & 23.2 & 86.9 & 1.54 & 99.63 & 99.50\\
500K & 32 & Window & 37.6 & 101.0 & 1.55 & 99.63 & 99.50\\
500K & 64 & Full score & 41.1 & 175.1 & 3.07 & 99.63 & 99.50\\
500K & 64 & Window & 70.1 & 202.4 & 3.10 & 99.63 & 99.50\\
\bottomrule
\end{tabular}
\end{table}

\begin{takeaway}
Selection work follows rows, not bytes: both collections hold about 13~GB of compressed rows per GPU, but a DEEP1B shard has 125M rows and a Cohere shard 12.5M. The \textbf{window plan} is faster at $K=100$K on DEEP1B, and the \textbf{full-score plan} in all four Cohere cells. The plan rule picks the faster plan on both from calibration times alone.
\end{takeaway}

\section{Libraries Beyond GPU Memory}
\label{sec:extension}
SweepANN keeps the whole compressed library in GPU memory, so aggregate GPU memory bounds the library it can scan. Eight GPUs hold two copies of the 500M-row DrugCLIP library, 1B logical rows of 774 bytes each, with 4,408~MiB free per GPU beside the library (Section~\ref{app:replicated1b}). If the compressed library does not fit in aggregate GPU memory, a compact candidate-generation index can stay on the GPU. Its candidate lists are then reranked with vectors read from host memory. Our refiner groups the requested vectors by source tile and shares each fetched vector among the queries that request it. It also bounds the number of vectors held on the GPU.

On one A40, the bounded refiner answers a request whose candidate buffer would not fit beside the index. The A40 has 47.70~GB of visible device memory. A 100M-row cuVS IVF-PQ index probing all 4,096 lists supplies 1.25M candidates per query for 32 queries, and the refiner keeps 500K of them. One FP16 buffer of the distinct candidate vectors would need 33.63~GB, or 53.67~GB with the resident index and its support allocations, more than the A40 holds. The refiner completes the request in 1,083.5~s at a sampled device-memory peak of 27.08~GB. This time covers search, refinement, delivery to a host consumer, and cleanup. The reranked answers reach 99.93\% mean and 99.73\% minimum recall against the exact FP32 top-500K of these rows. Appendix~\ref{app:ann-studies} gives its read pattern and compares it with keeping all candidates on the GPU.

\begin{takeaway}
Eight GPUs hold two copies of the DrugCLIP library, 1B logical rows. Beyond aggregate GPU memory, a GPU-resident candidate index feeds a refiner that reranks with vectors read from host memory. On one A40, one buffer of the candidate vectors would need 53.67~GB with the index, more than the GPU holds. The refiner peaks at 27.08~GB and reaches 99.93\% mean recall.
\end{takeaway}

\section{Related Work}
\label{sec:related}
\paragraph{Virtual screening.}
DrugCLIP casts pocket-to-molecule screening as dense retrieval~\cite{gao2023drugclip}, and its genome-wide study screens over 500M molecules~\cite{jia2026genome}. SparkleDock scales macromolecular docking to GPU supercomputers~\cite{meng2026sparkledock}. SweepANN serves the retrieval stage of Figure~\ref{fig:screening-pipeline}. That stage fixes the pool that docking then narrows, so a molecule it misses never reaches docking.

\paragraph{Top\nobreakdash-$K$ processing and large answers.}
Many top\nobreakdash-$K$ methods prune with a threshold that excludes no answer. The Threshold Algorithm stops once its $K$-th best score reaches an upper bound on unseen objects~\cite{fagin2001optimal,ilyas2008survey}, and initial thresholds speed up rank-safe WAND search at large $k$~\cite{kane2018split}. Relational engines push result limits into plans~\cite{carey1997enough} or run a top-$N$ query as a range selection with an estimated cutoff, restarting when too few rows qualify~\cite{donjerkovic1999probabilistic,chaudhuri1999evaluating}. TPUT and KLEE exchange thresholds among nodes~\cite{cao2004tput,michel2005klee}, and selection algorithms bound the target rank with sample pivots~\cite{floyd1975expected,ribizel2019approximate}. Dr.~Top-$k$ uses subrange delegates and their top\nobreakdash-$K$ cutoff to filter the remaining candidates~\cite{gaihre2021drtopk}. On GPUs, Faiss's WarpSelect filters values against the best $k$ so far into per-thread queues, with a native selection path for $k\le1{,}024$~\cite{johnson2019billion}. Bitonic top\nobreakdash-$K$ is the fastest of the compared GPU methods for $K\le256$, and radix selection is faster for larger $K$~\cite{shanbhag2018topk}. AIR top-$k$ and RadiK use radix selection, and RadiK scales to larger $K$ and batches~\cite{zhang2023parallelTopK,li2024radik}. cuVS selection carries payload IDs~\cite{cuvsSelectionDocumentation}. BBC reduces large\nobreakdash-$K$ collection and reranking costs in quantization-based search~\cite{yin2026bbc}, and TEngineDB-V runs large\nobreakdash-$K$ retrieval in an OLAP engine~\cite{wu2026tengine}.

SweepANN's cutoff shares properties with several of these methods. Like Dr.~Top-$k$, SweepANN filters with the $K$-th score of a subset of actual rows, so its cutoff excludes no answer. Like a conservative cutoff that guarantees at least $K$ qualifying rows, and unlike an estimated cutoff, it never restarts for too few rows. The cutoff comes from a sample of compressed rows scored once per batch. SweepANN keeps the sample's top\nobreakdash-$K$, the seed rows that set the cutoff, for the final merge. A strict filter is therefore exact, and every row is read once, whereas Dr.~Top-$k$ re-reads the subranges of its selected delegates. A weak cutoff lets more rows survive and can overflow the buffers, which costs recovery time but never changes the answer (Section~\ref{sec:capacity}). One exchange across GPUs fixes the cutoff, so the rest of the scan runs as a fixed, replayable schedule with batch-sized buffers.

\paragraph{Shared scans, batching, and placement.}
QPipe, cooperative scans, Crescando, and SharedDB share scans and operators among concurrent queries~\cite{harizopoulos2005qpipe,zukowski2007cooperative,unterbrunner2009predictable,giannikis2012shareddb}, and kNN joins pair each point with its $k$ nearest neighbors in another set~\cite{bohm2004knnjoin,xia2004gorder}. LEMP returns the matrix-product entries that exceed a threshold or rank in each row's top $k$~\cite{teflioudi2015lemp}. OPTIMUS chooses between blocked matrix multiplication and an index for exact inner-product search~\cite{abuzaid2019optimus}. TPU-KNN scans without an index at peak FLOP/s on TPUs. Its approximate top-$k$ guarantees recall in expectation under a random data layout~\cite{chern2022tpuknn}, whereas SweepANN's top\nobreakdash-$K$ is exact under its compressed scores. RUMMY~\cite{zhang2024rummy}, FusionANNS, BANG, PipeANN, and FlowANN organize search around data held in host memory or on SSDs~\cite{tian2025fusionanns,venkatasubba2025bang,guo2025pipeann,zhao2026flowann}. SweepANN instead keeps the compressed library resident and runs each batch as a block nested-loop join (Section~\ref{sec:architecture}). Each tile is read once per batch, and the memory left for answer state limits the batch size. Within this scan, SweepANN changes how often it selects and how many candidates it keeps between selections. Section~\ref{sec:extension} turns to libraries beyond GPU memory. OPTIMUS chooses between an index and a scan, whereas SweepANN chooses among plans for one scan.

\section{Discussion and Conclusion}
\label{sec:conclusion}
\paragraph{Limitations.}
SweepANN needs the compressed library to fit in aggregate GPU memory (Section~\ref{sec:extension}). The window plan is not the fastest plan in every setting, and the plan rule then picks another. When the full score matrix fits at the window plan's batch size, the full-score plan is faster than the window plan at the largest tested answer size (Section~\ref{sec:physical-plan-crossover}). At 100M rows, grouping alone obtains nearly all of the window plan's gain. At 500M rows on four GPUs, the widest grouped plan in the same workspace takes $1.41\times$ as long (Section~\ref{sec:batch-reuse-controls}). cuVS is the only other top\nobreakdash-$K$ primitive we measured, and it changes the plans' relative costs (Section~\ref{app:grouped}). The cutoff is fixed per batch, so a weak sample lets more survivors through, and an unfavorable seed-tile placement forces recovery. The local cutoff removed that recovery in the one placement tested (Section~\ref{app:replication}). Stable and Optimized were chosen on the 500M campaign, held-out pockets included. Capacities fitted instead on a protein-disjoint split overflow in a few batches (Section~\ref{app:replication}). The DrugCLIP library is shuffled by file, not by row, so its buffers hold about twice the survivors the law predicts. The law's capacities were timed on two row permutations (Section~\ref{app:replication}). On DEEP1B's 96-dimensional descriptors, the compressed scores give 97.1\% mean recall at $K=1$K and reach the 99\% target from $K=100$K. The compressed scan thus serves the large answers it was built for and leaves small answers to the indexes (Section~\ref{sec:general}).

\paragraph{Conclusion.}
Screening asks for answers of $K=500$K at 99\% mean recall. At that size, the tested indexes miss the target or lose most of their advantage over a scan. From $K=10$K on 25M to 500M rows, the GPU IVF indexes that reach the target take 8.2 to 24 times SweepANN's resident time. SweepANN therefore scans compressed vectors without an index, and a batch of queries shares each block read. The cost that grows with $K$ is then keeping the answers, so SweepANN chooses its plan by the memory beside the library first and by measured cost second. Where the score matrix fits, keeping every score is fastest at the largest tested $K$. Where it does not, the window plan is the fastest tested candidate. Its seed cutoff lets fixed-capacity buffers hold the candidates, and the retained seed keeps the answers exact under the compressed scores. Replay removes most launch overhead. More rows per GPU can push the score matrix out of memory while the window plan still fits. A larger $K$ instead enlarges the window plan's buffers, not the matrix. 

On four PRO 6000 GPUs, SweepANN answers the 500M-molecule campaign in 61.04~s resident at 99.63\% mean recall. There, the window plan is $1.41\times$ faster than the widest grouped plan in the same memory. Four GH200 GPUs serve the same campaign in a median 60.59~s. With one to eight GPUs, plan choices follow these two regularities on 25M to 500M rows of three embedding models. They also hold on the 1B rows of DEEP1B and on the 100M Cohere embeddings. On DEEP1B, the window plan at $K=100$K and 500K is 5.4 and 3.2 times faster in resident time than the full-score plan on the uncompressed FP32 rows, an exact scan. It needs 6.0 and 5.3 times less peak memory per GPU and loses 1.0 and 0.7 points of mean recall. Two copies of the DrugCLIP library, 1B logical rows and 774~GB, stay resident on eight GPUs. They answer a batch of 64 queries at $K=500$K in 155~ms resident at 99.59\% mean recall. At answer sizes where indexes lose their advantage, fast screening thus rests on keeping the answers in the memory beside the library, not on skipping rows.

\paragraph{Availability.}
The SweepANN code and the data-processing scripts will be released as open source.

\section*{Acknowledgments}\pdfbookmark[1]{Acknowledgments}{acknowledgments}
This work used Delta and DeltaAI at the National Center for Supercomputing Applications through allocation CIS260612 from the Advanced Cyberinfrastructure Coordination Ecosystem: Services \& Support (ACCESS) program, which is supported by U.S. National Science Foundation grants 2138259, 2138286, 2138307, 2137603, and 2138296.
\label{body:end}
\AddToHook{env/thebibliography/end}{\global\vfuzz=2pt\balance}
\bibliographystyle{ACM-Reference-Format}
\bibliography{references}
\clearpage
\nobalance\vfuzz=0.1pt
\appendix

\section{Experiment Index and Platforms}\label{app:experiment-index}\suppressfloats[t]
This appendix records the setup of every experiment. Table~\ref{tab:setup-summary} in Section~\ref{sec:setup} summarizes the platforms and main settings, and Table~\ref{tab:setups} gives the rows, GPUs, batch size $B$, answer size $K$, and queries of each experiment. Two CPU controls on 10M rows, run on other servers, close the appendix (Tables~\ref{tab:res-31a} and~\ref{tab:res-31b}).

\begin{table}[t]
\caption{Setup of each experiment. Rows, GPUs, batch size $B$, answer size $K$, and queries, listed by the section or figure that reports the experiment. GPUs: number and type, with P, G, H for PRO 6000, GH200, H200. HO: held-out queries (Section~\ref{sec:setup}), gene-disjoint for the other models (Section~\ref{sec:model-generalization}) and drawn from each collection's query pool for DEEP1B and Cohere (Section~\ref{sec:general}). The row-order control and Section~\ref{app:grouped} reuse the 100M sets. CPU collectors use one thread.}
\label{tab:setups}
\footnotesize
\renewcommand{\arraystretch}{0.84}
\begin{tabular}{@{}>{\raggedright\arraybackslash\hangindent=0.8em\hangafter=1}p{27mm}@{\hspace{4.5pt}}l@{\hspace{4.5pt}}l@{\hspace{4.5pt}}l@{\hspace{4.5pt}}l@{\hspace{4.5pt}}>{\raggedright\arraybackslash}p{12.5mm}@{}}
\toprule
Source & Rows & GPUs & $B$ & $K$ & Queries\\
\midrule
Sec.~\ref{sec:campaign}, Fig.~\ref{fig:app-replay}(a), Sec.~\ref{sec:quality} &500M&4P&64&500K&26,562\\
Sec.~\ref{sec:campaign}, GH200 &500M&4G&64&500K&26,562\\
Sec.~\ref{sec:campaign}, Sec.~\ref{app:grouped}, Sec.~\ref{sec:quality}, 256 proteins &500M&4P, 8P&64, 256&500K&256\\
Fig.~\ref{fig:app-replay}(b,c) &500M&4P&64&500K&5 batches\\
Fig.~\ref{fig:batch-reuse}(b) &100M&1, 2, 4P&1--64&500K&64 HO\\
Fig.~\ref{fig:batch-reuse}(a) &2M&1P&1, 8, 64&&\\
Fig.~\ref{fig:answer-size-cost} &500M&4P&64&1K--500K&64 HO\\
Fig.~\ref{fig:memory-boundary}(c) &500M&4P, 4G&16, 64&500K&64 HO\\
Fig.~\ref{fig:memory-boundary}(a,b) &500M&4P&8--64&500K&64 HO\\
Fig.~\ref{fig:app-plan-crossover}(a,b) &1M--50M&1P&32&500K&32 HO\\
Fig.~\ref{fig:app-plan-crossover}(c), Sec.~\ref{app:grouped} &100M&1P&64&500K&64 HO\\
Sec.~\ref{app:grouped}, campaigns &500M&8P, 4P&64, 256&500K&26,562\\
Sec.~\ref{app:grouped}, controls &500M&8P, 4P&64--256&500K&64--256 HO\\
Sec.~\ref{app:grouped}, paired 4P/8P &500M&4P, 8P&64&500K&64 HO\\
Sec.~\ref{app:grouped}, 1B logical rows &2$\times$500M&8P&64&500K&64 HO\\
Fig.~\ref{fig:app-seed-window}, Table~\ref{tab:capacity} &500M&4P&64&500K&campaign\\
Sec.~\ref{sec:capacity}, recovery exercise &500M&4P&64&100K&campaign\\
Fig.~\ref{fig:gpu-index-comparison}(a,b,d), Sec.~\ref{app:gpu-variants} (IVF) &25M&1P&32&1K--500K&32 HO\\
Sec.~\ref{app:gpu-variants}, CAGRA grid &25M&1P&32--256&100K, 500K&256 HO\\
App.~\ref{app:ann-studies}, CAGRA build &50M&1P&&&\\
Sec.~\ref{sec:campaign}, DrugCLIP script &25M&4P&4--64&100K&1K\\
Fig.~\ref{fig:gpu-index-comparison}(c), IVF-SQ8 500M &500M&8P&32&100K, 500K&32 HO\\
Sec.~\ref{app:gpu-variants}, IVF-SQ8 (4P) &500M&4P&32&100K, 500K&32 HO\\
Sec.~\ref{app:gpu-variants}, IVF on 100M &100M&4P&32&100K, 500K&32 HO\\
Sec.~\ref{app:gpu-variants}, best-batch &100M&4P&32--256&100K, 500K&256 HO\\
Sec.~\ref{sec:ann-factors}, PQ, RaBitQ &1M, 25M&1P, CPU&32&100--500K&32 HO\\
Sec.~\ref{sec:batch-reuse-controls}, $K=1$M &500M&4P&32&1M&64 HO\\
Sec.~\ref{app:grouped}, $K=1$M--5M &500M&8P&64&1M--5M&64 HO\\
Figs.~\ref{fig:index-limits}, \ref{fig:app-index-size}, Table~\ref{tab:ann-quality-budget} &1M--50M&CPU&&100, 100K&1K\\
Sec.~\ref{app:setups}, CPU controls &10M&CPU&&100, 100K&1K\\
Fig.~\ref{fig:app-cpu-collectors} &1M, 10M&CPU&&100--500K&32 HO\\
Fig.~\ref{fig:app-enrichment} &benchmarks&&&top 1\%&231\\
Fig.~\ref{fig:app-other-models} &100M, 250M&4P&32&100K, 500K&32 HO\\
Sec.~\ref{sec:model-generalization}, ConGLUDe &500M&4P, 4G&32&100K, 500K&32 HO\\
Sec.~\ref{sec:model-generalization}, ConPLex &250M&4H&32&100K, 500K&32 HO\\
Sec.~\ref{sec:model-generalization}, ConPLex &500M&8P&32&100K, 500K&32 HO\\
Sec.~\ref{sec:model-generalization}, ConPLex $B=64$ &500M&8P&64&500K&64 (32 HO)\\
Sec.~\ref{sec:extension} &100M&1 A40&32&500K&32\\
Sec.~\ref{sec:general}, DEEP1B &1B&8P&32, 64&1K--500K&64 HO\\
Sec.~\ref{sec:general}, DEEP1B dense &1B&8P&64&100K, 500K&64 HO\\
Sec.~\ref{sec:general}, DEEP1B batches &1B&8P&64--256&100K, 500K&256 HO\\
Sec.~\ref{sec:general}, Cohere &100M&8P&32, 64&100K, 500K&64 HO\\
Sec.~\ref{sec:general}, DEEP1B two and four GPUs &1B&2P, 4P&64&500K&64 HO\\
Sec.~\ref{app:grouped}, 500M rows &500M&4P&64&500K&64 HO\\
Sec.~\ref{app:grouped}, cuVS at 500M &500M&4P&64&500K&64 HO\\
Sec.~\ref{app:stages} &500M&4P, 4G&64&500K&64 HO\\
App.~\ref{app:seed-capacity-robustness}, row order &25M&1P&64&500K&64 HO\\
Sec.~\ref{app:replication}, permuted &500M&4P&64&500K&26,562\\
Sec.~\ref{app:replication}, local cutoff &500M&4P&64&500K&26,562\\
Sec.~\ref{app:replication}, fitted capacities &500M&4P&64&500K&26,562\\
\bottomrule
\end{tabular}
\end{table}

\paragraph{CPU controls on 10M rows.}
Two controls on other CPU servers support the IVF results of Section~\ref{sec:ann-query} on the 10M-row library. An exact flat scan is faster than IVF-Flat with a quarter of its lists, and a retrained coarse quantizer selects the default probe counts.

The flat-scan control ran Faiss on 24 threads of a second CPU server over the 1,000 queries in stored order. It used raw FP32 inner products, existing indexes, and three fresh repetitions per configuration. Warm-up, loading, and persistence lie outside the timed search. The quarter setting probes 4,096 of the 16,384 lists at both answer sizes. It is the 99\% setting of Table~\ref{tab:ann-quality-budget} only at $K=100$K. Recall is measured against the stored FP32 reference. An independent pass reread all 900.9M returned pairs with FP64 score intervals, and a separate exhaustive FP64 check covers only queries 0, 1, 998, and 999. All 18 repetition means reach 99\%, and Table~\ref{tab:res-31a} gives the per-query tails.

The quantizer control ran Faiss 1.13.2 on a 32-core CPU node. On 200 of the 1,000 queries, it selected the smallest probe count on a grid that meets 99\% mean recall and froze it before the other 800 ran. It did so for the original quantizers, the Faiss defaults of Section~\ref{sec:ann-build}. It also did so for one retrained with spherical $k$-means (20 iterations, at most 64 points per list, a fixed seed, 1,004.7~s). Both select the counts of Table~\ref{tab:ann-quality-budget}, and Table~\ref{tab:res-31b} gives the held-out points. Probing all 16,384 lists also meets the target. At the selected points, the probed lists hold 3.0\% and 24.6\% of the rows for IVF-Flat and 6.0\% and 24.6\% for IVF-SQ8 at $K=100$ and 100K, with either quantizer. Times cover the native search call, and the two quantizers differ in search time by at most 1.2\%.

\begin{table}[t]
\caption{Exact flat scan and IVF-Flat on a CPU. Faiss on 24 threads, 10M rows, 1,000 queries, three repetitions. ms: median time per query. Mean, Min.: mean and minimum recall (\%). $<$99: queries below 99\% in each repetition. Quarter: 4,096 of 16,384 lists.}
\label{tab:res-31a}
\footnotesize
\begin{tabular}{@{}lrrrrr@{}}
\toprule
Method & $K$ & ms & Mean & Min. & $<$99\\
\midrule
Flat scan & 100 & 127.5 & 99.999 & 99.00 & 0\\
Flat scan & 100K & 141.0 & 99.9998 & 99.998 & 0\\
IVF-Flat, quarter & 100 & 176.8 & 99.986 & 96.00 & 4\\
IVF-Flat, quarter & 100K & 160.7 & 99.866 & 98.417 & 17\\
IVF-Flat, all lists & 100 & 621.9 & 99.999 & 99.00 & 0\\
IVF-Flat, all lists & 100K & 562.8 & 99.9998 & 99.998 & 0\\
\bottomrule
\end{tabular}
\end{table}

\section{ANN Index Studies}\label{app:ann-studies}
This appendix holds the evidence that every tested index misses the 99\% target at large $K$ or loses most of its advantage over a scan (Sections~\ref{sec:ann-motivation} and~\ref{sec:published-baselines}). It first gives the fastest saved setting of every index sweep and the CPU studies of HNSW and one-bit RaBitQ codes. It then covers the CAGRA grids and builds, the IVF comparisons with SweepANN on 100M and 500M rows, and the read pattern of the bounded refiner of Section~\ref{sec:extension}.

\begin{table*}[t]
\caption{Fastest saved setting at each mean-recall target. CPU indexes on 10M rows (1,000 queries, the first 100 for HNSW at $K=100$K, ms per query) and GPU indexes on 25M rows (ms per batch of 32 calibration queries). Each cell gives setting: time (achieved mean recall, \%), and a dash means that no saved valid setting reaches the target. Settings are probes (IVF), \texttt{efSearch} (HNSW), or CAGRA preset (Table~\ref{tab:res-11}). Best: highest saved mean recall. Valid: valid of all saved points.}
\label{tab:res-01}
\scriptsize
\begin{tabular}{@{}llllllrr@{}}
\toprule
Index & $K$ & 90\% & 95\% & 99\% & 99.5\% & Best & Valid\\
\midrule
CPU IVF-Flat (probes) & 100 & 64: 3.62 (90.60) & 256: 7.43 (98.35) & 512: 10.59 (99.51) & 512: 10.59 (99.51) & $>$99.99 & 11/11\\
CPU IVF-Flat (probes) & 100K & 1,024: 16.20 (94.28) & 2,048: 26.85 (98.80) & 4,096: 45.23 (99.87) & 4,096: 45.23 (99.87) & $>$99.99 & 11/11\\
CPU IVF-SQ8 (probes) & 100 & 64: 7.34 (90.49) & 256: 11.66 (97.92) & 1,024: 22.87 (99.23) & -- & 99.31 & 11/11\\
CPU IVF-SQ8 (probes) & 100K & 1,024: 22.77 (94.28) & 2,048: 36.97 (98.76) & 4,096: 67.73 (99.73) & 4,096: 67.73 (99.73) & 99.81 & 11/11\\
CPU IVF-PQ48 (probes) & 100 & -- & -- & -- & -- & 36.62 & 11/11\\
CPU IVF-PQ48 (probes) & 100K & -- & -- & -- & -- & 78.50 & 11/11\\
CPU HNSW-Flat (ef) & 100 & 256: 0.13 (93.45) & 512: 0.20 (97.54) & 1,024: 0.45 (99.27) & 2,048: 1.57 (99.75) & 99.75 & 5/5\\
CPU HNSW-Flat (ef) & 100K & 100,000: 7,066.68 (99.72) & 100,000: 7,066.68 (99.72) & 100,000: 7,066.68 (99.72) & 100,000: 7,066.68 (99.72) & 99.72 & 1/1\\
CPU HNSW-SQ8 (ef) & 100 & 256: 0.20 (93.34) & 512: 0.43 (97.16) & -- & -- & 98.97 & 5/5\\
CPU HNSW-SQ8 (ef) & 100K & 100,000: 7,080.52 (99.56) & 100,000: 7,080.52 (99.56) & 100,000: 7,080.52 (99.56) & 100,000: 7,080.52 (99.56) & 99.56 & 1/1\\
\midrule
GPU IVF-Flat FP32 (probes) & 1K & 64: 28.93 (92.78) & 256: 110.06 (99.14) & 256: 110.06 (99.14) & 1,024: 407.77 (99.98) & 100 & 6/6\\
GPU IVF-Flat FP32 (probes) & 10K & 256: 111.17 (98.21) & 256: 111.17 (98.21) & 1,024: 410.86 (99.95) & 1,024: 410.86 (99.95) & 100 & 6/6\\
GPU IVF-Flat FP32 (probes) & 25K & 256: 110.50 (97.31) & 256: 110.50 (97.31) & 1,024: 408.70 (99.91) & 1,024: 408.70 (99.91) & 100 & 6/6\\
GPU IVF-Flat FP32 (probes) & 100K & 256: 111.87 (94.47) & 1,024: 413.38 (99.76) & 1,024: 413.38 (99.76) & 1,024: 413.38 (99.76) & $>$99.99 & 5/6\\
GPU IVF-Flat FP32 (probes) & 500K & 1,024: 435.79 (98.96) & 1,024: 435.79 (98.96) & 4,096: 496.07 ($>$99.99) & 4,096: 496.07 ($>$99.99) & $>$99.99 & 4/6\\
GPU IVF-Flat FP16 (probes) & 1K & 64: 13.11 (90.48) & 256: 51.14 (95.69) & -- & -- & 96.20 & 6/6\\
GPU IVF-Flat FP16 (probes) & 10K & 256: 50.52 (96.24) & 256: 50.52 (96.24) & -- & -- & 97.46 & 6/6\\
GPU IVF-Flat FP16 (probes) & 100K & 256: 52.37 (93.90) & 1,024: 199.92 (98.40) & -- & -- & 98.53 & 5/6\\
GPU IVF-Flat FP16 (probes) & 500K & 1,024: 246.58 (98.39) & 1,024: 246.58 (98.39) & 4,096: 335.56 (99.14) & -- & 99.14 & 4/6\\
GPU CAGRA (preset) & 1K & 0: 8.95 (99.92) & 0: 8.95 (99.92) & 0: 8.95 (99.92) & 0: 8.95 (99.92) & 99.98 & 6/6\\
GPU CAGRA (preset) & 10K & 26: 81.78 (98.93) & 26: 81.78 (98.93) & 3: 85.76 (99.39) & -- & 99.40 & 6/6\\
GPU CAGRA (preset) & 100K & 13: 4,268.69 (92.59) & 34: 8,099.86 (97.86) & -- & -- & 97.86 & 16/16\\
GPU CAGRA (preset) & 500K & -- & -- & -- & -- & 63.25 & 8/18\\
\bottomrule
\end{tabular}
\end{table*}
\begin{table}[t]
\caption{Original and retrained coarse quantizers. Faiss 1.13.2 on 32 cores, 10M rows, probes selected on 200 queries, 800 held-out queries. Flat, SQ8: IVF-Flat, IVF-SQ8. Probes: selected with either quantizer. ms: time per query. Mean, Min.: mean and minimum recall (\%). $<$99: queries below 99\%. Pairs: original / retrained.}
\label{tab:res-31b}
\footnotesize
\setlength{\tabcolsep}{3pt}
\begin{tabular}{@{}llrrrrr@{}}
\toprule
Index & $K$ & Probes & ms & Mean & Min. & $<$99\\
\midrule
Flat & 100 & 512 & 22.66 / 22.79 & 99.49 / 99.32 & 77 / 81 & 86 / 97\\
Flat & 100K & 4,096 & 186.78 / 188.92 & 99.87 / 99.85 & 98.565 / 98.434 & 12 / 13\\
SQ8 & 100 & 1,024 & 11.50 / 11.49 & 99.23 / 99.15 & 84 / 89 & 98 / 110\\
SQ8 & 100K & 4,096 & 48.50 / 48.22 & 99.73 / 99.72 & 98.555 / 98.432 & 13 / 16\\
\bottomrule
\end{tabular}
\end{table}

\paragraph{Saved index frontiers at four recall targets.}
Table~\ref{tab:res-01} lists, for every saved index sweep, the fastest saved setting that reaches 90\%, 95\%, 99\%, and 99.5\% mean recall. It analyzes saved sweeps only: no index was rebuilt and no query was rerun. The CPU rows are the 10M-row sweeps of Section~\ref{sec:ann-query}, with 1,000 queries (HNSW at $K=100$K: the first 100), timed in milliseconds per query. The GPU rows are the 25M-row calibration sweeps of Section~\ref{sec:published-baselines}, with 32 calibration queries at $B=32$, timed in milliseconds per batch. Their recall is the lowest of the three repetitions' query means. IVF-Flat times include query staging and delivery to the host, and CAGRA times are device search with resident queries. A dash means that no saved valid setting reaches the target.

At $K=500$K, the GPU IVF-Flat index already reaches 99.925\% with 2,048 of its 4,096 lists, in 846.2~ms per batch. Probing every list reaches 99.9999875\% in 496.1~ms and is the fastest saved setting that meets 99\%. Crossing a target on the calibration queries does not establish it on the held-out queries. At $K=1$K, 256 probes give 99.14\% on the calibration queries and 98.92\% on the held-out queries (Section~\ref{app:gpu-variants}). The saved exact flat scans take 214.6 and 302.6~ms per query on the CPU at $K=100$ and 100K. On the GPU, cuVS brute force takes 103.2, 106.6, 118.9, and 180.8~ms per batch at $K=1$K, 10K, 100K, and 500K on the held-out queries. Their query sets and endpoints differ from the table's.

\paragraph{The HNSW large-answer point.}
The HNSW-Flat~\cite{malkov2020hnsw} point of Section~\ref{sec:ann-query} (10M rows, $K=100$K, \texttt{efSearch} of 100,000) ran with Faiss's default bounded candidate queue. A paired control shows that a heap queue is far faster at identical recall. The point first ran with 32 OpenMP threads on the first 100 stored queries in batches of 10. The 100 queries took 706.668~s, 7,066.678~ms per query, at 99.72\% recall. The run set only \texttt{efSearch} and passed no per-call search parameters. In Faiss 1.13.2~\cite{johnson2019billion}, HNSW search then uses a bounded candidate queue by default, with capacity $\max(\texttt{efSearch},K)$, whose minimum extraction scans the candidate array. The alternative is a heap-based queue. The index file does not store this flag, and the run recorded no runtime value, so the setting is reconstructed from the source and its defaults. The latency therefore describes this Faiss configuration. The index occupies 33.4~GB and loaded in 8.6~s.

A paired four-thread control used the same 10M index, the same 100 queries, and $K=\mathtt{efSearch}=100{,}000$. It searched with the default bounded queue in 739.0~s and with a heap queue in 48.95~s, 15.1 times faster. Recall is identical (mean 99.72\%, minimum 99.01\%), and 20,000,000 returned pairs were checked independently. The ratio compares only these two four-thread runs, not the earlier 32-thread run, and Section~\ref{sec:ann-query} reports both four-thread times.

\paragraph{One-bit codes with rescoring.}
One-bit RaBitQ codes~\cite{gao2024rabitq} alone stay far below the 99\% target at large $K$. Rescoring with the FP32 vectors reaches it only from a candidate pool larger than twice the answer (Table~\ref{tab:res-09}). The experiment uses Faiss's native one-bit RaBitQ index on the first 25M rows, with the inner-product metric, unquantized queries, and a random rotation with a fixed seed. It is scored over the whole library on the CPU, so no list coverage limits recall. The codes take 2.6~GB and the FP32 vectors 76.8~GB.

Refinement rescores the top $C$ code candidates with the original FP32 vectors. On the 32 calibration queries of Section~\ref{sec:setup} (positions 0--31), Table~\ref{tab:res-09} gives the codes alone ($C=K$) and the pools $C=2K$, \ldots, $32K$. Rescoring $2K$ candidates gives 95.1\% and 98.4\% mean recall on these queries. The smallest $C$ among $K$, $2K$, \ldots, $32K$ reaching 99\% mean recall was frozen and run on the 32 held-out queries (positions 32--63), together with the codes alone. It gives 99.56\% at $K=100$K with $C=400$K and 99.96\% at $K=500$K with $C=2$M (minimum 97.49\% and 99.68\%). An independent CPU check read the 172.8M returned pairs and rescored the 34,992,582 distinct query-ID pairs with FP64 dot products, with a largest absolute difference of $1.223\times10^{-6}$. The runs measure recall only, so no serving time is reported. The pool that recovers the target grows with the answer, as Section~\ref{sec:ann-factors} describes for refinement.

\begin{table}[t]
\caption{One-bit RaBitQ codes with FP32 rescoring. Faiss's native one-bit index on the first 25M rows, scored on the CPU. Each cell gives mean (minimum) recall in \% of the FP32 top-$K$. $C$: candidates rescored with FP32 vectors ($C=K$: codes alone). Held-out values show the codes alone and the pool selected on the calibration queries, and a dash marks a pool not run on the held-out queries.}
\label{tab:res-09}
\small
\begin{tabular}{@{}rrrr@{}}
\toprule
$K$ & $C$ & Calibration & Held-out\\
\midrule
100K & 100K & 75.71 (64.38) & 75.70 (65.34)\\
100K & 200K & 95.12 (85.90) & --\\
100K & 400K & 99.61 (97.19) & 99.56 (97.49)\\
100K & 800K & 99.98 (99.78) & --\\
100K & 1.6M & $>$99.99 ($>$99.99) & --\\
100K & 3.2M & $>$99.99 ($>$99.99) & --\\
\midrule
500K & 500K & 82.14 (73.56) & 81.71 (73.96)\\
500K & 1M & 98.45 (93.79) & --\\
500K & 2M & 99.96 (99.53) & 99.96 (99.68)\\
500K & 4M & $>$99.99 (99.99) & --\\
500K & 8M & $>$99.99 ($>$99.99) & --\\
500K & 16M & $>$99.99 ($>$99.99) & --\\
\bottomrule
\end{tabular}
\end{table}

\paragraph{CAGRA parameter grid.}
The CAGRA calibration grid~\cite{ootomo2024cagra} of Section~\ref{sec:published-baselines} meets 99\% mean recall at $K=1$K and 10K and at neither larger answer size. It has 46 presets on the 25M rows, with the 32 calibration queries in one batch. The 23 search settings, 12 of an initial grid and 11 of a supplement, each ran on the NN-descent graph and on the IVF-PQ graph (graph degree 64, intermediate degree 128). Per builder, three settings target $K=1$K, three $K=10$K, eight $K=100$K, and nine $K=500$K. The settings vary the search algorithm (automatic or multi-kernel), \texttt{itopk\_size} from 4,096 to 930,016, and \texttt{search\_width} from 1 to 64. They also vary \texttt{max\_iterations} from 32 to 192, the hash-map fill rate (0.875 or 0.89), and, where set, the number of random samplings from 12 to 256. Ten presets at $K=500$K, five per builder, returned invalid outputs and are excluded. Table~\ref{tab:res-11} gives the fastest preset that meets 99\% and the best valid recall at each $K$, where recall is the lowest of three repetition means. The grid has no $K=25$K point.

\begin{table}[t]
\caption{CAGRA calibration grid. The 46 presets on 25M rows, 32 calibration queries in one batch. Fastest at 99\%: preset (graph): device time per batch (ms), mean recall. Best valid recall: recall: preset (graph), device time per batch (ms). Recall: lowest of three repetition means. N: NN-descent graph. I: IVF-PQ graph.}
\label{tab:res-11}
\footnotesize
\begin{tabular}{@{}rll@{}}
\toprule
$K$ & Fastest at 99\% & Best valid recall\\
\midrule
1K & 0 (N): 8.95, 99.92\% & 99.98\%: 1 (N), 26.78\\
10K & 3 (N): 85.76, 99.39\% & 99.40\%: 4 (N), 121.69\\
100K & none & 97.86\%: 14 (N), 8,100.57 and 34 (I), 8,099.86\\
500K & none & 63.25\%: 21 (N), 2,415.71\\
\bottomrule
\end{tabular}
\end{table}

\paragraph{A wider CAGRA grid on 25M rows.}
A second, wider CAGRA grid confirms, at every batch of 32 to 256 queries, that no configuration reaches 99\% mean recall at $K=100$K or 500K. It uses the first 25M rows on one PRO 6000 GPU and the NN-descent settings of Section~\ref{sec:published-baselines}. These are cuVS 26.04, FP16 vectors, inner product, 20 NN-descent iterations, graph degree 64, and intermediate degree 128, and the graph was built in 297.1~s. A 512-query set with an exact FP32 reference (TF32 off) extends the 64 queries of Figure~\ref{fig:gpu-index-comparison}. Calibration uses those 32 calibration queries and 224 new ones, and evaluation uses the 32 held-out queries and 224 new ones, 256 queries each.

The 140 search configurations were fixed before any measurement: for each batch of 32, 64, 128, and 256 queries, 17 at $K=100$K and 18 at $K=500$K. They cover cuVS's automatic algorithm choice, the multi-CTA and multi-kernel algorithms, and internal top-$k$ lists of 100,000 to 930,016 entries. They also cover automatic or explicit iteration limits of 32 to 8,192, search widths of 1 to 64, and 12 to 256 random samplings. Of the 140 configurations, 72 failed with explicit cuVS errors (32 at the public hash-map limit, 39 in the top-$k$ path, one at an RMM resource limit). The failed runs are every automatic and multi-CTA run and every multi-kernel run with an automatic or 512-to-8,192 iteration limit. Each completed configuration ran three timed repetitions. At $K=500$K, 20 of the 36 completed configurations returned fewer than $K$ valid IDs, one of them about 392,000 per query. With no configuration at the target on the calibration queries, no batch size was selected. For each batch size and $K$, the valid configuration with the highest calibration recall was frozen before any held-out run. It is multi-kernel search with a 930,016-entry list, search width 1, 32 iterations, and 256 random samplings at $K=100$K and 64 at 500K.

Table~\ref{tab:res-27} gives the held-out results. Every batch size returns the same recall, 97.88\% at $K=100$K and 63.27\% at 500K, and all 256 queries are below 99\% in every repetition. These configurations take 247 to 253 and 68 to 76~ms per query from resident queries. On the same queries, SweepANN's window plan at $B=32$ reaches 99.72\% and 99.80\% in 0.57 and 0.80~ms per query, transfer included. The CAGRA times start from queries already on the GPU in FP16, converted and uploaded once outside the timer. They end with sorted 32-bit IDs and FP32 scores on the device or in host memory. SweepANN ran the window plan as CUDA Graphs at $B=32$ ($r_0=4$, 2M-row tiles). It chose $A$ on the calibration queries, 409,600 at $K=100$K and 2,048,000 at 500K, at calibration recall 99.71\% and 99.79\%. Its times start from FP32 queries in host memory, include the transfer, and end with the answer on the GPU or in the owning host's memory. All 1.8432B saved CAGRA pairs passed an independent membership check with 32,768 source-score spot checks. That check does not cover every score bit, and the calibration arrays were not saved. All 1.2288B SweepANN pairs passed the CPU check.

FP16 storage alone does not hold CAGRA's recall below the target. On the 32 held-out queries of Figure~\ref{fig:gpu-index-comparison}, the frozen configurations reach 97.88\% and 63.26\% mean recall at $K=100$K and 500K, identically for all four batch sizes and three repetitions. To three decimals, these means are 97.880\% and 63.264\%, with minima 97.792\% and 63.150\%. An exhaustive scan of the same 25M FP16 vectors with FP32 accumulation in three 256-dimensional groups reaches 99.989\% and 99.993\% (minima 99.982\% and 99.987\%). The scan is an accuracy control, not a serving baseline.

\begin{table}[t]
\caption{Wider CAGRA grid against SweepANN. 256 held-out queries on 25M rows, one PRO 6000 GPU. Mean, Min.: recall (\%) against the exact FP32 top-$K$. Dev., Host: median time per query (ms) to the answer on the GPU and in host memory. CAGRA: the valid configuration with the highest calibration recall at each batch size, timed from queries already on the GPU. SweepANN: window plan at $B=32$, timed from FP32 queries in host memory.}
\label{tab:res-27}
\footnotesize
\setlength{\tabcolsep}{3pt}
\begin{tabular}{@{}lrrrrrr@{}}
\toprule
Plan & $B$ & $K$ & Mean (\%) & Min.\ (\%) & Dev. & Host\\
\midrule
CAGRA & 32 & 100K & 97.876 & 97.766 & 253.16 & 253.19\\
CAGRA & 64 & 100K & 97.876 & 97.766 & 249.04 & 249.08\\
CAGRA & 128 & 100K & 97.876 & 97.766 & 247.69 & 247.76\\
CAGRA & 256 & 100K & 97.876 & 97.766 & 247.30 & 247.37\\
SweepANN & 32 & 100K & 99.720 & 99.461 & 0.574 & 0.643\\
\midrule
CAGRA & 32 & 500K & 63.269 & 63.145 & 76.01 & 76.34\\
CAGRA & 64 & 500K & 63.269 & 63.145 & 70.85 & 71.15\\
CAGRA & 128 & 500K & 63.269 & 63.145 & 68.49 & 68.79\\
CAGRA & 256 & 500K & 63.269 & 63.145 & 67.50 & 67.81\\
SweepANN & 32 & 500K & 99.800 & 99.623 & 0.798 & 1.378\\
\bottomrule
\end{tabular}
\end{table}

\paragraph{CAGRA at 50M rows.}
On one PRO 6000, the NN-descent build of CAGRA fails at 50M rows whether the FP16 vectors start in device or in host memory. Both builds failed with a CUDA allocation of 76.8~GB. They used the first 50M rows, cuVS 26.04, NN-descent with 20 iterations, inner product, graph degree 64 and intermediate degree 128, and FP16 vectors rounded from the FP32 rows. With the vectors already on the device, the error came 17.8~s into the build call after 268.5~s of preparation. With the vectors in host memory, it came 20.1~s into the call after 265.8~s, with a host peak of 148.4~GB and no host-memory failure. These are times to the error, not build times, and no 50M index, search, or recall exists. The runs do not identify which buffer the allocation was for. Sampled GPU memory peaks, 91.25~GB for the process and 92.03~GB for the device, are lower bounds. A 1M-row build from host memory completes in 11.6~s.

\paragraph{Four-GPU IVF on 100M rows.}
On 100M rows over four GPUs, both native cuVS IVF indexes reach the 99\% target at 1,024 probes. Against SweepANN's frozen plans, they take 8.16 to 20.14 times as long resident and 5.63 to 16.45 times as long to the host. On the held-out queries, IVF-Flat's medians per batch are 426 and 483~ms resident at $K=100$K and 500K and 428 and 501~ms to the host. Its minimum recall is 99.18\% and 98.24\%. IVF-SQ8's are 173 and 241~ms resident and 174 and 255~ms to the host, with minimum recall 99.05\% and 98.21\%. SweepANN's frozen plans deliver the answers to the host in 26.0 and 45.4~ms, with minimum recall 99.45\% and 99.59\%. This is the 100M-row comparison that Section~\ref{sec:published-baselines} summarizes.

Both indexes split the first 100M rows into four 25M-row shards, one per PRO 6000 GPU. Each shard gets one native cuVS index with 4,096 lists, 20 k-means iterations, raw inner product, and TF32 disabled. IVF-Flat (cuVS 26.04) keeps the FP32 vectors without normalization or FP16 conversion and uses a k-means training fraction of 0.04. IVF-SQ8 (cuVS 26.10.00, the released binary) stores each residual to the list centroid as one unsigned byte per component and trains on at most 256 points per list. SweepANN's INT8 codes also hold one byte per component, but they quantize the original row in groups with FP16 scales. Neither the code bytes nor the scores therefore equal those of IVF-SQ8. Not counting the read of its rows from disk, each shard's index builds in 11.5 to 12.1~s. IVF-Flat takes 11.57 to 12.13~s, including the copy to the GPU, and IVF-SQ8 11.49 to 12.09~s. The four IVF-SQ8 build calls take 46.727~s in all, and the reads, hashing, and the four builds together 653.3~s. With only the index resident, each GPU has about 82.5~GB in use with IVF-Flat and 23.7 to 23.9~GB with IVF-SQ8. Across the four GPUs, the NVML readings are 82.45 to 82.51~GB and 23.73 to 23.86~GB, not query-lifecycle peaks. SweepANN keeps its compressed copy of the same 100M rows in about 19.4~GB per GPU, 19.35~GB at 774~B per row.

Each set of 32 queries forms one batch ($B=32$), which is copied to the four GPUs, searched on the four shards concurrently, and merged exactly on GPU~0 over peer-to-peer copies. The resident endpoint ends with the complete global answer on GPU~0 and includes the query copy, the native searches, the copies, and the merge. The host endpoint adds the final copy of the answer to the host. The queries are the 32 calibration (0--31) and 32 held-out (32--63) queries of the 25M runs, all used in earlier experiments. Recall is measured against the exhaustive FP32 top-$K$ over all 100M rows, with TF32 disabled. As in the 25M runs of Section~\ref{sec:published-baselines}, each point of the six-point probe grid ran three times on the calibration queries (Tables~\ref{tab:res-21a} and~\ref{tab:res-21c}). The fastest point by median host time whose three repetition means all reach 99\% was frozen before three held-out repetitions. This rule gives 1,024 probes for both indexes at both $K$. At $K=500$K, 16 probes return fewer than $K$ valid candidates after the global merge for either index, and that point counts as failed. At $K=100$K, IVF-Flat probing all 4,096 lists (0.5417~s resident) is faster than probing 2,048 (0.8543~s).

SweepANN ran on the same rows, queries, and GPUs with the same CPU affinity. It used $B=32$, $r_0=4$, $G=16$, $A=1{,}024{,}000$, and 2M-row tiles, and each 25M-row shard ends in a 1M-row tail. Each of its two methods ran three repetitions, with no overflow in any held-out repetition. The method with the faster median host time on the calibration queries was frozen before the held-out runs: the window plan at $K=100$K and the full-score plan at $K=500$K. At 500K, the unselected window plan takes 37.1~ms resident on the held-out queries. Table~\ref{tab:res-21b} gives both methods on the held-out queries. The table's ratios divide the IVF-Flat median by the SweepANN median of the same endpoint, and the two medians come from separate processes. The outputs differ in format: both indexes return FP32 scores with INT64 IDs (12~bytes per pair), and SweepANN returns FP16 scores with INT32 IDs (6~bytes per pair).

The first independent CPU check of the IVF-Flat outputs failed its fixed score tolerance (absolute and relative $2\times10^{-6}$) at an error of $7.70\times10^{-6}$, and we report that failure. A corrected check recomputed 158,349,773 distinct query-ID pairs of all four shards in FP64 and checked 355.2M merged pairs. All 13 occurrences outside the original tolerance come from one query-ID pair. They match an exact ascending-order IEEE fused-multiply-add evaluation bit for bit, and every score is within the full forward error bound (at most 0.0401 of it). No GPU result, time, or the 99\% threshold changed.

\begin{table}[t]
\caption{IVF-Flat probe grid on 100M rows. One cuVS index per 25M-row shard, four PRO 6000 GPUs, $B=32$. Res., Host: median time per batch (s) of three repetitions, to the answer on GPU~0 and in host memory. Mean, Min.: recall (\%) against the exhaustive FP32 top-$K$, identical in the three repetitions. Cal.: queries 0--31. Held-out: queries 32--63, run only at the frozen 1,024 probes. $\geq$99: all three repetition means reach 99\%. At $K=500$K, 16 probes failed because the merged answer held fewer than $K$ valid candidates.}
\label{tab:res-21a}
\footnotesize
\setlength{\tabcolsep}{4pt}
\begin{tabular}{@{}rlrrrrl@{}}
\toprule
Probes & Set & Res.\ (s) & Host (s) & Mean (\%) & Min.\ (\%) & $\geq$99\\
\midrule
\multicolumn{7}{@{}l}{$K=100$K}\\
16 & cal. & 0.0408 & 0.0423 & 56.036 & 28.868 & no\\
64 & cal. & 0.0490 & 0.0512 & 83.713 & 59.449 & no\\
256 & cal. & 0.1293 & 0.1319 & 97.408 & 87.350 & no\\
1,024 & cal. & 0.4339 & 0.4385 & 99.895 & 98.947 & yes\\
2,048 & cal. & 0.8543 & 0.8578 & 99.993 & 99.911 & yes\\
4,096 & cal. & 0.5417 & 0.5486 & 100 & 100 & yes\\
1,024 & held-out & 0.4264 & 0.4280 & 99.885 & 99.177 & yes\\
\midrule
\multicolumn{7}{@{}l}{$K=500$K}\\
16 & cal. & \multicolumn{5}{l}{failed (see caption)}\\
64 & cal. & 0.1136 & 0.1322 & 70.027 & 44.726 & no\\
256 & cal. & 0.1923 & 0.2126 & 93.906 & 77.686 & no\\
1,024 & cal. & 0.5012 & 0.5298 & 99.709 & 97.597 & yes\\
2,048 & cal. & 0.9238 & 0.9418 & 99.980 & 99.758 & yes\\
4,096 & cal. & 0.6044 & 0.6221 & 100 & 100 & yes\\
1,024 & held-out & 0.4832 & 0.5010 & 99.660 & 98.244 & yes\\
\bottomrule
\end{tabular}
\end{table}

\begin{table}[t]
\caption{IVF-Flat against SweepANN on 100M rows. Held-out queries 32--63 on four PRO 6000 GPUs, $B=32$: IVF-Flat at the frozen 1,024 probes and SweepANN's window and full-score plans ($*$: the plan frozen on the calibration queries). Res., Host: median time per batch (ms) of three repetitions. Mean, Min.: recall (\%) over the 32 queries against the exhaustive FP32 top-$K$ (SweepANN's three repetitions agree to the digits shown). Native / SweepANN: IVF-Flat median divided by the SweepANN median of the same endpoint.}
\label{tab:res-21b}
\footnotesize
\setlength{\tabcolsep}{3.5pt}
\begin{tabular}{@{}lrrrrrr@{}}
\toprule
Method & Res. & Host & Mean & Min. & \multicolumn{2}{c@{}}{Native / SweepANN}\\
 & (ms) & (ms) & (\%) & (\%) & res. & host\\
\midrule
\multicolumn{7}{@{}l}{$K=100$K}\\
IVF-Flat, 1,024 & 426.41 & 427.96 & 99.885 & 99.177 & -- & --\\
SweepANN window$*$ & 21.17 & 26.02 & 99.644 & 99.450 & 20.14 & 16.45\\
SweepANN full & 21.40 & 25.74 & 99.645 & 99.446 & 19.92 & 16.63\\
\midrule
\multicolumn{7}{@{}l}{$K=500$K}\\
IVF-Flat, 1,024 & 483.22 & 500.98 & 99.660 & 98.244 & -- & --\\
SweepANN window & 37.15 & 55.22 & 99.733 & 99.598 & 13.01 & 9.07\\
SweepANN full$*$ & 27.55 & 45.36 & 99.733 & 99.592 & 17.54 & 11.05\\
\bottomrule
\end{tabular}
\end{table}

\paragraph{Four-GPU IVF-SQ8 on the same 100M rows.}
Table~\ref{tab:res-21c} gives the IVF-SQ8 probe grid under the protocol above. Its held-out medians are 172.8~ms resident and 174.3~ms to the host at $K=100$K and 241.0 and 255.4~ms at $K=500$K, at 99.59\% and 99.49\% mean recall. Against the frozen SweepANN plans of Table~\ref{tab:res-21b}, the ratios are 8.16 resident and 6.70 to the host at $K=100$K and 8.75 and 5.63 at 500K. Against the other SweepANN plan they are 8.07 / 6.77 (full-score plan at 100K) and 6.49 / 4.62 (window plan at 500K). An independent CPU check read all 390 output files (21.31~GB). It recomputed the exact global merge of 355.2M pairs bit for bit from the raw per-shard native outputs and measured every returned ID against the FP32 reference. It checked the SQ8 scores themselves only on a bounded sample, the first two exported rows of every nonempty list. That check used FP64 dot products of the represented vectors and a tolerance of $2\times10^{-5}$. The check therefore does not recompute every native score, and it does not equate the SQ8 scores with the FP32 scores.

\begin{table}[t]
\caption{IVF-SQ8 probe grid on 100M rows. One cuVS index per 25M-row shard, four PRO 6000 GPUs, $B=32$. Res., Host: median time per batch (s) of three repetitions, to the answer on GPU~0 and in host memory. Mean, Min.: recall (\%) against the exhaustive FP32 top-$K$, identical in the three repetitions. Cal.: queries 0--31. Held-out: queries 32--63, run only at the frozen 1,024 probes. $\geq$99: all three repetition means reach 99\%. At $K=500$K, 16 probes failed because the merged answer held fewer than $K$ valid candidates.}
\label{tab:res-21c}
\footnotesize
\setlength{\tabcolsep}{4pt}
\begin{tabular}{@{}rlrrrrl@{}}
\toprule
Probes & Set & Res.\ (s) & Host (s) & Mean (\%) & Min.\ (\%) & $\geq$99\\
\midrule
\multicolumn{7}{@{}l}{$K=100$K}\\
16 & cal. & 0.0273 & 0.0287 & 55.745 & 29.331 & no\\
64 & cal. & 0.0395 & 0.0408 & 83.772 & 60.178 & no\\
256 & cal. & 0.0633 & 0.0649 & 97.275 & 87.329 & no\\
1,024 & cal. & 0.1669 & 0.1685 & 99.575 & 98.886 & yes\\
2,048 & cal. & 0.3158 & 0.3173 & 99.628 & 99.258 & yes\\
4,096 & cal. & 0.5458 & 0.5473 & 99.630 & 99.265 & yes\\
1,024 & held-out & 0.1728 & 0.1743 & 99.586 & 99.046 & yes\\
\midrule
\multicolumn{7}{@{}l}{$K=500$K}\\
16 & cal. & \multicolumn{5}{l}{failed (see caption)}\\
64 & cal. & 0.1062 & 0.1184 & 69.952 & 44.891 & no\\
256 & cal. & 0.1272 & 0.1412 & 93.846 & 77.727 & no\\
1,024 & cal. & 0.2332 & 0.2483 & 99.519 & 97.701 & yes\\
2,048 & cal. & 0.3784 & 0.3926 & 99.713 & 99.407 & yes\\
4,096 & cal. & 0.6190 & 0.6346 & 99.724 & 99.425 & yes\\
1,024 & held-out & 0.2410 & 0.2554 & 99.492 & 98.205 & yes\\
\bottomrule
\end{tabular}
\end{table}

\paragraph{Each method at its own best batch on 100M rows.}
When each method chooses its own batch, SweepANN's throughput on 100M rows is $9.90\times$ IVF-SQ8's to the host and $10.38\times$ resident at $K=100$K, and $4.85\times$ and $9.52\times$ at 500K. The comparison above fixes $B=32$ for every method. This second run keeps the four PRO 6000 GPUs and one 25M-row shard per GPU and lets each method choose its batch on 256 calibration queries. The tuned native cuVS IVF-SQ8 index has 4,096 lists and the probe count chosen at $B=64$ on the calibration queries, 1,024 at both $K$. It runs at $B\in\{32,64,128,256\}$, and SweepANN's window and full-score plans run at the same batches. Each method takes the fastest configuration whose per-repetition mean recall on the calibration queries stays at or above 99\%, frozen before the 256 held-out queries. Throughput is the 256 held-out queries over the aggregate time of the batches that serve them. Each value is the median of this per-repetition aggregate over three repetitions. Both endpoints start from FP32 queries in host memory and include the index's host-to-device input copies. The resident endpoint ends with the global output on the GPUs, and the host endpoint with the owned host result. The read-back of the raw shard outputs for the check and the disk writes lie outside the timer.

At $K=100$K, IVF-SQ8 is fastest at $B=32$, with 186.6 queries/s to the host and 188.4 resident (99.58\% mean and 98.00\% minimum recall). The window plan is fastest at $B=64$, with 1,847.0 and 1,956.0 queries/s (99.63\% and 99.20\%). At $K=500$K, both are fastest at $B=256$. The index serves 133.0 queries/s to the host and 143.9 resident (99.49\% and 96.22\%), and the full-score plan 644.6 and 1,370.1 (99.73\% and 99.36\%). Unrounded, the throughputs are 186.606 and 188.422 against 1,846.983 and 1,956.020 queries/s at $K=100$K and 132.995 and 143.851 against 644.632 and 1,370.100 at 500K. The unrounded ratios are $9.898\times$, $10.381\times$, $4.847\times$, and $9.524\times$. The unrounded recalls are 99.582\% and 98.004\% against 99.632\% and 99.200\% at $K=100$K, and 99.491\% and 96.216\% against 99.726\% and 99.361\% at 500K. With the same batch for both methods at $K=100$K, SweepANN's host throughput at $B=32$ is $6.89\times$ the index's with the full-score plan and $7.67\times$ with the window plan. At $B=64$ it is $8.20\times$ and $9.59\times$.

The index returns 12 bytes per pair and SweepANN 6. The index's scores were checked at the returned pairs that fall in a bounded sample of represented scores. Memory is measured as sampled total physical use for the index and as allocated and reserved auxiliary memory for SweepANN. This comparison covers the tuned IVF-SQ8 index only. Section~\ref{sec:published-baselines} reports these aggregates beside the fixed-$B=32$ medians on 32 queries. The next paragraph deploys the same index on all 500M rows over eight GPUs.

\paragraph{Eight-GPU IVF-SQ8 on 500M rows.}
On all 500M rows over eight PRO 6000 GPUs, IVF-SQ8 meets the target at 1,024 probes. It takes 532.2 and 688.4~ms per batch resident at $K=100$K and 500K, against 45.9 and 59.4~ms for SweepANN's calibrated plans (Table~\ref{tab:res-35}). The comparison ran on the eight-GPU server with one native cuVS IVF-SQ8 index (4,096 lists) per 62.5M-row shard. The index set was prepared on the main server in 7,667~s and then queried on the eight-GPU server. Preparation covers input reading and hashing, construction, serialization, and an independent verification. The 32 calibration queries select 1,024 probes at both $K$ for the index and, for SweepANN, the window plan at $K=100$K and the full-score plan at 500K. All six configurations, the index and both SweepANN plans at both $K$, then ran three timed repetitions of the 32 held-out queries at $B=32$.

To the host, the index takes 535.8 and 804.0~ms per batch and SweepANN's calibrated plans 47.5 and 114.5~ms. Table~\ref{tab:res-35} also gives SweepANN's other plan at each $K$. The lowest repetition mean recall is 99.53\% and 99.58\% for the index and 99.53\% and 99.64\% for SweepANN. The lowest single-query recall is 99.196\% (index, $K=500$K). Memory is the largest per-GPU reading under two different instruments. The index's total physical use, sampled every 50~ms, reaches 53.7 and 63.1~GiB. SweepANN's PyTorch allocator peak over capture, warmup, and the held-out runs reaches 45.6 and 49.5~GiB with the calibrated plans. The GPU run and an independent CPU check on the eight-GPU server validated all 920 native result arrays (4.1472B returned slots). Section~\ref{sec:published-baselines} gives the resulting ratios and the cost of preparing the index set.

\begin{table}[t]
\caption{IVF-SQ8 against SweepANN on 500M rows. 32 held-out queries on eight PRO 6000 GPUs, $B=32$, one IVF-SQ8 index per 62.5M-row shard. Sel.: selected on the calibration queries. Resident, Host: median time per batch (ms) of three repetitions. Mean: lowest repetition mean recall (\%). Min.: lowest single-query recall (\%). Mem.: largest per-GPU memory (GiB), NVML total for the index, PyTorch allocator peak for SweepANN.}
\label{tab:res-35}
\footnotesize
\setlength{\tabcolsep}{3pt}
\begin{tabular}{@{}rlcrrrrr@{}}
\toprule
$K$ & Method & Sel. & Resident & Host & Mean & Min. & Mem.\\
\midrule
100K & IVF-SQ8, 1,024 probes & yes & 532.2 & 535.8 & 99.53 & 99.34 & 53.7\\
100K & SweepANN window & yes & 45.9 & 47.5 & 99.53 & 99.28 & 45.6\\
100K & SweepANN full score & no & 51.0 & 52.8 & 99.53 & 99.27 & 48.9\\
500K & IVF-SQ8, 1,024 probes & yes & 688.4 & 804.0 & 99.58 & 99.20 & 63.1\\
500K & SweepANN window & no & 66.3 & 126.9 & 99.64 & 99.48 & 46.3\\
500K & SweepANN full score & yes & 59.4 & 114.5 & 99.64 & 99.47 & 49.5\\
\bottomrule
\end{tabular}
\end{table}

\paragraph{Four-GPU IVF-SQ8 on 500M rows.}
On the four PRO 6000 GPUs of the main server, the same index takes $9.8\times$ and $8.0\times$ as long as SweepANN's window plan to the host at $K=100$K and 500K. Resident, it takes $9.9\times$ and $8.9\times$ as long. This matched run uses the 32 calibration and 32 held-out queries of the eight-GPU run, $B=32$, and three timed repetitions. The eight trained IVF-SQ8 indexes are reused, two per GPU sharing one allocation pool, and searched at 1,024 probes. The answers are merged exactly on GPU~0 in chunks of four queries. SweepANN runs the window plan ($r_0=4$, $G=16$, $A=1{,}024{,}000$) at both $K$ with no plan selection, and every setting is fixed before the held-out queries. At $K=100$K and 500K, the index's held-out medians per batch are 838 and 1,006~ms to the host (837 and 960~ms resident). SweepANN's are 85.6 and 125.3~ms (84.1 and 108.2~ms resident). Mean recall is 99.53\% and 99.58\% for the index and 99.53\% and 99.64\% for SweepANN. The lowest single-query recall is 99.196\% for the index and 99.275\% for SweepANN, and no window overflows.

\paragraph{Bounded refiner read pattern.}
Reading contiguous row ranges, rather than only the requested rows, makes the bounded refiner of Section~\ref{sec:extension} faster. In a separate run on another A40 with a fixed candidate set, this variant lowers the median rerank time from 1,197.5 to 417.9~s. The median paired reduction is 65.02\%, with more bytes read in far fewer read calls. These runs did not control the page cache, so their rerank times are not compared with the 1,083.5-s request of Section~\ref{sec:extension}. On a PRO 6000, where all candidate vectors fit, keeping them on the GPU took a median of 25.86~s and bounded refinement 83.55~s, a median paired ratio of 3.24. Both times cover search, refinement, delivery to a host consumer, and cleanup.

\FloatBarrier

\section{Seed, Capacity, and Robustness}\label{app:seed-capacity-robustness}
This appendix holds the evidence for three conclusions of Sections~\ref{sec:seed-capacity}, \ref{sec:capacity}, and~\ref{app:replication}. Buffers sized by the random-order law hold on permuted rows, our file order needs an empirical margin, and an undersized buffer costs time but never changes the answer. The appendix states the law at the campaign's geometry, counts the survivors of every window of the file-ordered campaign, and evaluates capacities fitted on a calibration split of the queries. It then compares file order with row permutations of 25M and 500M rows, measures what recovery costs when a buffer overflows, and shows what the local cutoff saves under an unfavorable placement. The last paragraphs check the cost model's selection term on the PRO 6000, an A100, and a GH200.

\paragraph{The random-order law at the campaign's geometry.}
Under random row order (Proposition~\ref{prop:law}), a window's survivor count has a closed-form distribution, and a union bound over every window of the campaign sets the law's capacities. Let a window hold $N_w$ main-scan rows. Under a
uniform row permutation with fixed scores, a query's $N_0$ sample rows
are a uniform subset of the ranks among the $N_0+N_w$ rows of sample and
window. For distinct scores, the number $Y$ of window rows above the
sample's $K$-th score is then a negative hypergeometric count, the
beta-binomial variable with parameters $(N_w,K,N_0-K+1)$, with mean
$N_wK/(N_0+1)$ and variance
$N_wK(N_0-K+1)(N_0+N_w+1)/((N_0+1)^2(N_0+2))$. Ties can only lower $Y$,
so with ties this count bounds the strict survivor count from above. The
variance exceeds that of a binomial count with $p=K/(N_0+1)$ by the
factor $(N_0+N_w+1)/(N_0+2)$, which accounts for the shared random
cutoff. For the fullest windows the factor is 1.4375 under Stable (28M
rows, $N_0=64$M) and 1.9375 under Optimized (30M rows, $N_0=32$M).

The executor overflows when a count exceeds $A$, so the campaign bound
sums a geometric bound on each window's tail at $A+1$ over the 16 GPU
windows and the 26,562 queries. This union bound needs no independence
between queries or windows. The smallest capacities that bring it to
$10^{-6}$ are the 222,605 and 475,273 of Section~\ref{sec:seed-capacity},
with $\log_{10}$ bounds of $-6.0028$ and $-6.0017$. Exact rational arithmetic checks the distribution on 286 small cases, with 1,287 probabilities, 1,001 ratio checks, and 734 tail comparisons. It also checks the tie bound on 510 tie partitions with 753,208 sample assignments. The
production values agree at 80 and 120 decimal digits. These capacities
apply only to a row-permuted library.

\paragraph{Survivor census of every window.}
In file order, the fullest windows hold far more survivors than the law
predicts for random order. A counter pass covers all 26,562 queries in the campaign's geometry: four
125M-row shards, $B=64$, $K=500$K, and 2M-row tiles with one 1M-row tail
per GPU. For every seed and main tile, it counts the scores strictly
above the global seed's $K$-th score and those equal to it. Two independent integer reductions
agree on all 26,774,496 counters, and a 64,000,012-row oracle with two
queries checks the seed cutoff and every tile count.
Table~\ref{tab:res-02a} gives the geometry and the fullest window of each
configuration. Stable's fullest window, 505,512, is 2.31 times the law's
mean of 218,750 for its 28M-row windows, and Optimized's fullest holds
861,749 entries. Neither exceeds its capacity of 512,000 or 1,024,000,
though the campaign had informed the design. Table~\ref{tab:res-02b} gives the distribution per window position on two protein-disjoint splits of the queries. The calibration split has 5,322 queries and the evaluation split 21,240, and the next two paragraphs use them to fit and test a capacity.
Rows equal to the cutoff are few: in the evaluation split, a window holds
at most 9,021 of them under Stable and 15,577 under Optimized. The census
is a correctness count and adds no timing.

\begin{table}[t]
\caption{Survivor census of the file-ordered campaign (500M rows, four GPUs, $B=64$, $K=500$K): geometry and fullest window over all 26,562 queries. Seed rows: sample rows $N_0$ of all GPUs. Main rows: rows per window on each GPU, in scan order. $A$: production capacity.}
\label{tab:res-02a}
\footnotesize
\begin{tabular}{@{}lrlrr@{}}
\toprule
Config. & Seed rows & Main rows & Fullest & $A$\\
\midrule
Stable ($r_0=8$) &64M&28M, 28M, 28M, 25M&505,512&512,000\\
Optimized ($r_0=4$) &32M&30M, 30M, 30M, 27M&861,749&1,024,000\\
\bottomrule
\end{tabular}
\end{table}

\begin{table*}[t]
\caption{Strict survivors per window position in the file-ordered campaign, pooled over the four GPUs. Calibration: 5,322 queries of 1,991 proteins. Evaluation: 21,240 queries of 7,928 proteins. Rows: main rows $N_w$ of the window. Law: the random-order mean $N_wK/(N_0+1)$. SD: population standard deviation. p99: linearly interpolated 99th percentile. Fitted $A$: the calibration-fitted capacity of Table~\ref{tab:res-03}. Exc.: windows above it. No window exceeds the production capacities 512,000 and 1,024,000.}
\label{tab:res-02b}
\small
\begin{tabular}{@{}llrrrrrrrrr@{}}
\toprule
Config. & Split & Window & Rows & Law & Mean & SD & p99 & Max & Fitted $A$ & Exc.\\
\midrule
Stable & Cal. & 0 & 28M & 218,750 & 225,314 & 33,111 & 317,538 & 452,221 & 452,221 & 0\\
Stable & Cal. & 1 & 28M & 218,750 & 222,930 & 33,775 & 328,734 & 429,906 & 452,221 & 0\\
Stable & Cal. & 2 & 28M & 218,750 & 216,858 & 35,847 & 327,722 & 401,438 & 452,221 & 0\\
Stable & Cal. & 3 & 25M & 195,312 & 206,287 & 31,408 & 301,802 & 398,992 & 403,769 & 0\\
Stable & Eval. & 0 & 28M & 218,750 & 225,302 & 33,439 & 320,796 & 505,512 & 452,221 & 7\\
Stable & Eval. & 1 & 28M & 218,750 & 222,692 & 33,937 & 328,478 & 443,827 & 452,221 & 0\\
Stable & Eval. & 2 & 28M & 218,750 & 217,183 & 35,858 & 329,214 & 431,290 & 452,221 & 0\\
Stable & Eval. & 3 & 25M & 195,312 & 206,244 & 31,421 & 303,010 & 395,771 & 403,769 & 0\\
\midrule
Optimized & Cal. & 0 & 30M & 468,750 & 442,009 & 49,997 & 571,508 & 675,193 & 815,081 & 0\\
Optimized & Cal. & 1 & 30M & 468,750 & 438,901 & 58,040 & 603,380 & 815,081 & 815,081 & 0\\
Optimized & Cal. & 2 & 30M & 468,750 & 427,087 & 56,712 & 617,365 & 806,962 & 815,081 & 0\\
Optimized & Cal. & 3 & 27M & 421,875 & 400,805 & 58,537 & 545,704 & 689,064 & 733,573 & 0\\
Optimized & Eval. & 0 & 30M & 468,750 & 442,926 & 50,281 & 573,232 & 760,057 & 815,081 & 0\\
Optimized & Eval. & 1 & 30M & 468,750 & 439,543 & 57,867 & 603,805 & 816,838 & 815,081 & 1\\
Optimized & Eval. & 2 & 30M & 468,750 & 428,739 & 57,299 & 616,316 & 861,749 & 815,081 & 3\\
Optimized & Eval. & 3 & 27M & 421,875 & 401,742 & 58,284 & 547,156 & 784,949 & 733,573 & 1\\
\bottomrule
\end{tabular}
\end{table*}

\paragraph{Capacities fitted on the calibration split.}
A capacity sized from a sample of queries needs a margin: fitted on the
calibration split, it is exceeded on the evaluation split. For each
configuration, CV is the population coefficient of variation of the
strict counts of all seed tiles, pooled over the calibration queries and
the four GPUs. The fitted capacity of a window of $N_w$ main rows is $\lceil \mu(1+z\,\mathrm{CV})\rceil$ with $\mu=N_wK/(N_0+1)$, where $z$ is the smallest value that covers every calibration window. The fit was
saved before any evaluation count was read. It gives 452,221 for Stable
and 815,081 for Optimized, and 403,769 and 733,573 for the shorter final
windows (Table~\ref{tab:res-03}). Because $z$ is fitted to the
calibration maximum, CV cancels from the resulting capacities: the rule
restates the calibration maximum and does not predict it independently.
Among the 21,240 evaluation queries, 7 windows of 6 queries exceed the
fitted capacity under Stable and 5 windows of 5 queries under Optimized. The largest ratio of an evaluation count to its window's fitted capacity is 1.118 under Stable and 1.070 under Optimized, the latter in a final window. Fitted to the calibration 99th percentile instead of the
maximum, $z$ would be 1.174 and 0.705. These smaller capacities were
evaluated only offline, and no serving run used them. The production
capacities have no exceedance in either split.

\begin{table}[t]
\caption{Capacity fitted on the calibration split (5,322 queries, 1,991 proteins) and evaluated on the disjoint evaluation split (21,240 queries, 7,928 proteins, 339,840 windows per configuration) of the file-ordered campaign. CV: coefficient of variation of the calibration seed-tile counts. $z$: fitted multiplier.}
\label{tab:res-03}
\footnotesize
\begin{tabular}{@{}lrr@{}}
\toprule
 & Stable & Optimized\\
\midrule
Calibration seed tiles &170,304&85,152\\
Seed-tile mean (SD) &15,575.5 (6,708.3)&31,169.8 (12,251.5)\\
CV &0.4307&0.3931\\
Fitted $z$ &2.478&1.880\\
$A$, first three windows &452,221&815,081\\
$A$, final window &403,769&733,573\\
Evaluation windows above $A$ &7&5\\
Evaluation queries affected &6&5\\
Largest count / $A$ &1.118&1.070\\
Windows above 512,000 / 1,024,000 &0&0\\
\bottomrule
\end{tabular}
\end{table}

\paragraph{Fitted capacities over the whole campaign.}
Served with the fitted capacities, the whole campaign spends 19.6~s of
Stable's 102.7~s and 12.6~s of Optimized's 91.9~s of resident time in
overflowing batches (Table~\ref{tab:fitted-capacity}). We froze each configuration's
fitted capacity for its first three windows and used it as one scalar
for every window, the shorter final window included. Under
this rule the census predicts seven query-GPU flags in six batches for
Stable and four flags in four batches for Optimized. The fifth Optimized
exceedance is not flagged: it has 784,949 survivors in a final window,
above its fitted 733,573 but below 815,081. One pass per configuration served all 26,562 queries once, in stored order, on the main server's four PRO 6000 GPUs against the 500M rows. It ran at $B=64$ and $K=500$K, in 415 full batches and a two-query tail padded to 64. Both passes use the
default placement of evenly spaced seed tiles and the fixed cutoff, not
the local cutoff or the unfavorable placement. They match the
prediction: 6 of the 416 batches overflow under Stable and 4 under
Optimized. When a query overflows on a GPU, the batch's failed graphs are
cleaned up, the batch is rescanned without pruning, and the graphs are
recaptured, all inside the batch's measured interval. The times are measured with the survivor counters of the instrumented campaigns (Table~\ref{tab:res-24}). An observer copies the uncapped counts and flags at every window end, and its cost is inside the measured call.

The answers are stored as lossless deltas against the verified answers
of the four-GPU campaign. Of the 26.562B pairs the two runs return,
26,556,752,927 coincide with those answers and the other 5,247,073 were
recomputed with the unchanged scorer. An independent CPU pass
reconstructed every pair. Each configuration ran once, without an FP32 recall measurement.

\begin{table}[t]
\caption{One instrumented campaign pass per configuration with the capacity fitted on the calibration split, used for every window. The capacities are 452,221 for Stable and 815,081 for Optimized (500M rows in file order, four PRO 6000 GPUs, $B=64$, $K=500$K, 416 batches). Times in s. The times of overflowing batches include cleanup, the unpruned rescan, and graph recapture.}
\label{tab:fitted-capacity}
\footnotesize
\renewcommand{\arraystretch}{0.9}
\begin{tabular}{@{}lrr@{}}
\toprule
 & Stable & Optimized\\
\midrule
Overflowing batches (query-GPU flags) & 6 (7) & 4 (4)\\
Resident, to the host & 102.7, 126.7 & 91.9, 116.0\\
Other batches, resident & 83.1 & 79.4\\
Overflowing batches, resident & 19.6 & 12.6\\
\quad cleanup, rescan, recapture & 0.8, 5.7, 11.9 & 0.6, 3.8, 7.4\\
\bottomrule
\end{tabular}
\end{table}

\paragraph{Row order against the law.}
Only row-permuted layouts stay within the law's envelopes.
Table~\ref{tab:res-13} compares the law with the saved survivor counts.
For each case it gives the law's mean and standard deviation for the
largest window and the observed largest count. Two further columns come from the union bound, without independence. One is an upper bound on the expected largest count over all query-window pairs. The other is the count below which the largest one stays with probability 99\%.

A 25M-row control isolates row order. Rows keep their order within a
file (Section~\ref{sec:setup}). On the first 25M rows and one GPU, we compared this file order with one random row permutation fixed in advance. The runs use $B=64$, $K=500$K, 1M-row tiles, $r_0=4$, $G=4$, $A=1{,}024{,}000$, and the queries of Section~\ref{app:grouped}. With
$N_0=4$M, a query has seven windows of 4M, 3M, 3M, 3M, 3M, 4M, and 1M
rows. For random order the law gives the 1M-, 3M-, and 4M-row windows
standard deviations of 370, 758, and 935, around means of 125,000,
375,000, and 500,000. A union bound over the 64 queries' 448 query-window pairs
(128 of 4M rows, 256 of 3M, and 64 of 1M) keeps the largest count below
503,556 with probability at least 99\%. Per 64-query set the law expects
168.0M main-scan candidates (167,999,958). With permuted rows, the
calibration and held-out queries give 167,955,491 and 167,875,454
candidates, and their largest windows hold 502,207 and 501,937, inside
the envelope. In file order they give 168,394,588 (168.4M) and
170,740,199 candidates, and their largest windows hold 1,168,893 and
990,600. One calibration query gets a low cutoff in file order and
overflows in every repetition. Recovery keeps the answer correct, but the
median calibration batch takes 146.9 instead of 48.1~ms of resident time.
No held-out query overflows, and their batches take 47.6 and 48.0~ms in
the two orders. One permutation is a single draw, so the control tests
the envelopes but not their tail probability.

The other saved counts of Table~\ref{tab:res-13} are also above the
law's 99\% bounds. The peaks of the hash-ordered libraries of
Table~\ref{tab:model-pressure} are 1.13 to 2.19 times the law's means for
their 17M-, 26.5M- and 28M-row windows. For the 500M Stable census in
file order, the union bound evaluated at the observed 505,512 has a
$\log_{10}$ of $-37{,}730.4$. This value measures how far file order lies from the law, whose bound assumes random order.

\begin{table}[t]
\caption{The random-order law against observed survivor peaks. The cases are the 25M-row control in file order and permuted (calibration and evaluation queries), the hash-ordered ConGLUDe and ConPLex libraries of Table~\ref{tab:model-pressure} by rows and $K$, and the 500M Stable census in file order. Mean, SD: the law for the largest window. $E$max: upper bound on the expected largest count. 99\%: 99\% envelope of the largest count. Peak: observed largest count. The 500M row uses the 26,560 queries of an earlier census.}
\label{tab:res-13}
\footnotesize
\begin{tabular}{@{}lrrrrr@{}}
\toprule
Case & Mean & SD & $E$max & 99\% & Peak\\
\midrule
25M file, cal. &500,000&935&502,677&503,556&1,168,893\\
25M file, eval. &500,000&935&502,677&503,556&990,600\\
25M permuted, cal. &500,000&935&502,677&503,556&502,207\\
25M permuted, eval. &500,000&935&502,677&503,556&501,937\\
\midrule
ConGLUDe 100M, 100K &53,125&285&53,942&54,210&65,599\\
ConGLUDe 100M, 500K &265,625&633&267,437&268,031&300,671\\
ConGLUDe 250M, 100K &87,500&404&88,660&89,041&191,437\\
ConGLUDe 250M, 500K &437,500&899&440,073&440,917&527,121\\
ConPLex 100M, 100K &53,125&285&53,942&54,210&75,135\\
ConPLex 100M, 500K &265,625&633&267,437&268,031&341,190\\
ConPLex 250M, 100K &87,500&404&88,660&89,041&132,120\\
ConPLex 250M, 500K &437,500&899&440,073&440,917&552,202\\
\midrule
500M Stable, file &218,750&559&221,413&221,784&505,512\\
\bottomrule
\end{tabular}
\end{table}

\paragraph{The law at 500M rows.}
On a row permutation of all 500M rows, whole campaigns at the law's
capacities run without overflow. We drew one pseudorandom permutation fixed in advance (NumPy PCG64DXSM, seed 202610170100) and ran the whole campaign on it five times on the eight-GPU server of Section~\ref{sec:setup}. Each run is
one pass, without repetitions, over the 26,562 queries with $B=64$,
$K=500$K, $G=16$, and 2M-row tiles, in 416 batches, the last with two
real queries. Stable ($r_0=8$) ran on GPUs 4--7 with CPU cores 72--79 and Optimized ($r_0=4$) on GPUs 0--3 with cores 0--7. Each ran once at the law's capacity, 222,605 or 475,273, and once at the campaign's, 512,000 or 1,024,000. A fifth run at $r_0=2$ with the
law's capacity 1,011,487 also used GPUs 4--7. Table~\ref{tab:res-19}
lists the five runs.

\begin{table*}[t]
\caption{The five runs on one permutation of all 500M rows and two file-order runs of Optimized, all on the eight-GPU server (whole campaign, $B=64$, $K=500$K, $G=16$, one pass each). Max: largest window survivor count over all query-GPU pairs (\% of $A$). Mean: mean of the 106,248 query-GPU maxima (not recorded at $r_0=2$). Ovf.: overflowing batches. Times in s, summed over the 416 batches. Pos.: answer in permuted positions on the GPUs. Remap: host mapping to library IDs. Lib.\ ID: answer with library IDs on the GPUs, including the remap. Host: answer on the owning host. Recall in \%, mean / minimum, 64 held-out queries: served 500K answers, and the offline top 100K of the saved 500K. File order returns library IDs on the GPU.}
\label{tab:res-19}
\footnotesize
\setlength{\tabcolsep}{3.5pt}
\begin{tabular}{@{}llrrrrrrrll@{}}
\toprule
Run & $r_0$, $A$ & Max (\%) & Mean & Ovf. & Pos. & Remap & Lib.\ ID & Host & 500K & 100K offline\\
\midrule
Optimized, law & 4, 475,273 & 472,427 (99.40) & 468,363.6 & 0 & 58.227 & 355.494 & 413.721 & 488.576 & 99.636940625 / 99.2858 & 99.533015625 / 99.064\\
Optimized, campaign & 4, 1,024,000 & 472,427 (46.14) & 468,363.6 & 0 & 58.931 & 333.341 & 392.272 & 470.281 & 99.636940625 / 99.2858 & 99.533015625 / 99.064\\
Stable, law & 8, 222,605 & 221,425 (99.47) & 218,621.8 & 0 & 64.020 & 220.148 & 284.167 & 326.888 & 99.636950000 / 99.286 & 99.533015625 / 99.064\\
Stable, campaign & 8, 512,000 & 221,425 & 218,621.8 & 0 & 64.362 & 231.350 & 295.711 & 340.358 & 99.636953125 / 99.286 & 99.533015625 / 99.064\\
$r_0=2$, law & 2, 1,011,487 & 1,007,232 (99.6) & -- & 0 & 56.413 & 210.396 & 266.808 & 309.347 & 99.636971875 / 99.2854 & 99.533015625 / 99.064\\
\midrule
File order, run 1 & 4, 1,024,000 & -- & -- & 0 & -- & -- & 57.281 & 111.997 & 99.63825625 / 99.2854 & 99.533015625 / 99.064\\
File order, run 2 & 4, 1,024,000 & -- & -- & 0 & -- & -- & 57.394 & 127.996 & -- & --\\
\bottomrule
\end{tabular}
\end{table*}

The executor records, for every query and GPU, the largest survivor
count of its windows, and at equal $r_0$ these counts are identical at
both capacities. The fullest window of any
query on any GPU holds 221,425 entries at $r_0=8$ and 472,427 at $r_0=4$,
99.5\% and 99.4\% of the law's capacities, and 1,007,232 (99.6\%) at
$r_0=2$. Over the 106,248 query-GPU maxima, the counts range from 457,596
to 472,427 at $r_0=4$ (median 468,471, 95th percentile 470,228) and from
212,973 to 221,425 at $r_0=8$ (median 218,673, 95th percentile 219,666). They average 468,364 and 218,622. Taken over the four GPUs, the
26,562 per-query maxima at $r_0=8$ have a minimum of 213,347, a median of
219,043, and 95th and 99th percentiles of 219,960 and 220,282 (nearest
rank). At $r_0=2$ they range from 987,844 to 1,007,232 (median
1,000,320, 95th and 99th percentiles 1,003,471 and 1,004,535).

The seed tiles are fixed by position. On each 125M-row shard of 63 tiles, $r_0=4$ samples tiles 12, 25, 37, and 50, with $N_0=32$M sample rows. Each of the first three 16-tile windows then scans 30M main rows, and the last scans 27M. With $r_0=8$ the sample takes tiles 7, 14, 21, 28,
34, 41, 48, and 55, leaving 28M main rows per window (25M in the last)
and $N_0=64$M. The law's means for one full window, the row ratios
$K\times30\mathrm{M}/32\mathrm{M}=468{,}750$ and
$K\times28\mathrm{M}/64\mathrm{M}=218{,}750$, are close to the averages
above. Those averages are of maxima over windows and depend on each query's cutoff, FP16 ties, and the data, so the row ratios give their expected scale, not their exact values.

Each run returns 13,281,000,000 pairs. Their score bits were checked against one shared reference answer, and pairs outside it were rescored on the GPU from the permuted rows. CPU processes on the eight-GPU server and on the main server reread all pairs of every run. Recall is measured
for the 64 held-out queries (IDs 256--319) against the FP32 reference
computed with TF32 disabled. It covers the served 500K answers and,
offline, the top 100K of the saved 500K, chosen by descending score with
ties broken toward the smaller library ID. The offline lists have no
serving time. In all five runs the served answers reach 99.637\% mean
and 99.29\% minimum recall. On the same queries the first file-order
run's served answers reach 99.638\% mean recall (minimum 99.2854\%,
maximum 99.802\%), and the second file-order run has no recall record.

The GPU scan takes 64.0 and 64.4~s at $r_0=8$, 58.2 and 58.9~s at $r_0=4$, and 56.4~s at $r_0=2$. The executor's IDs are positions in the permuted order, so this positional time ends with the complete answer in permuted positions on the GPUs. The library-ID endpoint then copies the positions to the host, maps them to library IDs with a host lookup and range checks, copies the IDs back, and synchronizes all GPUs. This host remap adds 210 to 355~s per run, 85.9\% of the library-ID endpoint for the Optimized law run and 77.5\% for the Stable law run. It is not part of the search and is excluded from the scan time. The host endpoint adds the copy of scores and IDs to the owning host. The two file-order runs of Optimized on the same server return library IDs on the GPU, need no remap, and take 57.3 and 57.4~s. The library-ID and host endpoints of the permuted runs include the host remap, and these file-order times do not. At the same capacity, the positional time of 58.931~s is 2.68 to 2.88\% above the file-order 57.281 and 57.394~s. It also contains the permuted runs' survivor counters and time-varying load from the runs sharing the server. The counters capture one maximum after each compacted tile and read four small counts per batch. Between the configurations, the measured components follow the larger seed and the smaller windows of $r_0=8$. Over the campaign in the law runs, seed replay takes 17.303 against 8.684~s and main replay 36.546 against 39.752~s. The Stable runs' shorter remaps were measured on the other half of the server.

\paragraph{A second permutation of all 500M rows.}
An independently generated permutation also completes the campaign at the law's capacity without overflow or recovery. The answer is in permuted positions after 59.2~s, and the complete answer with library IDs after 244.7~s. The run used four RTX PRO 6000 GPUs of the fourth server of Section~\ref{sec:setup} (AMD EPYC 9B14, driver 580.95.05) and a permutation of all 500M rows drawn with seed 202610170159. Optimized ran the whole campaign once (26,562
queries, $B=64$, $r_0=4$, $G=16$) at the law's capacity
$A=475{,}273$. The fullest window, 472,652, fills
99.45\% of $A$, against 472,427 on the first permutation. All
13.281B returned pairs passed the GPU check and an independent CPU
read-back on another machine. On 64 fixed queries the answers reach
99.6384\% mean and 99.2844\% minimum FP32 recall, and all 64 queries are
above 99\%. Truncated offline to 100K, the saved 500K answers reach
99.5330\% mean and 99.0640\% minimum recall, with no 100K serving time.

The positional output, measured as host wall time, takes 59.180~s: query copy 0.287, seed replay 8.673, seed merge and cutoff 4.487, cutoff broadcast 0.047, main replay 40.566, and final merge with the overflow read-back 5.070~s. The host remap to library IDs takes 185.560~s. It covers a device-to-host copy, a memory-mapped lookup in the 2~GB map from permuted positions to library IDs, conversion to int32, a host-to-device copy, and synchronization. The complete answer with library IDs is resident
after 244.740~s and on the owning host after 286.507~s. Library loading
(80.758~s) and setup with graph capture (2.277~s) lie outside these
intervals. The closest run on the first permutation, with seed 202610170100, is Optimized at the law's capacity in Table~\ref{tab:res-19}. It has the same $B$, GPU type and count, $A$, $r_0$, $G$, tiles, scorer configuration, and query bytes. It took 58.227~s positional and
413.721~s complete (488.576~s to the owning host). Against that run, the complete endpoint here is 40.8\% lower and the positional time 1.6\% higher, on a different server and driver with a different seed. The 61.04~s file-order campaign has no remap, and the complete endpoint here, which includes the remap, is about four times as long. 

\paragraph{Further permutations and placements, counts only.}
Fourteen further campaigns at the law's capacities overflow no window.
They ran on the fourth server (500M rows, 26,562 queries, $B=64$, $K=500$K, 416 batches each) and count survivors without timing. Four more row permutations, seeds 202610170160 to 202610170163, ran under Stable and Optimized. Table~\ref{tab:counts-cells} writes these seeds by their last three digits. The seed-159 permutation ran under three seed-tile placements with the fixed and with the local cutoff. The fullest window
holds 221,174 of Stable's 222,605 entries and 473,052 of Optimized's
475,273, and the fixed and local cutoffs give the same maximum under each
placement (Table~\ref{tab:counts-cells}). These runs share their queries,
so they describe the observed runs and are not independent trials of the
overflow probability.
\begin{table}[tb]
\caption{Fullest window over the campaign in the fourteen counts-only runs at the law's capacities, 222,605 for Stable and 475,273 for Optimized (500M rows, 26,562 queries, $B=64$, $K=500$K). Seeds are written by their last three digits. Seed 159 is the second permutation, run with Optimized under three seed-tile placements, where the fixed and local cutoffs give the same maximum. No window overflows.}
\label{tab:counts-cells}
\small
\begin{tabular}{@{}lrr@{}}
\toprule
Permutation, placement & Stable & Optimized\\
\midrule
Seed 160 &220,719&473,052\\
Seed 161 &220,992&472,650\\
Seed 162 &220,594&472,769\\
Seed 163 &221,174&472,613\\
Seed 159, default placement &&472,652\\
Seed 159, placement seed 71 &&472,927\\
Seed 159, placement seed 72 &&472,814\\
\bottomrule
\end{tabular}
\end{table}

\paragraph{Forced recovery of a 500M batch.}
Recovering a batch in which every query overflows on every GPU takes
2,826.5~ms of resident time. On the main server, a 500M-row control forces all 64
queries of one batch to overflow on each of the four GPUs, raising 256
query-by-GPU flags. The batch is recovered by releasing the window plan's
graphs and state, rescanning all 500M rows without pruning, and
recapturing the window graphs. The timing record attributes 0.114~s to
releasing state, 0.884~s to the unpruned recomputation, and 1.663~s to
recapture. All 32M returned pairs pass the score-bit check. A first version of this control could not recover: with the captured window state still allocated, the rescan requested 246~MiB while 152~MiB was free on GPU~0. The separate seed-71 campaign recovery
of Section~\ref{app:replication} already released its graph and query workspaces and its rejected outputs before rescanning. It also freed the delivered output before recapture. Its 106 recovered batches and
348.2~s of resident time therefore do not depend on the failed memory
handling of that first version.

\paragraph{Local cutoff under the unfavorable placement: all nine runs.}
The local cutoff removes every overflow of the unfavorable placement and
brings its time close to that of the default placement.
Table~\ref{tab:res-24} lists the nine instrumented campaigns behind the
paragraph of Section~\ref{app:replication}, three repetitions of each
of three settings. The settings are the seed-71 placement that overflows
under the fixed cutoff, the same placement with the local cutoff of
Section~\ref{app:grouped}, and the campaign's default placement with the
local cutoff. All use Optimized's $r_0=4$, $G=16$, $A=1{,}024{,}000$,
$B=64$, $K=500$K, four PRO 6000 GPUs, and the campaign's saved query
order. The resident and host endpoints include recovery (cleanup, the
unpruned rescan of the overflowing batch, and recapturing the CUDA
Graphs). The initial setup is excluded and listed separately. Every
overflowing batch logs two timed events, so 212 events are 106 recovery
cycles. The survivor counters run in all nine campaigns and add to these times, whereas the 348.2~s and 61.04~s of Sections~\ref{app:replication} and~\ref{sec:campaign} come from uninstrumented runs. All 119.529
billion retained pairs were reconstructed on the CPU against the
compressed reference, with exact score multisets and legal boundary ties. The score bits of the 15,992,299 pairs absent from the reference were recomputed on the GPU. This is a selection-preservation check, not a new FP32 recall
measurement.

\begin{table}[t]
\caption{Nine instrumented 500M-row campaigns ($B=64$, $K=500$K, four PRO 6000 GPUs, Optimized's parameters, saved query order). The settings are the seed-71 placement with the fixed and with the local cutoff, and the campaign's default placement with the local cutoff. Res., Host: campaign time to the GPUs and to the owning host, in s, recovery included. Setup: initial setup, excluded from both. Ovf.: overflowing batches (of 416). Max: fullest window of any query on any GPU. Clean, Rescan, Recap.: recovery cleanup, unpruned rescans, and graph recapture, in s.}
\label{tab:res-24}
\scriptsize
\setlength{\tabcolsep}{2.2pt}
\begin{tabular}{@{}lrrrrrrrrr@{}}
\toprule
Setting & Rep. & Res. & Host & Setup & Ovf. & Max & Clean & Rescan & Recap.\\
\midrule
seed 71, fixed & 1 & 390.047 & 415.028 & 2.24 & 106 & 1,219,067 & 14.0 & 100.1 & 196.7\\
seed 71, fixed & 2 & 390.052 & 415.206 & 1.81 & 106 & 1,219,067 & 14.6 & 100.2 & 196.4\\
seed 71, fixed & 3 & 388.559 & 412.250 & 1.81 & 106 & 1,219,067 & 14.5 & 99.8 & 196.2\\
seed 71, local & 1 & 78.195 & 104.588 & 1.77 & 0 & 982,679 & 0.0 & 0.0 & 0.0\\
seed 71, local & 2 & 79.757 & 107.326 & 1.75 & 0 & 982,679 & 0.0 & 0.0 & 0.0\\
seed 71, local & 3 & 79.647 & 105.337 & 1.75 & 0 & 982,679 & 0.0 & 0.0 & 0.0\\
campaign, local & 1 & 77.172 & 104.017 & 1.75 & 0 & 765,660 & 0.0 & 0.0 & 0.0\\
campaign, local & 2 & 78.637 & 103.428 & 1.76 & 0 & 765,660 & 0.0 & 0.0 & 0.0\\
campaign, local & 3 & 79.365 & 105.767 & 1.75 & 0 & 765,660 & 0.0 & 0.0 & 0.0\\
\bottomrule
\end{tabular}
\end{table}

\paragraph{Held-out check of the selection term.}
On one PRO 6000, the selection term of the cost model (Section~\ref{sec:cost-model}) predicts all 16 held-out timings within the 20\% tolerance fixed in advance. Its interpolation of $L(n,K)$ was fitted on 24 synthetic selection cases and frozen before validation. Eight held-out cases, each timed at two endpoints, give the 16 predictions, with a median absolute error of 6.37\% and a maximum of 7.81\%, an underprediction. The 64 saved outputs,
729.6M pairs, were checked on the CPU. The check covers the selection
primitive only. The filter-and-compact term $F(T)$, communication, and
plan-level predictions were not validated.

\paragraph{The selection term on an A100.}
On an A100, the selection term needs its own calibration. A model fitted to the A100's own training cases meets the 20\% criterion at every held-out endpoint, and the PRO 6000 model applied unchanged meets it at none. The check ran on one A100 of NCSA Delta  (6~min~18~s). The inputs are the same 32 saved
score banks, whose bits match the originals, with the same case order,
seeds, shapes, and timing protocol. The protocol takes three rounds of seven samples per endpoint and case and reports the CUDA-event median. All 160 saved input and output arrays passed hash checks. An independent CPU pass reread all 729.6M returned pairs, checking score bits, unique IDs within range, the exclusion of padding, and the tie-aware top-$K$ boundary. The
24 training cases cross $B$ of 16 and 64, $K$ of 100K and 500K, and
widths $W$ of 1M, 2M, and 4M scores per query, each with and without
padding (the last $\lfloor(W-K)/4\rfloor$ scores set to $-\infty$). The
eight held-out cases use $B=32$ and widths of 1.5M and 3M. For each
answer size, padding regime, and endpoint, the model is a monotone upper envelope of the A100's training medians. It interpolates bilinearly in $B$ and width without extrapolation, and its predictions were frozen before any held-out timing. Table~\ref{tab:res-20} gives the 16 held-out
predictions, CUDA-graph replay and eager execution for each case. All
fall within the 20\% criterion fixed in advance, with a median absolute
error of 0.5668\% and a maximum of 1.1094\%. The PRO 6000 predictions
applied unchanged meet the criterion at none of the 16 endpoints, with
errors up to 50.2099\%. Over all 64 endpoint timings the A100 takes 1.51
to 3.17 times as long as the PRO 6000 (median 1.76). Both the GPU and the software differ (Torch 2.13.0 with CUDA 13.0, Triton 3.7.1, and NumPy 1.26.4 on the A100). The experiment times a standalone selection and ID-gather primitive on the same synthetic score banks.

\begin{table}[t]
\caption{Held-out selection cases on one A100, $B=32$: CUDA-event
median (Meas.), the prediction from the A100's training cases (A100)
and the PRO 6000 prediction applied unchanged (PRO), in ms, each with
its signed error (prediction $-$ measured) / measured, in \%. $W$:
scores per query. Pad: last $\lfloor(W-K)/4\rfloor$ scores are
$-\infty$. G: graph replay. E: eager.}
\label{tab:res-20}
\footnotesize
\setlength{\tabcolsep}{4pt}
\begin{tabular}{@{}llllrrrrr@{}}
\toprule
$K$ & $W$ & Pad & Mode & Meas. & A100 & Err. & PRO & Err.\\
\midrule
100K & 1.5M & no & G & 1.127 & 1.130 & $+$0.27 & 0.585 & $-$48.11\\
100K & 1.5M & no & E & 1.170 & 1.172 & $+$0.17 & 0.588 & $-$49.73\\
100K & 1.5M & yes & G & 1.089 & 1.094 & $+$0.45 & 0.558 & $-$48.77\\
100K & 1.5M & yes & E & 1.131 & 1.137 & $+$0.55 & 0.563 & $-$50.21\\
100K & 3M & no & G & 1.998 & 1.980 & $-$0.91 & 1.100 & $-$44.94\\
100K & 3M & no & E & 2.040 & 2.024 & $-$0.80 & 1.103 & $-$45.96\\
100K & 3M & yes & G & 1.922 & 1.932 & $+$0.49 & 1.066 & $-$44.54\\
100K & 3M & yes & E & 1.964 & 1.976 & $+$0.64 & 1.069 & $-$45.56\\
500K & 1.5M & no & G & 1.280 & 1.290 & $+$0.79 & 0.750 & $-$41.43\\
500K & 1.5M & no & E & 1.323 & 1.332 & $+$0.74 & 0.757 & $-$42.75\\
500K & 1.5M & yes & G & 1.248 & 1.262 & $+$1.11 & 0.729 & $-$41.61\\
500K & 1.5M & yes & E & 1.293 & 1.306 & $+$1.00 & 0.736 & $-$43.07\\
500K & 3M & no & G & 2.250 & 2.245 & $-$0.26 & 1.329 & $-$40.95\\
500K & 3M & no & E & 2.292 & 2.287 & $-$0.25 & 1.339 & $-$41.58\\
500K & 3M & yes & G & 2.172 & 2.159 & $-$0.58 & 1.279 & $-$41.11\\
500K & 3M & yes & E & 2.212 & 2.203 & $-$0.42 & 1.288 & $-$41.77\\
\bottomrule
\end{tabular}
\end{table}

\paragraph{The selection term on a GH200.}
On a GH200, unlike on the A100, the PRO 6000 model also transfers
unchanged. The A100 check was repeated on one GH200  with the same 32 saved
score banks, case order, seeds, shapes, and timing protocol. All 160
saved arrays passed hash checks, and an independent CPU pass reread all
729.6M returned pairs. A model built in the same way from the GH200's 24
training medians predicts all 16 held-out endpoints within the 20\%
criterion fixed in advance (median absolute error 0.96\%, maximum
2.34\%). The PRO 6000 predictions applied unchanged
also meet it at all 16 endpoints (median 6.28\%, maximum 13.61\%). Over
the 64 endpoint timings the GH200 takes 0.80 to 1.66 times as long as
the PRO 6000 (median 0.98). Hardware and software both differ. The check times synthetic selections.

\FloatBarrier

\section{Campaign, Batching, and Delivery}\label{app:campaign-delivery}
This appendix holds the measurements behind two conclusions of Sections~\ref{sec:campaign} and~\ref{sec:batch-reuse-controls}. The window plan serves the 500M-row campaign at 99.63\% mean recall, and it stays faster than the grouped and per-tile plans within the same memory on four and eight GPUs. The appendix first completes the campaign with the per-GPU memory, DrugCLIP's public script against matched SweepANN runs, and two checks of the returned answers. The controls follow: the top one million rows per query, and grouped spans and the selection primitive at 500M rows. The appendix closes with the eight-GPU runs and their host delivery, pockets of further proteins, and the larger answers on eight GPUs.

\paragraph{Per-GPU memory.}
GPU~0 reaches its peak workspace in the final merge. Under Optimized
at $B=64$, the four named terms of $W_{\mathrm{query}}$
(Section~\ref{sec:seed-capacity}) take 1.04~GB of its 2.57-GB peak
workspace, so $W_{\mathrm{other}}\approx 1.54$~GB on both platforms.
Counted from the code, the final merge holds 0.97~GB of retained state
(no score tile) and 1.60~GB of merge tensors, 2.572387~GB in all,
within 4~kB of the peak. GPU~0 peaks 62~MB above the other GPUs. In the
timed PRO 6000 runs, its allocator holds 4.26~GB beyond the 96.75-GB
shard, another 0.81~GB is in use outside the allocator, and 158~MiB is
free (306~MiB on GH200).

\paragraph{DrugCLIP's public script: batch sweep.}
Among the tested batch sizes, the released DrugCLIP screening script~\cite{gao2023drugclip,drugclipCode} is fastest at 32 queries (Table~\ref{tab:res-18a}). At commit 1097a07, where \texttt{screening\_chunk.py} matches the cited revision, it ran unchanged on the
four PRO 6000 GPUs of the main server. It scanned the first 25M rows of
768-dimensional FP32 vectors for the 1,000 original pocket queries at
$K=100$K. Only its command-line batch size was varied, with one run per
batch size. Each run is timed from process invocation to the observed
exit, which includes the script's input, normalization, and output.
Each exit was observed within 1.08~s of the last check that the process
was alive. Recall is measured on eight queries chosen in advance,
against the exact FP32 top-100K over all 25M rows. For each run, an
independent CPU reader checked all 100M returned pairs and 2.5B
structural pairs.

\begin{table}[t]
\caption{DrugCLIP's public script, run unchanged, on 25M rows (four
PRO 6000 GPUs of the main server, 1,000 queries, $K=100$K): one run per
batch size. Time: from process invocation to the observed exit.
Recall: eight preselected queries against the exact FP32 top-100K.}
\label{tab:res-18a}
\small
\begin{tabular}{@{}rrrr@{}}
\toprule
Batch & Time (s) & Mean recall (\%) & Min.\ (\%)\\
\midrule
4 & 187.122434 & 99.999875 & 99.999\\
32 & 172.435461 & 99.999750 & 99.999\\
64 & 199.176245 & 99.999750 & 99.999\\
\bottomrule
\end{tabular}
\end{table}

\begin{table}[t]
\caption{SweepANN on the rows, queries, and GPUs of
Table~\ref{tab:res-18a} ($B=4$, $r_0=1$, $G=2$): three passes of 250
batches after one setup, in s. The first pass is delivered 7.3614~s
after main entry. Pass: whole pass, including writing the answers
(Write). Resident, Host: sums of the per-batch intervals to the GPUs
and to the owning host. Recall: mean over the eight queries of
Table~\ref{tab:res-18a}, minimum 99.524\% in every pass.}
\label{tab:res-18b}
\small
\begin{tabular}{@{}rrrrrr@{}}
\toprule
Pass & Pass (s) & Write (s) & Resident (s) & Host (s) & Recall (\%)\\
\midrule
1 & 1.6206 & 0.1172 & 1.1958 & 1.2430 & 99.672875\\
2 & 1.6193 & 0.1172 & 1.1957 & 1.2371 & 99.672500\\
3 & 1.6281 & 0.1185 & 1.1967 & 1.2387 & 99.672750\\
\bottomrule
\end{tabular}
\end{table}

\paragraph{Matched SweepANN runs.}
On the same rows, queries, and GPUs, SweepANN made three passes over all 1,000 queries after one setup (Table~\ref{tab:res-18b}). It scanned its compressed INT8 copy of the rows at the script's default batch size, $B=4$. The other settings are $r_0=1$, $G=2$, $A=1{,}024{,}000$, and 2M-row tiles, with four 6.25M-row shards that each end in a 250K-row tail. Its first pass is delivered 7.3614~s after
main entry. That span includes loading the INT8 rows quantized
beforehand, initialization, graph capture, and warm-up, and it excludes
interpreter start-up and the quantization. An independent CPU process checked all 400M returned and reference pairs
of SweepANN.

The same configuration also ran at $B=32$ and $B=64$, three whole
passes of 1,000 queries each after one setup, with the same eight
quality queries and FP32 reference (Table~\ref{tab:res-18c}). The
$B=64$ pass pads the last batch (40 real queries), and $B=32$ pads to
8. On the eight queries, SweepANN's minimum recall is 99.524\% in every
pass at every batch size, and every checked query of both systems
exceeds 99\%. Recall of both systems was measured on these eight
queries, not on all 1,000. The script's times run from process
invocation to the observed exit, 172.435~s at its fastest tested batch
of 32, while SweepANN's columns are first-delivery, per-pass, and
resident intervals. 

\begin{table}[t]
\caption{SweepANN on the 25M rows, 1,000 queries, and four GPUs of
Table~\ref{tab:res-18a} at three batch sizes: median of three passes,
in s. First: from main entry to the first persisted delivery, including
loading the INT8 rows, initialization, capture, and warm-up, and
excluding interpreter start-up and the prior quantization. Pass: whole
pass, including writing the answers. Resident: sum of the per-batch
intervals to the GPUs. Recall: mean over the eight quality queries,
range of the three passes.}
\label{tab:res-18c}
\footnotesize
\begin{tabular}{@{}rrrrl@{}}
\toprule
$B$ & First (s) & Pass (s) & Resident (s) & Recall (\%)\\
\midrule
4 & 7.361440 & 1.620594 & 1.195791 & 99.672500--99.672875\\
32 & 5.855880 & 0.386205 & 0.254053 & 99.672500--99.672750\\
64 & 6.138808 & 0.327835 & 0.218450 & 99.672625--99.672750\\
\bottomrule
\end{tabular}
\end{table}

\paragraph{Selection check over every query.}
For every one of the 26,562 queries, the window plan returns the score
multiset of an unpruned reference plan. The re-run of
Section~\ref{sec:quality} runs the Optimized window plan ($r_0=4$,
$G=16$, $A=1{,}024{,}000$) and the unpruned reference plan over all
queries. The scoring kernel is fixed to one configuration (block sizes
32, 64, and 256 and eight warps) for both plans and all tails.
Each of the 13,281,000,000 returned pairs had its FP16 score bits
recomputed from the complete unpruned tile score matrices. A separate CPU
process reopened the 26,562,000,000 pairs returned by the two plans. It
checked file hashes, finite scores, valid distinct IDs, equal score
multisets, and the rank of the $K$-th score from whole-library counts
of the scores above and equal to it. The number of library rows that
share a query's $K$-th score is 665 at minimum, 5,215 at the median,
and 27,757 at most. The number of returned rows tied at the $K$-th
score is 1, 2,385, and 20,979 at the same three points. No batch
overflowed. A 40,000,012-row oracle with the same fixed kernel
configuration checks normal execution, a forced overflow that raises
256 flags, recovery, and recapture and replay. The score arrays saved
by the original campaign have the same score multisets for all 26,562
queries. Their IDs were not all recomputed. The re-run adds no
timing and no FP32 recall measurement.

\paragraph{TF32-disabled reference.}
A reference recomputed with TF32 disabled leaves the campaign's mean
recall on the 64 held-out queries of the 500M controls (query IDs
256--319) at 99.638\%. We recomputed the exact top-$K$ over all 500M deployed rows with FP32 queries and vectors. TF32 was disabled explicitly in the matrix multiplication, in cuDNN, and by the driver override. Every returned pair was gathered from an unpruned FP32 score
matrix and checked by counting. A CPU process reopened all 8.064
billion saved per-GPU pairs and rebuilt the 32M final reference pairs,
keeping every row tied at the boundary with row-ID tie ordering. The
new lists overlap the original reference lists by 99.9997\% on average
(minimum 99.999\%). Under either reference, the campaign's per-query
recall on these queries lies between 99.29\% and 99.80\%. Per-query
recall changes by $-1.25\times10^{-7}$ on average and by at most
$4\times10^{-6}$ in either direction, two IDs of $K=500{,}000$. For
rows in both lists, the per-query mean absolute score difference lies
between $5.0\times10^{-8}$ and $1.8\times10^{-7}$, and the per-query
maximum between $4.2\times10^{-7}$ and $1.4\times10^{-6}$. The code of the original 621M reference uses FP32 tensors but does not disable TF32, and its setting was not recorded.

\paragraph{Answers of one million on 500M rows.}
With a window capacity of $A=1{,}920{,}000$, SweepANN returns the top
$K=1$M of all 500M rows without overflow, and every answer contains
the strict FP32 top-500K (Table~\ref{tab:res-22}). The runs serve the 64 fixed held-out queries (IDs 256--319) as two batches of $B=32$ on the four PRO 6000 GPUs of the main server. They use four 125M-row shards, $r_0=4$, $G=16$, and 2M-row tiles, with one measured pass per batch and method. The second capacity, $A=937{,}500$, is the random-order mean
$\lceil 30\text{M}\cdot 1\text{M}/(32\text{M}+1)\rceil$ itself, with no
margin for overflow risk or for file order. With it both batches
overflowed, and each took 2.4~s with recovery and graph recapture
included. The flags in the table count the
GPU and query pairs of a batch in which any window overflowed, not
windows or distinct queries. The full-score method at $B=32$ failed with CUDA out
of memory: allocating a $32\times125$M FP16 score matrix requested
7.45~GiB with 4.11~GiB free on GPU~0.

The resident endpoint is the host wall time from the call with the
queries on the host to the complete answer on the GPUs after all
devices synchronize. It covers the copy to the GPUs, the seed graphs, the seed merge and cutoff exchange, the main graphs, the final merge, the overflow readback, and any recovery (the failed attempt, cleanup, full rescan, and recapture). The host endpoint adds the copy of scores
and IDs into NumPy arrays and ends before they are written to files.
Both exclude loading the rows (214.655~s), the setup (1.35 and
1.36~s), the warm-up, and the checks. A GPU check covered every
physical tile, including each shard's final 1M-row tail, and all 192M
retained pairs. An independent CPU process reread and hashed all 12
primary array files and checked every returned pair. Three methods completed: the eager unpruned reference and the window plan at both capacities. For each of the 64 queries, every returned 1M answer of these methods contains all 500K IDs of the strict FP32 top-500K computed with TF32 disabled (mean and minimum 100\%). This measures containment of the existing
reference, not Recall@1M. The eager reference was not timed, and the
measurement covers these 64 queries, not the 26,562.

\begin{table}[t]
\caption{Answers of $K=1$M over 500M rows (four PRO 6000 GPUs of the
main server, 64 held-out queries 256--319 in two batches of $B=32$, one
pass per batch). Res., Host: resident and host endpoints in ms,
including recovery and recapture. Flags: per batch, the number of GPU
and query pairs in which any window overflowed (at most 128).
Containment of the FP32 top-500K: 100\% for every query of the three
completed methods.}
\label{tab:res-22}
\footnotesize
\setlength{\tabcolsep}{4pt}
\begin{tabular}{@{}lrrrrrl@{}}
\toprule
Method & $A$ & Batch & Res.\ (ms) & Host (ms) & Flags & Recovery\\
\midrule
Window & 937,500 & 0 & 2,396.88 & 2,457.01 & 66 & yes\\
Window & 937,500 & 1 & 2,384.55 & 2,451.00 & 57 & yes\\
Window & 1,920,000 & 0 & 124.74 & 184.16 & 0 & no\\
Window & 1,920,000 & 1 & 125.24 & 181.34 & 0 & no\\
Eager unpruned & -- & 0, 1 & \multicolumn{4}{l}{completed, not timed}\\
Full score & -- & -- & \multicolumn{4}{l}{CUDA out of memory}\\
\bottomrule
\end{tabular}
\end{table}

\paragraph{Main server: calibration rows and caps.}
On the held-out queries of the main server, the window plan is
$1.413\times$ faster to resident output than the three-tile span, the
widest grouped span within its caps, and $1.308\times$ faster to the
host. It is $2.241\times$ faster than the per-tile plan to resident
output.
Table~\ref{tab:res-07} gives every run of this 500M-row control of
Section~\ref{app:grouped}, calibration rows included. Spans of one to
four tiles were declared in advance. On the calibration queries, one
to three tiles stay within the caps, and three, the fastest of them,
was chosen before any held-out run. Four tiles exceed the reserved cap
by 62,914,560 bytes on GPUs 1--3 and were not run on the held-out
queries. The caps are 2.57~GB allocated and 4.26~GB reserved on GPU~0,
and 2.51~GB allocated with 2.66, 3.69, and 2.66~GB reserved on GPUs 1,
2, and 3. All 928M returned pairs were rescored from the compressed
rows with the production arithmetic and checked by an independent CPU
process. Normal runs did not overflow. Loading the library took
91.4~s in this run with a warm disk cache and 371.4~s in the first
attempt, and the serving times exclude it.

\begin{table}[t]
\caption{Grouped spans against the window plan on 500M rows (four
PRO 6000 GPUs of the main server, $B=64$, $K=500$K): median (range) of
three repetitions in ms per batch, to device output (Resident) and to
the host. $S$: span of $S$ tiles, $S=1$ is the per-tile plan. Caps:
within the window plan's allocated and reserved peaks on every GPU.
Cal.: calibration queries 128--191. HO: held-out queries 256--319.}
\label{tab:res-07}
\small
\begin{tabular}{@{}llcrr@{}}
\toprule
Set & Plan & Caps & Resident & Host\\
\midrule
Cal. &Window&yes&148.2 (146.6--150.5)&199.5 (198.3--200.2)\\
Cal. &$S=1$&yes&330.2 (329.0--352.3)&383.3 (380.4--406.2)\\
Cal. &$S=2$&yes&239.0 (235.8--240.3)&293.6 (292.9--293.7)\\
Cal. &$S=3$&yes&209.9 (208.1--209.9)&264.9 (263.2--266.6)\\
Cal. &$S=4$&no&194.8 (194.1--195.2)&247.0 (247.0--251.6)\\
\midrule
HO &Window&yes&146.2 (145.1--147.7)&200.2 (199.2--200.6)\\
HO &$S=1$&yes&327.6 (326.4--329.5)&382.0 (380.2--382.8)\\
HO &$S=3$&yes&206.6 (206.1--209.9)&261.9 (258.7--262.1)\\
\bottomrule
\end{tabular}
\end{table}

\paragraph{Eight-GPU server, four GPUs: all spans and caps.}
On four GPUs of the eight-GPU server, the window plan is $1.483\times$
faster than the three-tile span and $1.368\times$ faster than the
four-tile span at the resident endpoint on the held-out queries.
Table~\ref{tab:res-06a} lists every plan measured in this control of
Section~\ref{app:grouped}: four PRO 6000 GPUs, 500M rows, $B=64$,
$K=500$K, 2M-row tiles, CPU cores 0--7, and three repetitions. Resident
time ends with the complete answer on GPU~0, and host time adds the
copy to the host. The caps are the window plan's peak allocated
workspace and peak reserved memory above the library on each GPU of
this server (Table~\ref{tab:res-06b}). The original grid declared spans
of 1, 2, 3, 4, 6, 8, 12, 16, and 32 tiles and the full score matrix,
with calibration queries 128--191 and held-out queries 256--319. The
six-tile span failed allocation during setup, so the larger spans were
not run. Five tiles were not in this grid. Calibration chose three
tiles under the caps, four under physical memory, and the window plan
under every rule. A separately declared query set, calibration queries
512--575 and held-out queries 576--639, then measured four and five
tiles and the window plan. Calibration chose five tiles under physical
memory, and the window plan is $1.314\times$ faster on its held-out
queries. All 928M and 640M returned pairs of the two query sets were
rescored from the compressed rows on the GPU and checked by CPU
processes on this server and on the main server. No held-out run
overflowed.

\begin{table}[t]
\caption{Every plan measured on four GPUs of the eight-GPU server
(500M rows, $B=64$, $K=500$K), ms per 64-query batch. First query set:
calibration 128--191, held-out 256--319. Second: calibration 512--575,
held-out 576--639. Caps: within the window plan's allocated and
reserved peaks on every GPU. Cal.: calibration median (host time).
HO: held-out median (range). HO host: held-out host median. Spans not
chosen on calibration have no held-out run.}
\label{tab:res-06a}
\footnotesize
\begin{tabular}{@{}llcrrr@{}}
\toprule
Query set & Plan & Caps & Cal. & HO & HO host\\
\midrule
First &Window&yes&140.7 (230.2)&140.5 (138.4--140.8)&245.9\\
First &$S=1$&yes&330.2 (420.3)&330.7 (329.9--330.9)&435.9\\
First &$S=2$&yes&239.3 (329.6)&--&--\\
First &$S=3$&yes&208.0 (300.3)&208.4 (205.5--208.8)&313.7\\
First &$S=4$&no&192.2 (284.4)&192.3 (190.6--192.6)&297.4\\
First &$S=6$&--&\multicolumn{3}{l}{allocation failed during setup}\\
\midrule
Second &Window&yes&141.1 (244.2)&140.7 (139.0--141.3)&245.9\\
Second &$S=4$&no&194.7 (296.4)&194.2 (192.3--194.3)&299.0\\
Second &$S=5$&no&186.1 (281.8)&184.9 (182.1--185.5)&284.3\\
\bottomrule
\end{tabular}
\end{table}

\begin{table}[t]
\caption{Peak workspace of each plan on four GPUs of the eight-GPU
server, GB: allocated above the library (equal on all GPUs for the
grouped plans) and reserved above the library on GPUs 0, 1, and 2
(GPU 3 equals GPU 1). The window plan's values are the caps of
Table~\ref{tab:res-06a}.}
\label{tab:res-06b}
\small
\begin{tabular}{@{}lrrrr@{}}
\toprule
Plan & Allocated & Resv.\ GPU 0 & Resv.\ GPU 1 & Resv.\ GPU 2\\
\midrule
Window &2.57 / 2.51\textsuperscript{a}&4.25&2.66&3.69\\
$S=1$ &1.73&3.23&1.95&2.97\\
$S=2$ &1.98&3.48&2.20&3.23\\
$S=3$ &2.24&3.73&2.45&3.48\\
$S=4$ &2.50&4.00&2.72&3.75\\
$S=5$ &2.75&4.25&2.97&4.00\\
\bottomrule
\multicolumn{5}{@{}l}{\textsuperscript{a}GPU 0 / GPUs 1--3.}
\end{tabular}
\end{table}

\paragraph{Full-campaign grouped controls on the eight-GPU server.}
Over complete campaigns, the three-tile span takes $1.49\times$ and $1.23\times$ as long as the window plan to the GPUs and to the owning host. The per-tile plan takes $2.36\times$ and $1.95\times$ as long (Table~\ref{tab:res-26}). These campaigns ran on four PRO 6000 GPUs
of the eight-GPU server ($B=64$, $K=500$K, all 26,562 queries in file
order, one pass each). The per-tile plan and the three-tile span of
Section~\ref{app:grouped} kept their frozen parameters and ran under
the window plan's two frozen per-device memory caps. Both plans satisfy
both caps without overflow. Each campaign returns 13.281B pairs. The two grouped outputs were checked by byte-for-byte identity of all 832 candidate files, with each other and with a shared reference output verified on the main server. This check needed no second full read. The window campaign was
verified separately. It differs from that reference in 2,731,239
boundary pairs, and its strict recall differs on 52 of the 64 queries
only beyond the displayed precision. Against the TF32-off FP32
reference, available for the 64 held-out queries 256--319, all three
campaigns serve 99.638\% mean and 99.285\% minimum recall at
$K=500$K. Cut to their top 100K, the saved answers reach 99.533\% and
99.064\%, and no campaign was timed at $K=100$K. These campaigns make
no new parameter selection.

A second complete window-plan pass on the same GPUs passed the same
checks, with all 13.281B pairs read on the main server and 2,731,230
boundary pairs differing from the shared reference. It took 57.394~s
resident and 127.996~s to the owning host, with the same served recall
and the same top-100K recall of the saved answers at the displayed
precision. The resident times of the two passes differ by 0.20\% and
the host times by 14.28\%. The ratios in Table~\ref{tab:res-26} use the first pass.

\begin{table}[t]
\caption{Three complete 500M-row campaigns on four PRO 6000 GPUs of the
eight-GPU server ($B=64$, $K=500$K, 26,562 queries, one pass each).
Res., Host: time to the GPUs and to the owning host, in s. Res.\
$\times$, Host $\times$: ratio to the window plan. Ovf.: overflowing
batches. Mean, Min.: recall on the 64 held-out queries against the
TF32-off FP32 reference at $K=500$K.}
\label{tab:res-26}
\footnotesize
\setlength{\tabcolsep}{3pt}
\begin{tabular}{@{}lrrrrrrr@{}}
\toprule
Plan & Res.\ (s) & Host (s) & Res.\ $\times$ & Host $\times$ & Ovf. & Mean & Min.\\
\midrule
Window & 57.281 & 111.997 & 1.00 & 1.00 & 0 & 99.638\% & 99.285\%\\
Three-tile span & 85.193 & 137.743 & 1.49 & 1.23 & 0 & 99.638\% & 99.285\%\\
Per-tile plan & 135.358 & 218.826 & 2.36 & 1.95 & 0 & 99.638\% & 99.285\%\\
\bottomrule
\end{tabular}
\end{table}

\paragraph{Selection primitive.}
With cuVS selection in place of PyTorch's top\nobreakdash-$K$ operator,
the per-tile and window plans become slower and the grouped plans faster. The window plan's workspace then exceeds its caps. A first
control uses the 100M-row setting of Section~\ref{app:grouped} (one
GPU, $B=64$, $K=500$K). It replaces every PyTorch top\nobreakdash-$K$
call in the per-tile, 7M-row grouped, and window plans with cuVS
selection, keeping their calling convention (32-bit positions become
64-bit for the gathers). Of two cuVS algorithms, radix selection with
an extra pass was faster on the calibration queries for all plans. On
the held-out queries, the per-tile plan takes 410.3~ms of resident
time per batch and the window plan 147.3~ms, against 391.3 and
143.4~ms with PyTorch's operator in the same run. The grouped plan
becomes faster, 134.3~ms instead of 145.9~ms, but allocates 2.43~GB, more than the budget. The window plan with cuVS allocates 2.51~GB.
Every repetition of every plan returns the per-tile plan's score
multiset. A selector fused into the plans was not measured.

A second control repeats the comparison on all 500M rows and the four
PRO 6000 GPUs of the main server ($B=64$, $K=500$K) for the window plan
and the three-tile grouped plan of Section~\ref{sec:batch-reuse-controls}.
Each plan ran five timed repetitions on the 64 held-out queries with
PyTorch's top\nobreakdash-$K$ and with cuVS radix selection (11-bit,
extra pass). One scorer configuration, frozen on the calibration
queries, serves all four cases. It differs from that of the earlier
run of Section~\ref{sec:batch-reuse-controls}, so the two runs are not
pooled. In Table~\ref{tab:res-34}, cuVS speeds the grouped plan by
$1.06\times$ and slows the window plan. The caps are the PyTorch
window plan's workspace peaks, and with cuVS the window plan exceeds
them (2.64 against 2.57~GB allocated, 4.32 against 4.26~GB
reserved). Every output passed the GPU and CPU pair checks, and
every query of every repetition reaches at least 99.28\% recall against
the FP32 reference.

\begin{table}[t]
\caption{PyTorch's top\nobreakdash-$K$ against cuVS radix selection on
500M rows (four PRO 6000 GPUs of the main server, $B=64$, $K=500$K)
in the window plan and the three-tile grouped plan. Resident, Host:
median of five repetitions per 64-query batch, in ms. Alloc.,
Reserv.: workspace peaks per GPU, in GB. Mean, Min.: FP32 recall (\%)
of the 64 held-out queries.}
\label{tab:res-34}
\footnotesize
\setlength{\tabcolsep}{3pt}
\begin{tabular}{@{}llrrrrrr@{}}
\toprule
Selection & Plan & Resident & Host & Alloc. & Reserv. & Mean & Min.\\
\midrule
PyTorch & Window & 144.8 & 197.7 & 2.57 & 4.26 & 99.64 & 99.29\\
PyTorch & Grouped & 210.7 & 263.9 & 2.24 & 3.73 & 99.64 & 99.29\\
cuVS & Window & 150.4 & 201.0 & 2.64 & 4.32 & 99.64 & 99.29\\
cuVS & Grouped & 198.0 & 246.8 & 2.30 & 3.75 & 99.64 & 99.28\\
\bottomrule
\end{tabular}
\end{table}

\paragraph{Eight-GPU calibration and held-out runs.}
On eight GPUs, the window plan beats the fastest grouped span within
its per-GPU memory caps, three tiles, at every batch size
(Table~\ref{tab:res-23a}). The eight PRO 6000 GPUs of the eight-GPU server hold the 500M rows as eight 62.5M-row shards (2M-row tiles, per-shard $r_0=4$, $G=16$, $A=1{,}024{,}000$). The seed sample thus has 64M rows against 32M on four GPUs. Calibration ran every plan, the
window plan, grouped spans of 1 to 16 tiles, and the full-score plan,
at $B=64$, 128, and 256 on calibration batches, with three timing
repetitions each. It froze one plan per memory budget before any
held-out batch. The caps cover allocated and reserved memory and are
checked on each GPU separately, so a plan whose peak is lower overall
can still exceed the cap of one GPU. Outside the caps, the fastest
plan is a 15-tile span at $B=64$ and the full-score plan at $B=128$
and 256. Table~\ref{tab:res-23a} gives the held-out medians at both
endpoints.

The campaign plan, the window plan at $B=256$, was frozen on resident
throughput before the held-out runs. It is not the best batch for the
host endpoint: at $B=64$ the window plan delivers about 237 queries per
second to the host against 208 at $B=256$. Table~\ref{tab:res-23b}
compares four and eight GPUs on the same server and the same 64
held-out queries (256--319) at $B=64$, with the same fixed scoring
kernel on every GPU. Eight GPUs take 103.6~ms per batch resident and
270.3~ms to the host, $1.36\times$ faster resident than four and
$0.91\times$ as fast to the host. The two configurations also differ in their sample (32M against 64M rows), and co-located background work on the host was not excluded.

\paragraph{The eight-GPU campaign.}
The window plan ran at $B=256$, frozen on the calibration batches. One pass over all 26,562 queries at $K=500$K on the eight GPUs took 40.288~s resident (659 actual queries per second). The answers reached the owning host after 132.296~s (201 per second). It ran 104 batches, the last with
194 queries and 62 padding slots, after one graph capture, with no
overflow and both per-device cap rules satisfied. All 13.281B returned
pairs passed the GPU check against the reference and independent CPU
checks on the eight-GPU server and on the main server. Of these pairs,
13,274,766,711 match the verified reference, and the other 6,233,289
boundary pairs, which the reference does not hold, were rescored on
the GPU. The checks verify selection
under the compressed scores, not raw FP32 scores. Against the TF32-off
FP32 reference, available only for the 64 held-out queries 256--319,
the served answers reach 99.638\% mean and 99.284\% minimum recall at
$K=500$K. Cut to their top 100K, the saved answers reach the same
99.533\% and 99.064\% as the four-GPU campaigns.

Against the four-GPU $B=64$ campaign on the same server (Table~\ref{tab:res-26}), eight GPUs at their best resident batch are faster resident and slower to the host. The two campaigns also differ in batch size, and the matched $B=64$ held-out run of Table~\ref{tab:res-23b} isolates the GPU count.

A full-score campaign at $B=256$,
outside both frozen window caps, ran one verified pass of 104 batches
with zero overflow. It took 39.493~s resident and 121.635~s to the
owning host and served 99.638\% mean and 99.285\% minimum recall on the
64 held-out queries. Its peak Torch allocation is 86.26~GB and its
reserved peak 91.65~GB per device, against the window campaign's 58.66
and 65.60~GB, with raw rows of 48.375~GB per device.
Section~\ref{app:grouped} compares the two campaigns.

\begin{table}[t]
\caption{Plans on eight PRO 6000 GPUs of the eight-GPU server (500M
rows, $K=500$K, held-out batches): median of three repetitions per
batch, in ms, and the implied queries per second. Held-out queries
differ by $B$: 256--319, 256--383, and 256--511. Grouped $S$: span of
$S$ tiles. Full: the full-score plan. Caps: inside or outside the
window plan's per-device memory caps.}
\label{tab:res-23a}
\footnotesize
\setlength{\tabcolsep}{3pt}
\begin{tabular}{@{}rlrrrrl@{}}
\toprule
$B$ & Plan & Res.\ (ms) & Host (ms) & Res.\ q/s & Host q/s & Caps\\
\midrule
64 & window & 103.610 & 270.257 & 617.7 & 236.8 & inside\\
64 & grouped 3 & 119.954 & 296.062 & 533.5 & 216.2 & inside\\
64 & grouped 15 & 95.814 & 269.232 & 668.0 & 237.7 & outside\\
128 & window & 199.499 & 636.911 & 641.6 & 201.0 & inside\\
128 & grouped 3 & 231.224 & 664.698 & 553.6 & 192.6 & inside\\
128 & full & 193.164 & 628.495 & 662.7 & 203.7 & outside\\
256 & window & 389.018 & 1231.590 & 658.1 & 207.9 & inside\\
256 & grouped 3 & 448.507 & 1288.595 & 570.8 & 198.7 & inside\\
256 & full & 380.899 & 1217.040 & 672.1 & 210.3 & outside\\
\bottomrule
\end{tabular}
\end{table}

\begin{table}[t]
\caption{Window plan at $B=64$ on the eight-GPU server, held-out
queries 256--319, three repetitions: four GPUs (125M rows per shard,
32M-row sample) against eight (62.5M rows per shard, 64M-row
sample).}
\label{tab:res-23b}
\footnotesize
\begin{tabular}{@{}rrrrr@{}}
\toprule
GPUs & Res.\ (ms) & Host (ms) & Res.\ q/s & Host q/s\\
\midrule
4 & 140.501 & 245.927 & 455.5 & 260.2\\
8 & 103.610 & 270.257 & 617.7 & 236.8\\
\bottomrule
\end{tabular}
\end{table}

\paragraph{Host-delivery gap of the eight-GPU campaign.}
The 92.007~s between the two endpoints of the window campaign above is 69.55\% of the host endpoint. It is exactly the interval holding the two sequential device-to-host output conversions after the resident timestamp. Writing the NPY files, hashing, and the numerical check
come after the host timestamp and lie outside both endpoints. The resident subsegments sum to 40.273~s, leaving 0.015~s unnamed. They are query copy and dispatch 0.198~s, seed replay and barrier 8.318~s, seed merge and cutoff read 7.013~s, cutoff broadcast 0.024~s, main replay and barrier 17.107~s, and final merge and overflow read 7.613~s. Across the 104 batches, the delivery-only interval has a
median of 939.3~ms (range 692.3 to 1,128.0~ms). It moves
768,000,000~B per full batch and 582,000,000~B for the 194-query tail,
79,686,000,000~B in total, at 0.866~GB/s. That rate includes host
allocation, first touch, CUDA delivery, and synchronization, so it is
not a DMA bandwidth.

An earlier diagnostic on the same GPU family at $B=64$ checked every
byte and compared five delivery paths for 192~MB at $K=500$K and
384~MB at $K=1$M. Their median times are 86.840 and 179.940~ms for the
original conversion path, 13.749 and 27.199~ms for a reused pageable
destination, and 3.483 and 6.823~ms for a fresh pinned destination. A
reused pinned destination takes 3.439 and 6.787~ms, and a reused
pinned destination followed by an independent NumPy copy 46.397 and
126.308~ms. This last path is a preallocated pinned buffer followed by a fresh CPU copy that owns its memory. It is faster than the original conversion path at both sizes and leaves selection and every returned byte unchanged. The campaign above used the original conversion path,
and no campaign was run with the faster one.

\paragraph{Pockets of further proteins.}
On the 256 further pockets of Section~\ref{app:grouped}, both campaigns keep every query above 99\% recall (Table~\ref{tab:res-28}). The pockets were drawn by a rule fixed before any result of the new queries was read. A protein is
the UniProt accession of a pocket's AlphaFold model, and the 26,562
stored pockets come from 9,919 proteins. The rule excludes 1,436
proteins in all. The first group, 440 proteins, is every protein with a pocket among the first 512 stored pockets. These pockets hold the union of all calibration and held-out queries of the eight-GPU runs at $B=64$, 128, and 256. The others are the 128 proteins of an earlier tuning
and confirmation query set, the 473 proteins that set had itself
excluded, and the proteins of the 512 queries of
Section~\ref{app:gpu-variants}. From the 8,483 remaining proteins, it
draws 256 uniformly without replacement and one stored pocket of each
uniformly (PCG64, seed 202610171427). Of these, 14 proteins and seven
pockets also occur among the first 1,000 stored pockets, the workload
of Section~\ref{sec:ann-motivation} and of the released-script run of
Section~\ref{sec:campaign}. The separation is by accession, with no
sequence-homology clustering, and all 26,562 pockets, these included,
were served in both campaigns.

 The reference is the exact FP32 top-500K
over the 500M rows with TF32 disabled, computed in four serial parts. It holds 128M reference pairs, and all 32.256B partial-result pairs were read back on the CPU. The four-GPU answers come from the re-run of
Section~\ref{sec:quality} ($B=64$) and the eight-GPU answers from the
window campaign of Section~\ref{app:grouped} ($B=256$). Neither ran
again, and no configuration changed. Against this reference, the
four-GPU re-run reaches 99.63\% mean and 99.32\% minimum recall and the
eight-GPU window campaign 99.63\% and 99.33\%. Their total hit counts
differ by 75, 127,529,741 against 127,529,816 of the 128M reference
pairs. The four-GPU re-run, with 75 fewer hits, is lower on 126
queries, equal on 16, and higher on 114, by at most 29 rows. The two runs differ in GPU count, batch size, and sample.

\begin{table}[t]
\caption{Recall (\%) of 256 pockets from 256 further proteins against
the exact FP32 top-500K over 500M rows (TF32 disabled), for the
four-GPU re-run and the eight-GPU window campaign: mean, minimum, 1st
and 5th percentiles, median, maximum, and the number of queries below
99.5\% (none is below 99\%).}
\label{tab:res-28}
\footnotesize
\setlength{\tabcolsep}{3pt}
\begin{tabular}{@{}lrrrrrrr@{}}
\toprule
Output & Mean & Min. & P1 & P5 & Median & Max. & ${<}99.5$\\
\midrule
Four GPUs, $B=64$ & 99.633 & 99.325 & 99.388 & 99.482 & 99.647 & 99.826 & 19\\
Eight GPUs, $B=256$ & 99.633 & 99.330 & 99.389 & 99.480 & 99.646 & 99.825 & 19\\
\bottomrule
\end{tabular}
\end{table}

\paragraph{Larger answers on eight GPUs: full record.}
Table~\ref{tab:res-36} is the full record of the nine runs behind Table~\ref{tab:largek-8gpu} of Section~\ref{app:grouped}. They use all 500M rows, 62.5M-row shards, $B=64$, $K=1$M, 2M, and 5M, the 64 held-out queries, and five timed repetitions per plan. Beyond the medians, it gives
the range of the five repetitions, memory in GiB with the allocated
increment above the library, and the lowest mean recall of any single
repetition. Resident time ends with the
synchronized complete selection and merge on the GPUs. Host time also
covers the copy of scores and IDs into independently retained host
arrays. Loading, setup and capture, warm-up, writing the files, the
check scans, and the CPU checks lie outside both endpoints. Memory is
the largest per-GPU PyTorch allocator value over setup and capture,
warm-up, and serving (lifecycle), library included. The library alone
takes 45.053~GiB on every GPU. Counters reset after capture see
only serving and understate the earlier allocation peak. No run raised
an overflow flag or fell back to $B=32$, and the plans were fixed in
advance with no automatic plan selection. An earlier attempt at
$K=1$M stopped because the run was interrupted, not for lack of
memory.

\begin{table*}[t]
\caption{Answer sizes of $K=1$M to 5M on 500M rows (eight PRO 6000
GPUs, $B=64$, 64 held-out queries), the full record of
Table~\ref{tab:largek-8gpu}. Resident, Host: medians with [min, max]
over five repetitions, in ms. Alloc., Res., Above: lifecycle allocated
and reserved peaks of the largest GPU, library included, and the
allocated increment above the library, in GiB. Mean, Low rep., Min.:
mean recall, lowest mean of a single repetition, and minimum per-query recall, in \%.}
\label{tab:res-36}
\footnotesize
\setlength{\tabcolsep}{2.5pt}
\begin{tabular}{@{}rlllrrrrrr@{}}
\toprule
$K$ & Plan & Resident & Host & Alloc. & Res. & Above & Mean & Low rep. & Min.\\
\midrule
1M & full & 116.72 [115.58, 123.62] & 467.35 [464.56, 488.10] & 53.815 & 56.215 & 8.762 & 99.6714 & 99.6714 & 99.3460\\
1M & window & 151.92 [151.62, 152.61] & 495.12 [492.55, 502.41] & 49.295 & 52.309 & 4.242 & 99.6716 & 99.6716 & 99.3463\\
1M & per-tile & 288.11 [286.73, 288.78] & 628.17 [626.70, 631.88] & 48.033 & 50.785 & 2.980 & 99.6714 & 99.6714 & 99.3460\\
2M & full & 152.50 [152.13, 156.65] & 832.89 [829.74, 853.51] & 55.126 & 59.906 & 10.073 & 99.7112 & 99.7112 & 99.4216\\
2M & window & 228.11 [227.37, 229.02] & 922.35 [917.37, 925.73] & 53.537 & 58.578 & 8.484 & 99.7112 & 99.7112 & 99.4204\\
2M & per-tile & 502.49 [501.98, 503.19] & 1,187.01 [1,186.32, 1,436.11] & 50.775 & 56.273 & 5.722 & 99.7112 & 99.7112 & 99.4216\\
5M & full & 258.53 [255.81, 310.73] & 1,976.88 [1,948.89, 2,520.41] & 59.060 & 71.020 & 14.007 & 99.7583 & 99.7583 & 99.5108\\
5M & window & 483.19 [480.73, 487.70] & 2,035.75 [1,986.26, 2,038.72] & 66.263 & 75.004 & 21.210 & 99.7583 & 99.7583 & 99.5106\\
5M & per-tile & 850.24 [847.03, 850.84] & 2,257.54 [2,244.51, 2,506.40] & 58.166 & 69.633 & 13.113 & 99.7583 & 99.7583 & 99.5108\\
\bottomrule
\end{tabular}
\end{table*}

\section{Other Embedding Models}\label{app:other-models}
This appendix covers the runs behind two conclusions of Section~\ref{sec:model-generalization}. The same plans serve ConGLUDe and ConPLex at more than 99.6\% mean recall on 100M and 250M rows, and the full-score plan is fastest at $K=500$K in every tested setting of these models. It first describes how those runs were made: the second executor, the model inputs, the populations and references, and the measurement. It then gives the survivor load that the window plan's buffers carry and the full record of the ConPLex runs on 500M rows for both query sets.

\paragraph{Second executor.}\phantomsection\label{app:matched-models}
The other two models run on a second executor that accepts any vector
dimension. Unlike the DrugCLIP executor of Section~\ref{sec:execution}, it fuses compaction into the scoring kernel and sends the cutoff between GPUs by peer-to-peer copy rather than through the host. It runs a captured merge on the root GPU and captures its recovery path with preallocated buffers. It also stores a 64-bit ID per row, 1~GB per
125M-row shard. The DrugCLIP campaign uses the original executor and
was not run with the second.

\paragraph{Model inputs.}
The model adapters keep each model's released preprocessing and
trained projections. ConGLUDe's protein input is the observed
structure, which can cover only a fragment of the target protein. Its
molecular vector is the 256-dimensional protein-screening output of
the released molecule encoder, normalized on its own. ConPLex applies
its released sequence preprocessing, which keeps the first 1,022 amino
acids of longer sequences, and then the trained target projection.
Both molecular encoders process the same molecule records. Zero outputs
are kept as produced.

\paragraph{Populations and references.}
The populations take source chunks of 100K records in a fixed
hash-based priority order. The first 1,000 chunks form the 100M
population and the first 2,500 the 250M population. The 500M
population for ConGLUDe extends the same order, and the one for
ConPLex also begins with the 100M population. For ConPLex on 500M
rows, all three plans and the FP32 reference read the same frozen
500M-row population. None of these populations has a row-wise
correspondence to the DrugCLIP library. Row IDs are one-based
source-corpus line numbers, and shards keep the chunk order.

ConGLUDe's 250M reference combines the exact top\nobreakdash-$K$ of
the 100M prefix with an exhaustive scan of the remaining 150M rows. The
ConPLex embeddings, references, and compressed libraries were produced
on NCSA Delta and copied to the main server and to the eight-GPU server.

\paragraph{Measurement and calibration queries.}
Each combination of model, scale, answer size, and plan runs in a
fresh process allocating only that plan's buffers, the window plan's
captured recovery path included. Five timed repetitions follow warmup.
Resident time follows Section~\ref{sec:setup}, with library loading,
capture, warmup, and host copies timed apart. Peak allocation is the
maximum per-GPU allocation that Torch records, graph pools included,
and no other memory is counted.

One ConPLex calibration query vector is all zeros, so every candidate
ties for it. The FP32 reference breaks these ties by ascending row ID,
while the output semantics accept any $K$ distinct IDs with the
correct scores. The query's strict-ID recall is therefore zero at both
sizes, although its answer is valid under score equivalence. We keep
the query in the calibration summary and apply the criterion of 99\%
mean recall, fixed in advance, to the 31 nonzero calibration queries
(ties at the $K$-th score still count). All 32 held-out queries are
nonzero and were not used to choose any configuration.

\paragraph{Survivor load.}
\begin{table}[t]
\caption{Survivor load of the other two models on held-out queries. Pass: mean fraction of main-scan rows above the cutoff. Peak: most survivors in any window ($G=16$) of any query and GPU, as \% of $A=1{,}000{,}000$. With the 32M-row sample, $K/N_0$ is 0.31\% at $K=100$K and 1.56\% at 500K (Section~\ref{sec:seed-capacity}).}
\label{tab:model-pressure}
\centering
\footnotesize
\renewcommand{\arraystretch}{0.9}
\begin{tabular}{@{}lrrrr@{}}
\toprule
 & \multicolumn{2}{c}{$K=100$K} & \multicolumn{2}{c}{$K=500$K}\\
\cmidrule(lr){2-3}\cmidrule(l){4-5}
Model / rows & Pass (\%) & Peak (\%) & Pass (\%) & Peak (\%)\\
\midrule
ConGLUDe / 100M & 0.28 & 6.56 & 1.43 & 30.07\\
ConGLUDe / 250M & 0.34 & 19.14 & 1.56 & 52.71\\
ConPLex / 100M & 0.30 & 7.51 & 1.53 & 34.12\\
ConPLex / 250M & 0.33 & 13.21 & 1.60 & 55.22\\
\bottomrule
\end{tabular}
\end{table}

On the held-out queries at 100M and 250M rows, no window of either model fills more than 55.22\% of its buffer (Table~\ref{tab:model-pressure}). A separate diagnostic run records each shard's cutoff, its total survivor count, and the most survivors in any window.
Table~\ref{tab:model-pressure} reports the last two for the held-out
queries. The diagnostic runs return the score multisets of the measured
runs, and no buffer overflowed in them.

\paragraph{ConPLex on 500M rows: both query sets.}
Section~\ref{sec:model-generalization} gives the held-out medians of
ConPLex on 500M rows. Table~\ref{tab:res-29} adds the ranges of the
five repetitions, and its caption gives the calibration medians. The runs use the eight-GPU server of Section~\ref{sec:setup} and the second executor, with 2M-row tiles, $r_0=4$, $G=16$, and $A=1{,}000{,}000$ on each of the eight 62.5M-row shards. They run at $B=32$ on the 32 calibration and 32 held-out queries of the 100M and 250M runs.
Each plan ran in its own process as described above. One output per
plan, query set, and $K$ was kept. All 115.2M returned pairs passed the
selection check on the GPU and an independent readback on the CPU. As at 100M and 250M, the all-zero calibration query has zero
strict-ID recall at both sizes. The calibration means over the other
31 queries are 99.63\% and 99.69\% for every plan, and the means over
all 32 are 96.51\% and 96.57\%. Host times vary much more than
resident times across repetitions, and at $K=100$K the host ranges of
the full-score and window plans overlap.

\begin{table}[t]
\caption{ConPLex on 500M rows with the second executor (eight PRO 6000
GPUs, $B=32$). Resident, Host: median (range) of five repetitions per
32-query batch, in ms. Mean, Min.: recall (\%) of the 32 held-out
queries. GB: peak Torch allocation per GPU.
Calibration resident medians of the per-tile, full-score, and window
plans: 96.9, 68.2, and 68.1~ms at $K=100$K and 203.9, 109.1, and
160.7~ms at 500K.}
\label{tab:res-29}
\footnotesize
\setlength{\tabcolsep}{2.5pt}
\begin{tabular}{@{}llrrrrr@{}}
\toprule
$K$ & Plan & Resident & Host & Mean & Min. & GB\\
\midrule
100K & Per-tile & 97.2 (97.1--97.3) & 124.9 (105.9--138.2) & 99.607 & 98.903 & 66.5\\
100K & Full score & 68.3 (68.3--68.4) & 89.2 (76.4--109.3) & 99.607 & 98.903 & 69.6\\
100K & Window & 68.2 (68.1--68.4) & 84.3 (75.9--125.5) & 99.608 & 98.908 & 66.9\\
\midrule
500K & Per-tile & 203.9 (203.5--204.0) & 413.6 (375.3--425.4) & 99.676 & 99.292 & 69.2\\
500K & Full score & 109.2 (108.9--109.3) & 328.4 (286.5--358.5) & 99.676 & 99.292 & 72.0\\
500K & Window & 160.9 (160.6--161.1) & 373.3 (334.8--409.4) & 99.676 & 99.290 & 70.3\\
\bottomrule
\end{tabular}
\end{table}

\vfuzz=2pt\balance
\end{document}